\documentclass[a4paper,fleqn]{cas-sc}
\usepackage[numbers]{natbib}
\usepackage{color}
\usepackage{natbib}
\usepackage{graphicx}
\usepackage{amsmath}
\usepackage{amssymb}
\usepackage[super]{nth}

\begin{document}
\shorttitle{ISMC in 2D}
\shortauthors{Steinberg \& Heizler}
\title [mode = title]{Multi-Frequency Implicit Semi-analog Monte-Carlo (ISMC) Radiative Transfer Solver in Two-Dimensions (without Teleportation)}
\author[1]{Elad Steinberg}
\ead{elad.steinberg@mail.huji.ac.il}
\address[1]{Racah Institute of Physics, The Hebrew University, 9190401 Jerusalem, Israel}
\author[1]{Shay I. Heizler}[orcid=0000-0002-9334-5993]
\ead{shay.heizler@mail.huji.ac.il}
\begin{keywords}
Radiative Transfer \sep Boltzmann Equation \sep Monte-Carlo schemes
\end{keywords}
\ExplSyntaxOn
\keys_set:nn { stm / mktitle } { nologo }
\ExplSyntaxOff
\begin{abstract}
    We study the multi-dimensional radiative transfer phenomena using the ISMC scheme, in both gray and multi-frequency problems. Implicit Monte-Carlo (IMC) schemes have been in use for five decades. The basic algorithm yields teleportation errors, where photons propagate faster than the correct heat front velocity. Recently [Poëtte and Valentin, J. Comp. Phys., 412, 109405 (2020)], a new implicit scheme based on the semi-analog scheme was presented and tested in several one-dimensional gray problems. In this scheme, the material energy of the cell is carried by material-particles, and the photons are produced only from existing material particles. As a result, the teleportation errors vanish, due to the infinite discrete spatial accuracy of the scheme. We examine the validity of the new scheme in two-dimensional problems, both in Cartesian and Cylindrical geometries. Additionally, we introduce an expansion of the new scheme for multi-frequency problems. We show that the ISMC scheme presents excellent results without teleportation errors in a large number of benchmarks, especially against the slow classic IMC convergence.
\end{abstract}
\maketitle
\section{Introduction}

The physics of radiative transfer plays an important role in many physical phenomena, such as modeling the radiation through astrophysical objects (e.g. supernova), or in high-energy density physics laboratory astrophysics~\cite{castor2004}. Specifically, radiative heat (Marshak) waves that propagate in media that can range from optically thick to ultra-thin regimes~\cite{marshak,ps,hammerrosen}. Each limit creates its different computational difficulties for getting an accurate enough solution of the radiation profile. 

The equation that governs the physics of radiative transfer is the radiative transport equation (RTE), which is the Boltzmann equation for X-ray photons~\cite{pomraning1973,castor2004}. This equation is an integro-differential equation and the specific intensity is a function of time, space, direction and frequency, and thus, achieving an accurate solution in the general case is complicated. This is the reason that simpler approximations have been used for many years instead of the full RTE solution, where the most well-known is the local-thermodynamic-equilibrium (LTE) diffusion approximation (also called the Eddington approximation)~\cite{pomraning1973}. However, the diffusion assumption forces the particle's distribution to be isotropic (or close to isotropic). In many physical scenarios, when far from LTE, this assumption breaks. For this reason, massive attempts have been made to solve the exact RTE physics using several numerical approximations. 

The most well-known approximations are: (1) The spherical harmonics approximation (or the $P_N$ approximation), where the specific intensity is derived using spherical harmonic sets of coupled moment equations~\cite{pomraning1973}, (2) The discrete ordinates method (the $S_N$ method), where the specific intensity is solved in finite discrete directions~\cite{pomraning1973}. Both of these approximations are deterministic methods (i.e., solve the RTE numerically), and tend to the exact Boltzmann solution when the number of the moments/directions goes to infinity. A different approach that yields the exact Boltzmann behavior is through a statistical Monte-Carlo algorithm, that calculates the physical events of the photons explicitly, such as absorption, advection, scattering and black-body emission. The most well-known algorithm in this family of solutions, is the implicit-Monte Carlo (IMC) scheme, proposed by~\citet{IMC}. We note that IMC is exact only when the physical quantities such as the opacity and the heat capacity are temperature-free. In those cases, there is no actual need for an implicit algorithm. In real world physical cases, where the opacity and the heat-capacity are temperature-dependent, the Monte-Carlo algorithm is not really ``exact", and converges to the exact solution only when $\Delta t \to 0$ and $\Delta x\to 0$.

In the IMC algorithm, the implicitization of the black-body emission term enables the use of relatively large time-steps. The implicitization replaces the physical process of absorption and then self-emission, with an ``effective'' scattering term, yielding faster convergence. Although there have been many modifications addressing several IMC issues during the years~\cite{Gentile2001,Densmore2011,teleportation,teleportation2,Gentile2010,Gentile2011,Gentile2012,Gentile2014,MCCLARREN_URBATSCH,shi_2020_1,shi_2020_3,shi_2019,gael2} (and in the review~\citet{four_decades}), the basic algorithm remained the same, due to the simple logic that stands behind a Monte-Carlo simulation of particles.

Although it has great popularity, IMC suffers from some well-known problems~\cite{teleportation,teleportation2,ISMC}. One of the most well-known problem is {\em{teleportation}}, in sufficiently opaque materials. This problem is caused due to the finite discretization of both the spatial mesh and the time-step. Suppose that the photons arrive to a cold, and thus opaque material, they are absorbed immediately in the outer part of the cell. However, in the next step, the warmed cell emits from all of the regions of the cell (uniformly, as a first simple choice), though physically, the heat wave should not reach the end of the cell. This causes the heat wave to propagate faster than in reality, i.e. {\em{``teleportates''}}. One very bothersome issue is that the amount of {\em{teleportation}} increases at small time-steps, i.e. smaller time-step (at a fixed spatial resolution) lead to a greater teleportation error \citep{ISMC}, and this creates a hard convergence-check issue.

Many solutions have been offered to deal with this problem, most of them use tilts, i.e. some spatial estimator for the profile of the material temperature, based on spatial gradients for determining the distribution of the black-body emitted particles~\cite{Densmore2011,teleportation,teleportation2,gael2,shi_2020_3}. The intrinsic noise in Monte-Carlo methods, causes numerical difficulties in their implementation. A different approach was proposed by~\citet{SMC} (which is called semi-analog Monte-Carlo - SMC). The basic idea in SMC is to use two kinds of ``particles'', radiation particles (like in IMC) and material-particles (which cannot propagate by themselves), that ``carry'' the energy of the material. A material particle can transform to radiation (due to black-body radiation), while the radiation particle transforms to a material particle when absorbed. In this way, emission is determined by the exact spatial distribution of the material energy. However, this method is an explicit scheme and thus enforces small time-steps.
Recently,~\citet{ISMC} published an important milestone in Monte-Carlo radiative transfer modeling, called implicit semi-analog Monte-Carlo (ISMC). In their work, they took the basic idea of SMC, and derived an implicitization of it. In this manner, we can gain from large time-steps like IMC (and thus, reasonable simulation times), while avoiding teleportation errors like in SMC. \citet{ISMC} showed in simple one-dimensional gray problems that the new method does not suffer from teleportation both in spatial resolution and small time-steps.

In this work, we expand the check to two-dimensions, both in XY and RZ geometries, both in gray and multi-frequency (multi-group) benchmarks. In addition, we do a full check, both in 1D and 2D, checking the strength of ISMC in both optically thin, optically thick and combined problems. Specifically, we are interested in seeing whether ISMC succeeds to avoid the teleportation errors of IMC, in 2D, with regard to both spatial resolution and time-step, since convergence is hard to achieve in multi-dimensions problems. 
We run a large amount of well-known benchmark problems to validate the ISMC algorithm:
\begin{enumerate}
   \item One-dimensional problems:
   \begin{enumerate}
     \item Gray problems:
     \begin{itemize}
       \item Su \& Olson (1997)~\cite{suol97}.
       \item Marshak-wave (2016)~\cite{teleportation,ISMC}.
       \item Olson (2019)~\cite{olson2019}.
     \end{itemize}
     \item Multi-frequency (multi-group problems):
      \begin{itemize}
       \item Olson (2020)~\cite{olson2020}.
       \item Densmore et al. (2012)~\cite{Densmore,teleportation2}.
     \end{itemize}
   \end{enumerate}
   \item Two-dimensional problems:
   \begin{enumerate}
     \item Gray problems:
     \begin{itemize}
       \item McClarren \& Hauck XY hohlraum (2010)~\cite{MCCLARREN,MCCLARREN2}.
       \item McClarren \& Urbatsch RZ hohlraum (2009)~\cite{MCCLARREN_URBATSCH,shi_2020_1,shi_2020_3,shi_2019}.
       \item Graziani's ``crooked-pipe'' problem (2000)~\cite{Gentile2001,teleportation}.
     \end{itemize}
     \item Multi-frequency (multi-group problems):
      \begin{itemize}
       \item Olson (2020)~\cite{olson2020}.
     \end{itemize}
   \end{enumerate}
\end{enumerate}
Large teleportation errors are seen in the problems that include opaque regions when we use the IMC implementation, as opposed to the ISMC method, which is teleportation free, both in spatial resolution and small time-steps. 

First, we introduce a brief summary of the ISMC algorithm in \S~\ref{sec:ISMC}. Afterwards in \S~\ref{1d_tests} and in \S~\ref{2d_tests} we present the benchmark results of the one-dimensional and the two-dimensional test, respectively. We finish with a short discussion in \S~\ref{discussion}.

\section{Brief Summary of the ISMC Method}
\label{sec:ISMC}
In this section we introduce briefly the SMC linearization in the ISMC algorithm, as taken from~\cite{ISMC}. First we introduce the gray case, and then we present our generalization to the multi-frequency case.

\subsection{Gray MC}
\label{sec:ISMC_gray}
The frequency independent (gray) radiative transfer equation (RTE), and the coupled energy-balance for the material energy are:
\begin{subequations}
\label{eq:rad_transfer}
\begin{eqnarray}
    \frac{1}{c}\frac{\partial I(\boldsymbol{r},\boldsymbol{\Omega},t)}{\partial t}+\boldsymbol{\Omega\cdot\nabla} I(\boldsymbol{r},\boldsymbol{\Omega},t)+(\sigma_a+\sigma_s)I(\boldsymbol{r},\boldsymbol{\Omega},t)&=&\sigma_a B(T)+\sigma_s\int_{4\pi}\frac{I(\boldsymbol{r},\boldsymbol{\Omega}',t)}{4\pi}\boldsymbol{d\Omega}'\\
    \frac{\partial e(T)}{\partial t}&=&c \sigma_a\left(\int_{4\pi}I(\boldsymbol{r},\boldsymbol{\Omega}',t)\boldsymbol{d\Omega}'-4\pi B(T)\right)
    \end{eqnarray}
\end{subequations}
where $I(\boldsymbol{r},\boldsymbol{\Omega},t)$ is the radiation intensity for a unit space $\boldsymbol{r}$ and unit direction $\boldsymbol{\Omega}$ at time $t$, $\sigma_a$ is the absorption cross-section (opacity), $\sigma_s$ is the scattering cross-section, $e(T)$ is the thermal energy per unit volume of the medium, $T$ is the material temperature, $B(T)=aT^4/4\pi$ is the frequency integrated Planck function, $a$ is the radiation constant and $c$ is the speed of light. 

In order to derive the semi-analog Monte Carlo (SMC) equations~\citep{SMC}, we define $\eta(e)\equiv 4\pi B(e)/e$, and rewrite Eqs.~\ref{eq:rad_transfer} as:
\begin{subequations}
\label{eq:smc}
\begin{eqnarray}
    \frac{1}{c}\frac{\partial I(\boldsymbol{r},\boldsymbol{\Omega},t)}{\partial t}+\boldsymbol{\Omega\cdot\nabla} I(\boldsymbol{r},\boldsymbol{\Omega},t)+(\sigma_a+\sigma_s)I&=&\frac{\sigma_a \eta(e) e}{4\pi}+\sigma_s\int_{4\pi}\frac{I(\boldsymbol{r},\boldsymbol{\Omega}',t)}{4\pi}\boldsymbol{d\Omega}'\\
    \frac{\partial e(T)}{\partial t}&=&c \sigma_a\left(\int_{4\pi}I(\boldsymbol{r},\boldsymbol{\Omega}',t)\boldsymbol{d\Omega}'-\eta(e) e\right).
    \end{eqnarray}
\end{subequations}
Eqs.~\ref{eq:smc} suggest that $I$ and $e$ can both be evolved in a MC method, since the equations are linear in $I$ and $e$, assuming an explicit value for $\eta(e)$. In the semi-analog MC method, particles can either be radiation photons or material particles. Radiation photons are transported with a velocity $c$, and material particles are stationary.~A radiation photon has a chance to be either scattered and change its direction, or be absorbed and turn into a material particle.~A material particle has a chance to transform into a radiation photon, i.e. material energy can propagate {\em{only}} by radiation. 

These equations discretize both the radiation field and the material, where the latter property is crucial to prevent teleportation errors. The discrete nature of the material particles generates a so-called infinite spatial resolution of the heat front location, due to the discrete positions of the particles. The energy or temperature of a numerical cell determines only the opacities and the heat capacities. Unfortunately, the explicit nature of the equations imposes a severe restriction on the time step in order to obtain stability~\cite{SMC_stable}.

Recently,~\citet{ISMC} have presented a new implicitization of the SMC equations, that increases dramatically the stability. They suggest to replace the explicit value of $\eta$ with its estimated value at the end of the time step, $\eta^{n+1}$. We present here a short derivation of the ISMC discretization, for the convenience of the experienced IMC reader, where we follow the equivalent steps of the classic derivation of~\citet{IMC}. We note that in ISMC, the implicitization is on $\eta$ where in IMC it is on the equilibrium radiation energy density, $u_r\equiv aT^4$.~\citet{IMC} have derived the IMC scheme, where $u_r$ was centered in time, $u_r^{\gamma}=\alpha u_r^{n+1}+(1-\alpha) u_r^n$. We focus here in the most stable state, i.e. fully implicit scheme ($\alpha=1$), thus we use the notation $\eta^{n+1}$, instead of $\eta^\gamma$. The new equations now read:
\begin{subequations}
\begin{eqnarray}
    \frac{1}{c}\frac{\partial I(\boldsymbol{r},\boldsymbol{\Omega},t)}{\partial t}+\boldsymbol{\Omega\cdot\nabla} I(\boldsymbol{r},\boldsymbol{\Omega},t)+(\sigma_a+\sigma_s)I(\boldsymbol{r},\boldsymbol{\Omega},t)&=&\frac{\sigma_a \eta^{n+1}(e) e}{4\pi}+\sigma_s\int_{4\pi}\frac{I(\boldsymbol{r},\boldsymbol{\Omega}',t)}{4\pi}\boldsymbol{d\Omega}'\\
    \frac{\partial e(T)}{\partial t}&=&c \sigma_a\left(\int_{4\pi}I(\boldsymbol{r},\boldsymbol{\Omega}',t)\boldsymbol{d\Omega}'-\eta^{n+1}(e) e\right).
    \end{eqnarray}\label{eq:ismc_temp}
\end{subequations}
In order to estimate $\eta^{n+1}$, we write down its time derivative:
\begin{equation}
\frac{\partial\eta}{\partial t}=\frac{4\pi}{e}\frac{\partial B}{\partial e}\frac{\partial e}{\partial t}-\frac{4\pi B}{e^2}\frac{\partial e}{\partial t}.\label{eq:eta_dt}
\end{equation}
Defining $\beta\equiv\frac{\partial aT^4}{\partial e}$ as the ratio between radiation and material heat capacities (the same as the definition in \citet{IMC}) and $\zeta\equiv\beta-\eta$, we rewrite Eq.~\ref{eq:eta_dt}:
\begin{equation}
\label{eta_eq}
\frac{\partial\eta}{\partial t}=\zeta\frac{1}{e}\frac{\partial e}{\partial t}=\zeta c\sigma_a\left(\frac{1}{e}\int_{4\pi}I(\boldsymbol{r},\boldsymbol{\Omega}',t)\boldsymbol{d\Omega}'-\eta^{n+1}\right),
\end{equation}
where in the last transition we used Eq.~\ref{eq:ismc_temp}b. Equation ~\ref{eta_eq} is the equivalent (in the implicitization procedure) of eq. 1.4b in~\citet{IMC}.

Discretizing the time step yields:
\begin{equation}
    \eta^{n+1}-\eta^n=\Delta t\zeta c \sigma_a\left(\frac{1}{e}\int_{4\pi}I(\boldsymbol{r},\boldsymbol{\Omega}',t)\boldsymbol{d\Omega}'-\eta^{n+1}\right),
\end{equation}
which is the ISMC equivalent to the radiation source-free eq. 1.7b in~\citet{IMC}, with $\zeta$ that replaces $\beta$ and $\eta$ that replaces $u_r$ (again, assuming $\alpha=1$). However, it is important to note that as in the classic IMC implementation, we use the values of $\zeta$, and $\sigma_a$ from the beginning of the time step. A simple manipulation gives the estimate of $\eta^{n+1}$ at the end of the time step:
\begin{equation}
\label{eq:eta_n1}
    \eta^{n+1}=\eta^{n}\chi+(1-\chi)\frac{1}{e}\int I(\boldsymbol{r},\boldsymbol{\Omega}',t)\boldsymbol{d\Omega}',
\end{equation}
where $\chi\equiv\frac{1}{1+\zeta^{n}c\sigma_{a}\Delta t}$, in analogy to the fleck factor in IMC (with $\zeta^{n}$ that replaces $\beta^{n}$ in IMC)~\citep{IMC}. Equation ~\ref{eq:eta_n1} is the ISMC equivalent to the source-free eq. 1.15b in~\citet{IMC} (again, assuming $\alpha=1$).

Plugging Eq.~\ref{eq:eta_n1} into Eq.~\ref{eq:ismc_temp}, we have that the linear (in $I$ and $e$) implicit-SMC (ISMC) equations are:
\begin{subequations}
\begin{eqnarray}
    \frac{1}{c}\frac{\partial I(\boldsymbol{r},\boldsymbol{\Omega},t)}{\partial t}+\boldsymbol{\Omega\cdot\nabla} I(\boldsymbol{r},\boldsymbol{\Omega},t)+(\sigma_a+\sigma_s)I(\boldsymbol{r},\boldsymbol{\Omega},t)&=&\frac{\sigma_a\eta^n\chi e}{4\pi}+\left(\sigma_s+\left(1-\chi\right)\sigma_a\right)\int_{4\pi}\frac{I(\boldsymbol{r},\boldsymbol{\Omega}',t)}{4\pi}\boldsymbol{d\Omega}'\\
    \frac{\partial e(T)}{\partial t}&=&c \sigma_a\chi\left(\int_{4\pi}I(\boldsymbol{r},\boldsymbol{\Omega}',t)\boldsymbol{d\Omega}'-\eta^{n}  e\right),
    \end{eqnarray}\label{eq:ismc}
\end{subequations}
with $(1-\chi)\sigma_a$ as an ISMC equivalent to the ``effective'' scattering term, that results from the implicit MC scheme.
The main points of a time advancement Monte-Carlo schemes are as follows:
\begin{itemize}
    \item At the beginning of the time step the temperature in a cell is calculated from $e$, which is equal to the sum of energies from all the material particles in a given cell.
    \item The values for $\eta$ and $\chi$ are calculated using the opacities and $\beta$ (heat capacities) at the beginning of the time step.
    \item If there are external radiation sources/boundary conditions, new radiation photons are created.
    \item Radiation photons are transported with a velocity $c$ and with a chance per unit time of $c(\sigma_a+\sigma_s)$ a collision occurs, which distributes exponentially. Once there is a collision, it is scattered with a probability of $(\sigma_s+(1-\chi)\sigma_a)/(\sigma_a+\sigma_s)$ and it is absorbed and transformed into a material particle with a chance $\sigma_a\chi/(\sigma_a+\sigma_s)$.
    \item A material particle is held stationary and with a chance per unit time of $c\sigma_a\chi\eta$, it changes into a radiation photon.
\end{itemize}
\subsection{Multi-frequency}
\label{sec:ISMC_freq}
We now bring our attention to expand the ISMC method to include frequency dependence. For clarity we show Eq.~\ref{eq:ismc_temp} once more with frequency dependence assuming only elastic scattering~\cite{pomraning1973}:
\begin{subequations}
\begin{align}
    \frac{1}{c}\frac{\partial I_\nu(\boldsymbol{r},\boldsymbol{\Omega},t)}{\partial t}+\boldsymbol{\Omega\cdot\nabla} I_\nu(\boldsymbol{r},\boldsymbol{\Omega},t)&=
    \begin{aligned}[t]
    &-(\sigma_{a,\nu}+\sigma_{s,\nu})I_\nu(\boldsymbol{r},\boldsymbol{\Omega},t)+\frac{\sigma_{a,\nu}b_\nu \eta^{n+1} e}{4\pi}\\
    &+\sigma_{s,\nu}\int_{4\pi}\frac{I_\nu(\boldsymbol{r},\boldsymbol{\Omega}',t)}{4\pi}\boldsymbol{d\Omega}'\\
    \end{aligned}\\
    \frac{\partial e(T)}{\partial t}&=c \int_0^{\infty}\left(\sigma_{a,\nu'}\left(\int_{4\pi}I_{\nu'}(\boldsymbol{r},\boldsymbol{\Omega}',t)\boldsymbol{d\Omega}'-\eta^{n+1}b_{\nu'} e\right)d\nu'\right),
    \end{align}\label{eq:ismc_freq_temp}
\end{subequations}
where $\nu$ denotes per unit frequency and $b_{\nu}\equiv B_\nu/B=4\pi B_\nu/aT^4$ where $B_\nu=\frac{2h\nu^3}{c^2}\left(\exp{(h\nu/k_BT)}-1\right)^{-1}$ is the Planck function ($h$ is the Planck constant and $k_B$ is the Boltzmann constant).

Following similar steps as in \S \ref{sec:ISMC_gray}, we yield:
\begin{equation}
    \eta^{n+1}=\eta^{n}\chi+(1-\chi)\frac{1}{e}\int_0^{\infty}\int_{4\pi}\sigma_{a,\nu'}I_{\nu'}(\boldsymbol{r},\boldsymbol{\Omega}',t)\boldsymbol{d\Omega}'d\nu'
\end{equation}
where now $\chi\equiv\frac{1}{1+\zeta^{n}c\sigma_{a,p}\Delta t}$ and
the Planck opacity is defined as $\sigma_{a,p}\equiv\int_0^{\infty}\sigma_{a,\nu}b_\nu d\nu$.

The final frequency dependent ISMC equations are:
\begin{subequations}
\begin{align}
    \frac{1}{c}\frac{\partial I_\nu(\boldsymbol{r},\boldsymbol{\Omega},t)}{\partial t}+\boldsymbol{\Omega\cdot\nabla} I_\nu(\boldsymbol{r},\boldsymbol{\Omega},t)&=
    \begin{aligned}[t]
    &-(\sigma_{a,\nu}+\sigma_{s,\nu})I_\nu(\boldsymbol{r},\boldsymbol{\Omega},t)+\frac{\sigma_{a,\nu}b_\nu\eta^n\chi e}{4\pi}\\
    &+\sigma_{s,\nu}\int_{4\pi}\frac{I_\nu(\boldsymbol{r},\boldsymbol{\Omega}',t)}{4\pi}\boldsymbol{d\Omega}'+\left(1-\chi\right)\frac{\sigma_{a,\nu}b_\nu}{\sigma_{a,p}}\int_0^{\infty}\int_{4\pi}\frac{I_
    {\nu'}(\boldsymbol{r},\boldsymbol{\Omega}',t)\sigma_{a,\nu'}}{4\pi}\boldsymbol{d\Omega}'d\nu'\\
    \end{aligned}\\
    \frac{\partial e(T)}{\partial t}&=c \chi\left(\int_0^{\infty}\int_{4\pi}\sigma_{a,\nu'}I_\nu'(\boldsymbol{r},\boldsymbol{\Omega}',t)\boldsymbol{d\Omega}'d\nu'-\sigma_{a,p}\eta^{n}  e\right).
    \end{align}\label{eq:ismc_freq}
\end{subequations}
We see that evolving the frequency dependent ISMC equations is similar to the frequency dependent IMC equations. When a material particle is evolved into a radiation photon, its new frequency is selected in the same manner as in the emission case in IMC.

The main points of a time advancement Monte-Carlo schemes are as follows:
\begin{itemize}
    \item At the beginning of the time step the temperature in a cell is calculated from $e$, which is equal to the sum of energies from all the material particles in a given cell.
    \item The values for $\eta$ and $\chi$ are calculated using the values at the beginning of the time step.
    \item If there are external radiation sources/boundary conditions, new radiation photons are created with their corresponding frequency distribution.
    \item Radiation photons are transported with a velocity $c$ and with a chance per unit time of $c(\sigma_{a,\nu}+\sigma_{s,\nu})$ a collision occurs. Once there is a collision, it is scattered with a probability of $(\sigma_{s,\nu}+(1-\chi)\sigma_{a,\nu})/(\sigma_{a,\nu}+\sigma_{s,\nu})$ and it is absorbed and transformed into a material particle with a chance $\sigma_{a,\nu}\chi/(\sigma_{a,\nu}+\sigma_{s,\nu})$. If there is a scattering event, with a probability $\sigma_{s,\nu}/(\sigma_{s,\nu}+(1-\chi)\sigma_{a,\nu})$ it is an elastic scattering and only a new angle is sampled, else also a new frequency is sampled, as explained below.
    \item A material particle is held stationary and with a chance per unit time of $c\sigma_{a,p}\chi\eta$ it changes into a radiation photon with a frequency sampled from the distribution probability $\int_0^{\infty} \sigma_{a,\nu}b_\nu d\nu/\sigma_{a,p}$ (due to the cell temperature and Plank distribution), this is also the distribution for sampling a new frequency in an ``effective'' scattering event.
\end{itemize}
\section{1D Tests}
\citet{ISMC} have tested the ISMC algorithm in 1D Marshak waves problem, which are relatively optically thick~\cite{teleportation}. In this section we present several 1D problems that include some well-known benchmarks, checking the validity of the ISMC in both optically thin and thick scenarios. In specific, we present the multi-group benchmarks that was offered lately by~\citet{olson2020} and by Densmore et al.~\cite{Densmore,teleportation2}.
\label{1d_tests}
\subsection{Su \& Olson 1997}
In~\citet{suol97}, the authors presented a non-equilibrium gray transport test, which has a semi-analytical solution. In this test, the heat capacity of the material is set to be a radiation-like power-law dependency, $C_V = \alpha T^3$ (where $\alpha=4a/\epsilon$ and $\epsilon$ is set to be unity since the dimensionless time is scaled like $\epsilon c \sigma_{t} t$), and the opacity is set to unity. With this choice, the diffusion approximation of this problem yields a linear diffusion equation. The choice of a linear opacity enables a solution with the explicit SMC algorithm of~\citet{SMC}. 
\begin{figure}
(a)\includegraphics*[width=7.5cm]{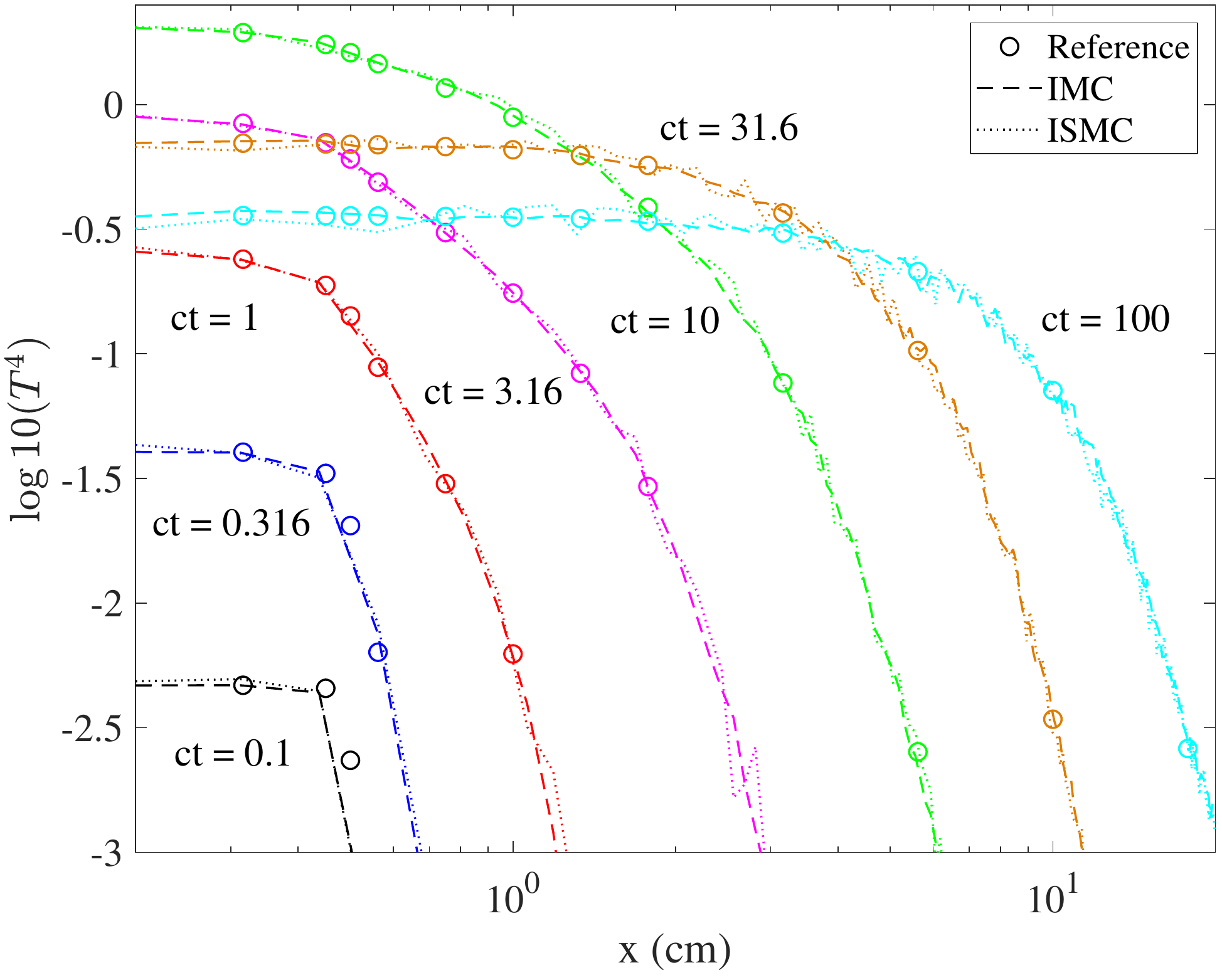}
(b)\includegraphics*[width=7.5cm]{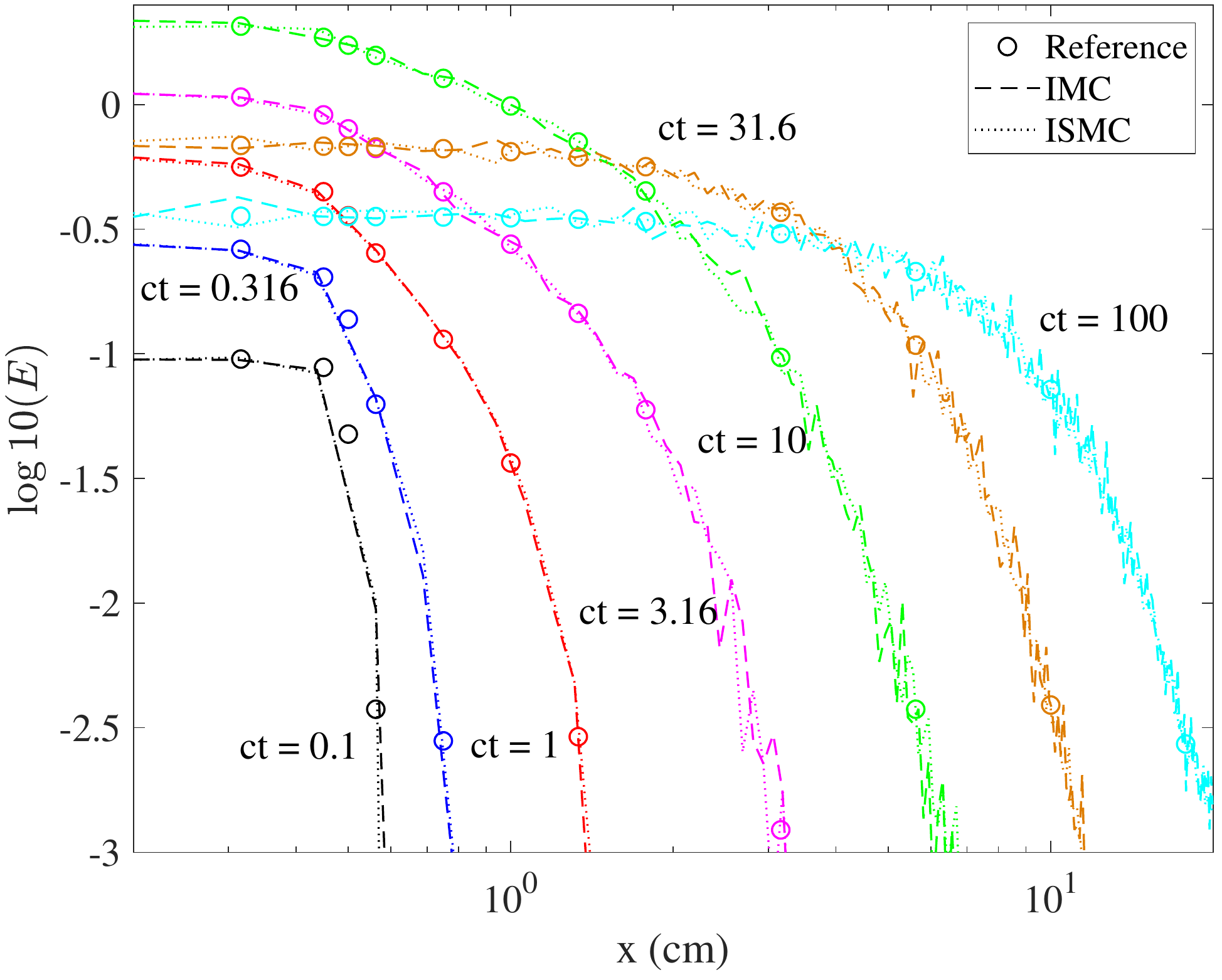}
\caption{(a) The material energy at different times for the Su \& Olson test problem. (b) The radiation energy density at different times for the Su \& Olson test problem.}
\label{fig:su_olson}
\end{figure}

A radiation source term $Q=a$ is applied in the region $0\leqslant x \leqslant 0.5$ until a dimensionless time of 10, with a reflective boundary condition at $x=0$. In early times, this problem represents an optically thin problem, that tends to be more opaque as the heat wave propagates in later times. We took the case of fully absorbing medium (the case of scattering-included medium yields qualitatively similar results). For both IMC and the ISMC we use a spatial resolution of 160 equally spaced cells, create $10^3$ new particles each time step, limit the total number of particles to be $2\cdot10^4$ and set a constant dimensionless time step of $c\Delta t = 10^{-14}$. 

In Fig.~\ref{fig:su_olson} we present the material (a) and radiation (b) energies for different times compared with the analytical reference solution. For all times, both methods give a satisfactory result. However, it can be noticed that as the time progress, the number of particles per cell goes down and the ISMC exhibits a noisier results in the material energy than IMC due to the discreteness of its material field. The radiation energy densities in both methods have comparable level of noise, since both methods have a discretized radiation field and a comparable number of photons.

\subsection{Marshak Wave}
We repeat the Marshak wave problem presented in~\citet{ISMC}, emphasizing the teleportation error that occurs in IMC, and its solution using the ISMC algorithm. This problem is relatively opaque and IMC codes exhibit teleportation error in it, if they are run with low spatial resolution. We test our implementation for both IMC and ISMC codes comparing to~\cite{ISMC}, as a preparation for 2D.
\begin{figure}
(a)\includegraphics*[width=7.5cm]{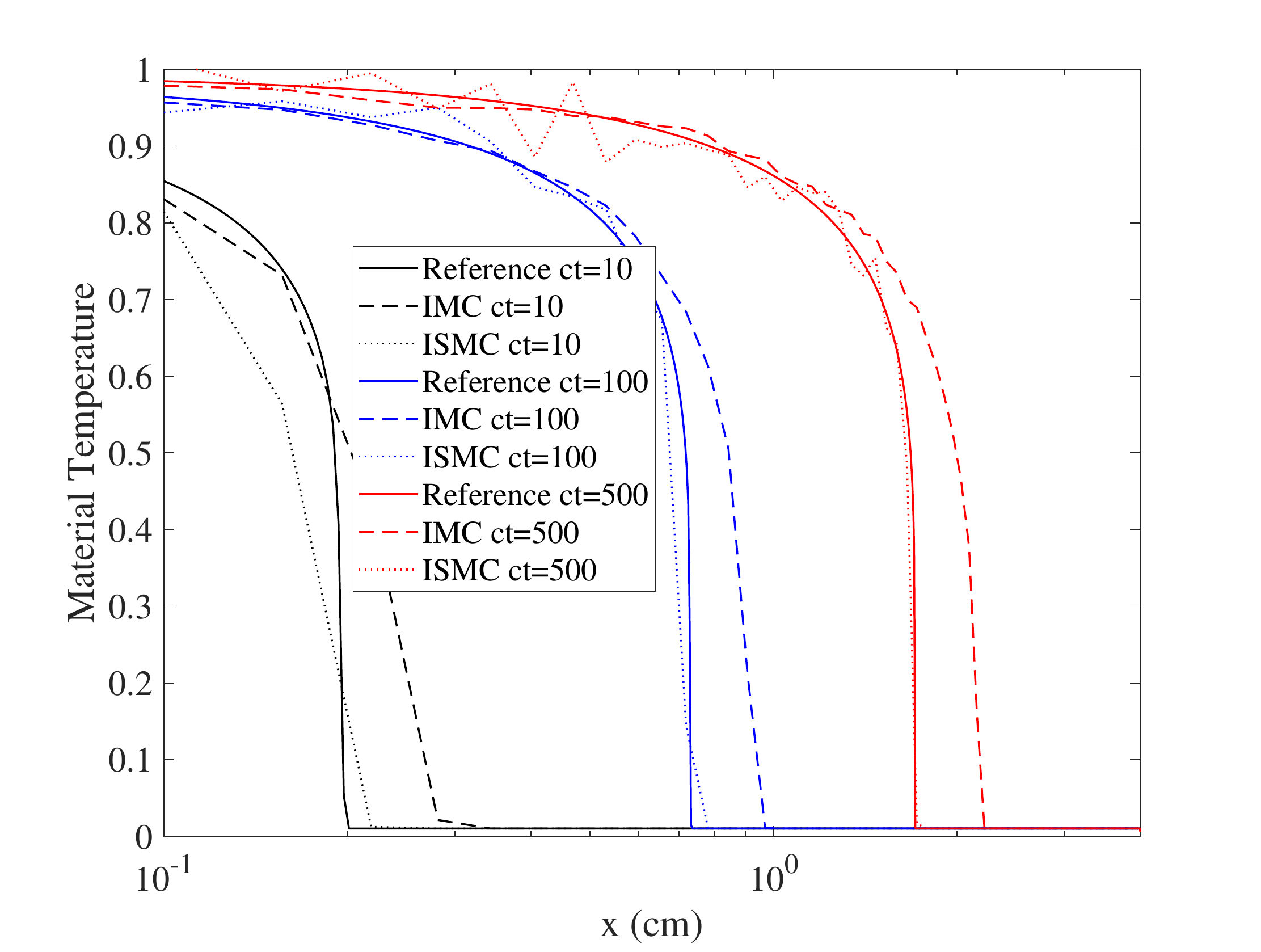}
(b)\includegraphics*[width=7.5cm]{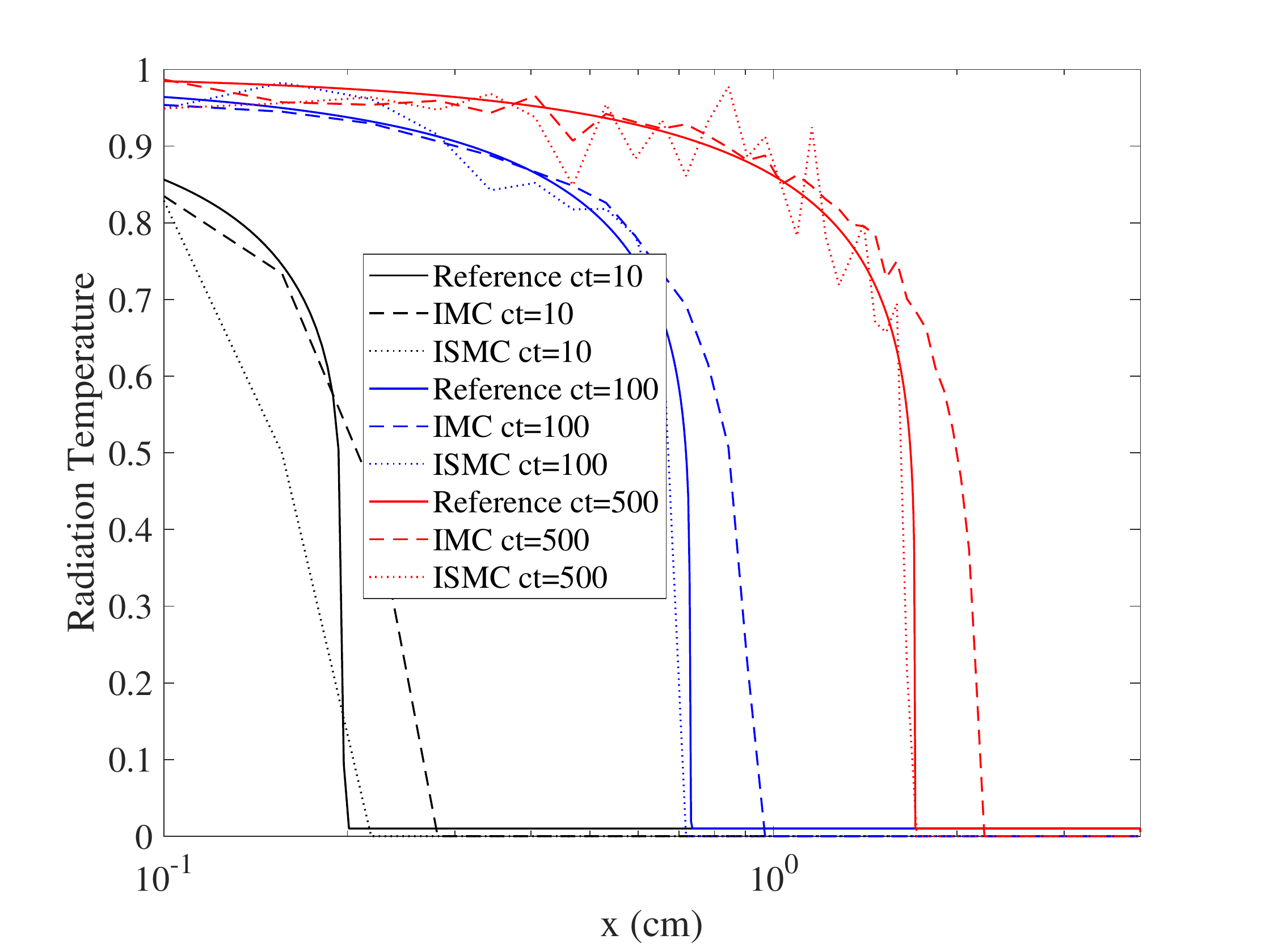}
\caption{(a) The material temperature at different times for the Marshak wave test problem. (b) The radiation energy density at different times for the Marshak wave test problem.}
\label{fig:gold}
\end{figure}

In this problem, a black body heats the left boundary of the domain with a normalized temperature of unity, the opacity is set to be $\sigma_a = 10\cdot T^{-3}$ and the heat capacity is $C_V=7.14a$. The initial temperature
is set to be $T(t=0)=0.01$ and we run the simulation until a time $ct=500$. For both IMC and the ISMC we use {\em{low}} spatial resolution of 64 equally spaced cells, create $2\cdot10^3$ new particles each time step, limit the total number of particles to be $2\cdot10^4$ and set a constant time step of $c\Delta t = 0.03$. 
\begin{figure}
(a)\includegraphics*[width=7.5cm]{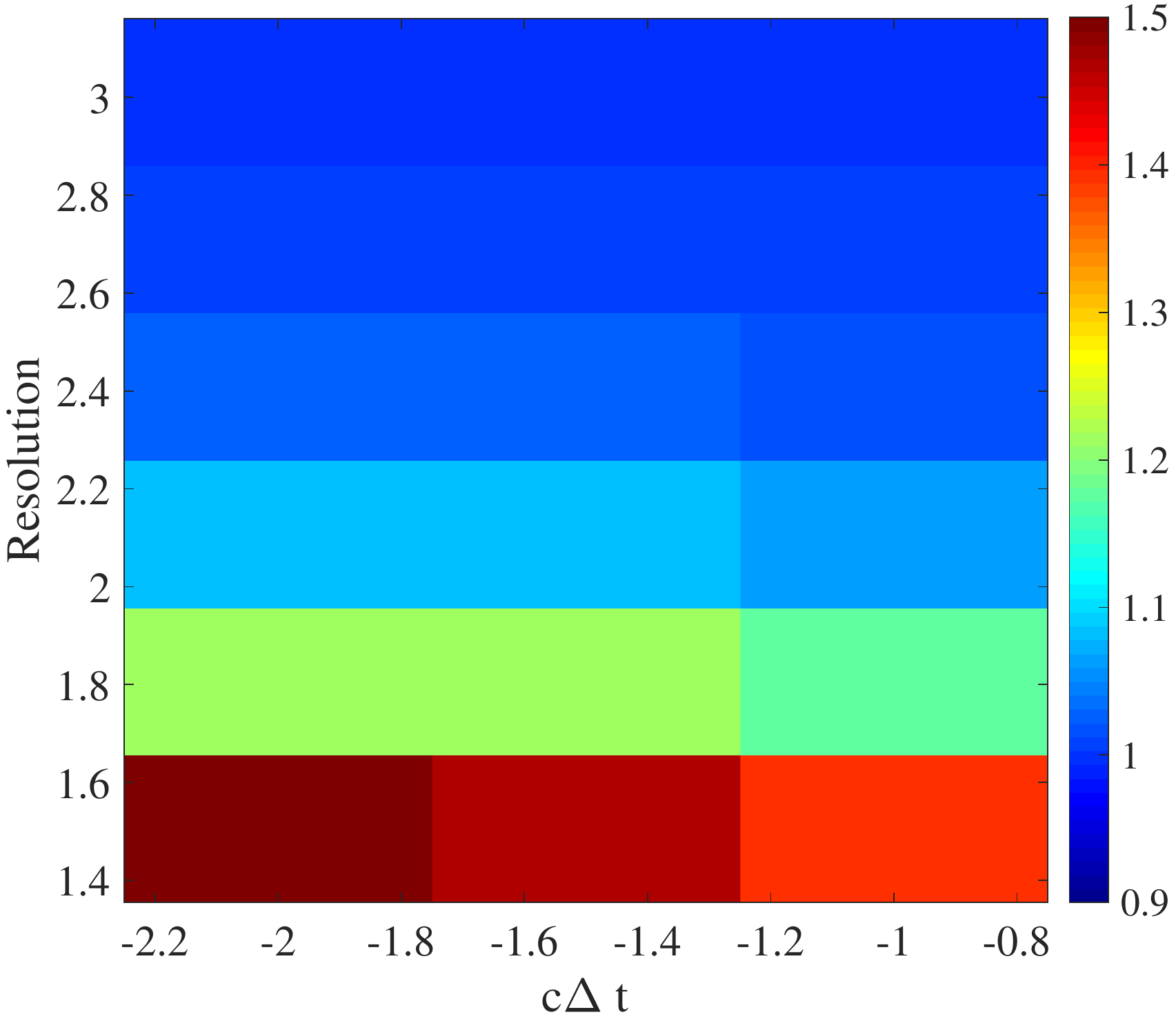}
(b)\includegraphics*[width=7.5cm]{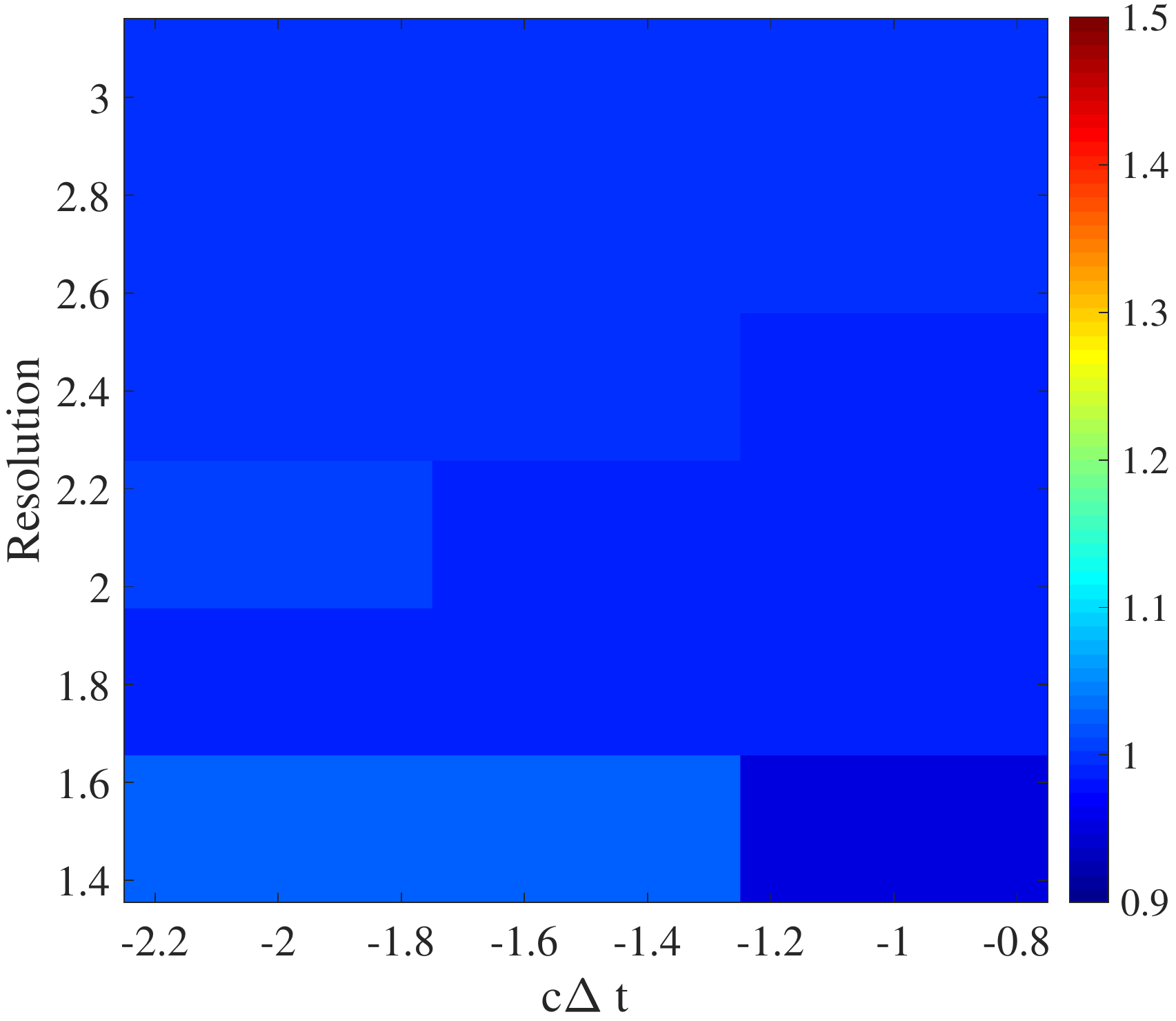}
\caption{The ratio of the position of the Marshak wave to the position obtained from the diffusion approximation at time $ct=500$ for the (a) IMC scheme, and (b) ISMC scheme. The x axis is the log of the time step and in the y axis is the log of the resolution.}
\label{fig:gold_convergence}
\end{figure}

Our results are compared to a solution obtained using diffusion approximation solver, which yields the correct behavior in this optically thick regime~\cite{ISMC}. Fig.~\ref{fig:gold} shows the material and radiation temperature. 
It can be seen that the IMC results exhibit a teleportation error, and the front of the wave advances too fast, due to the low spatial resolution, while ISMC results are very close to the diffusion reference solution. Nevertheless, we note the ISMC yields noisier result, both in radiation and material temperature, using the same amount of particles as IMC (ISMC require larger amount of particles for getting smooth solutions). This is due to the discrete nature of the absorption-emission process.
\begin{figure}
(a)\includegraphics*[width=7.5cm]{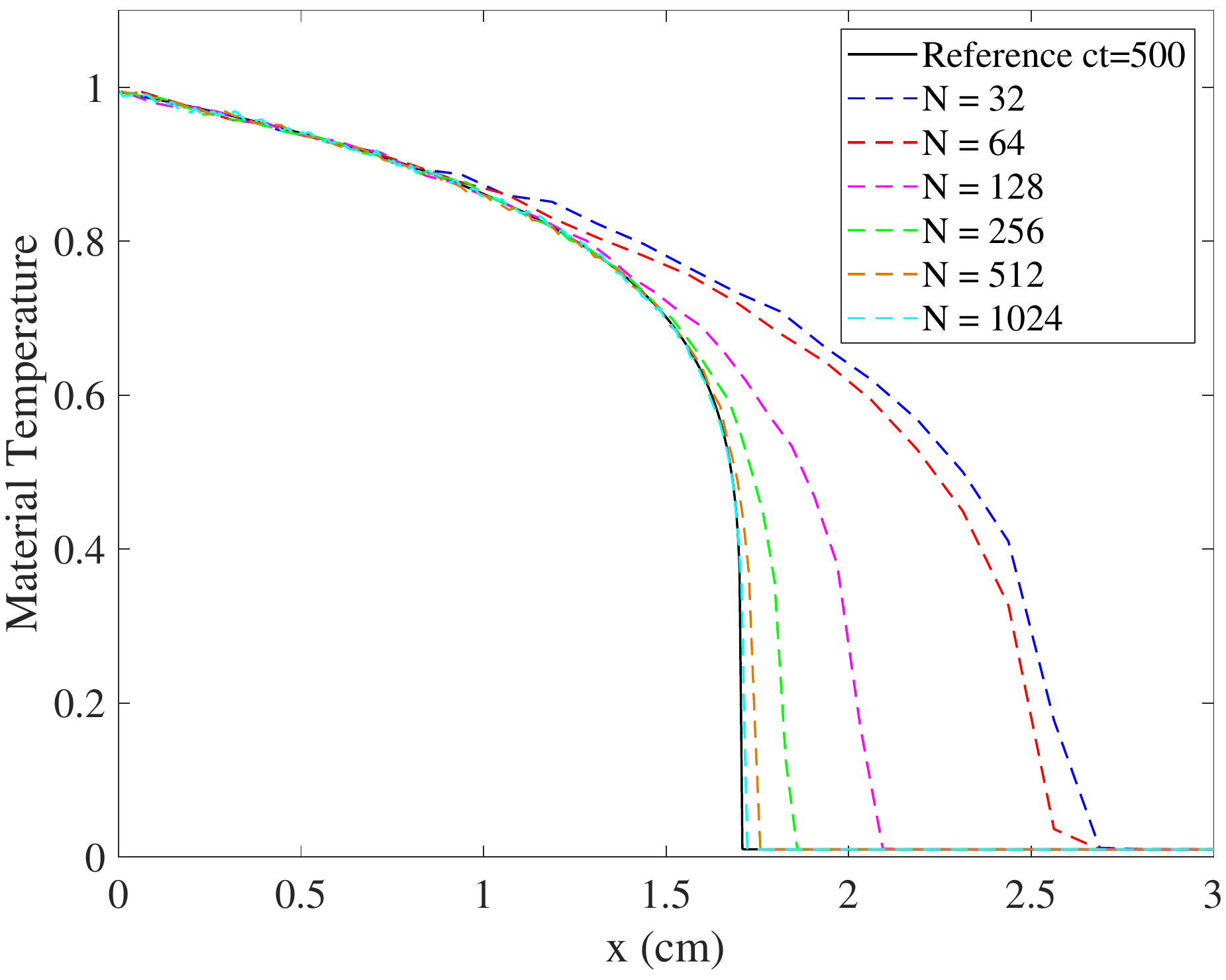}
(b)\includegraphics*[width=7.5cm]{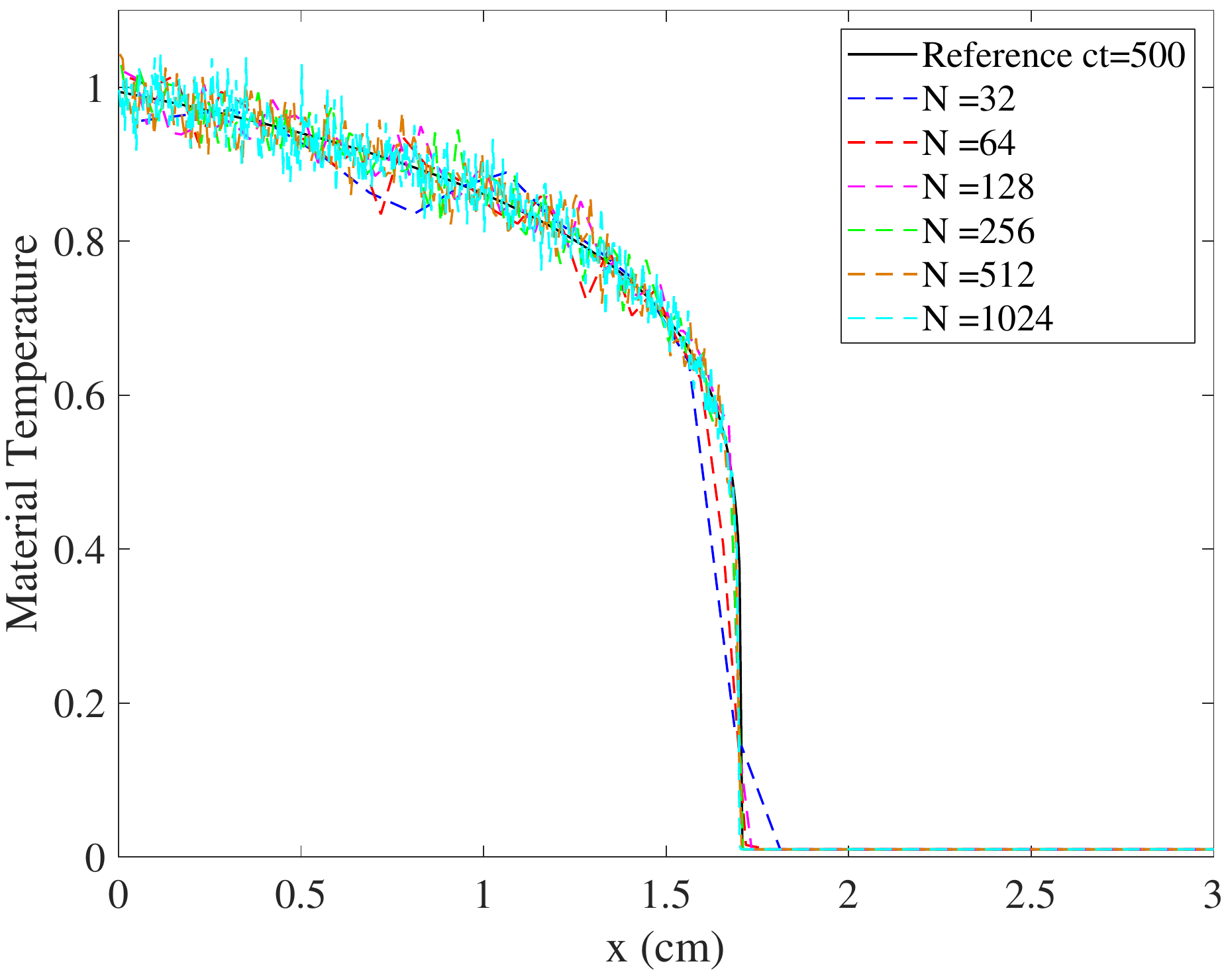}
\caption{The material temperature for the Marshak wave with different resolutions using a time step of $c\Delta=0.01$ for (a) IMC scheme, and (b) the ISMC scheme.}
\label{fig:gold_dx_convergence}
\end{figure}

In order to check the convergence in the position of the Marshak wave, we vary the resolution and time step of both schemes. Fig.~\ref{fig:gold_convergence} shows the ratio of the position of the Marshak wave (defined as the first position where $T<0.1$) to the position obtained from the diffusion approximation at time $ct=500$ for the IMC and ISMC schemes respectively. Clearly, the variance of the heat-front position in the ISMC varies with few \%, while in the IMC, it varies with tens of \%. In order to achieve convergence with IMC the spatial resolution had to be increased by approximately an order of magnitude. In addition, the ISMC shows the correct position using the smallest time step and the lowest resolution, while for the IMC scheme the minimum time step that achieves convergence is resolution dependent. 
\begin{figure}
(a)\includegraphics*[width=7.5cm]{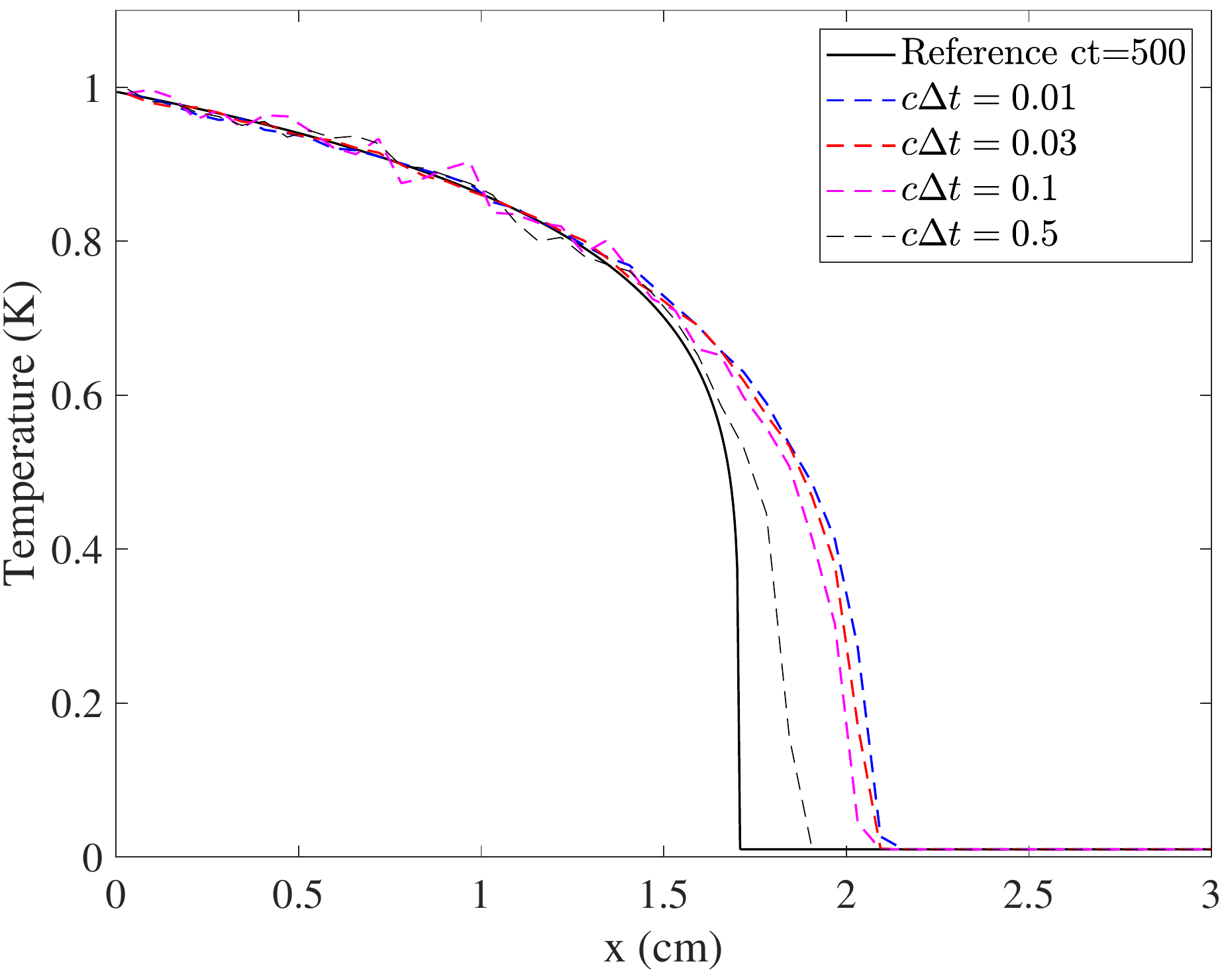}
(b)\includegraphics*[width=7.5cm]{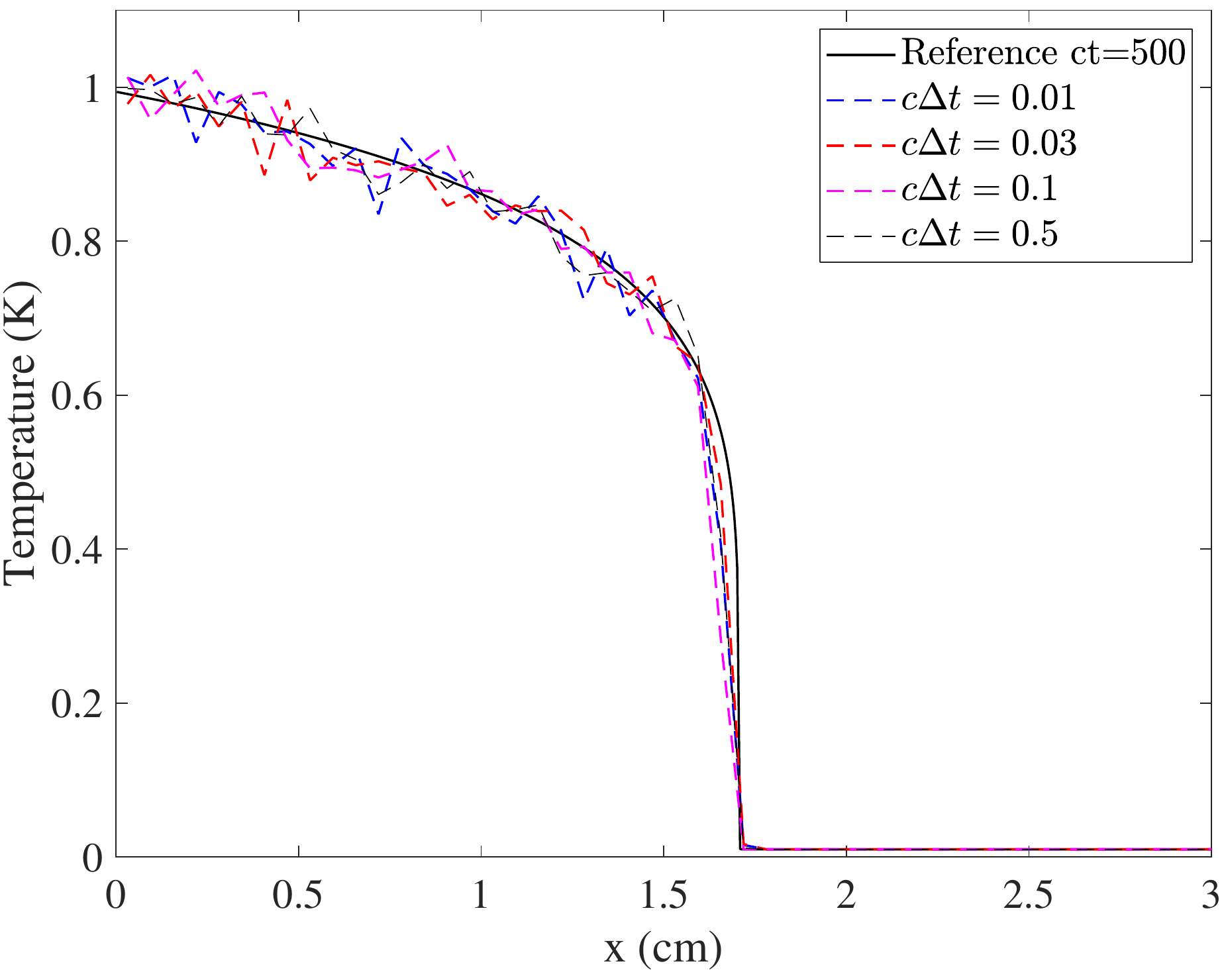}
\caption{The material temperature for the Marshak wave with different time steps for a resolution of $N=64$ cells for (a) IMC scheme, and (b) the ISMC scheme.}
\label{fig:gold_dt_convergence}
\end{figure}

Fig.~\ref{fig:gold_dx_convergence} shows the Marshak wave using different resolutions and using a constant time step size of $c\Delta t=0.01$. For the IMC scheme, the correct position of the wave is only achieved for very high spatial resolution. In contrast, in the ISMC scheme, the position of the Marshak wave is resolution independent, and the correct position is achieved with very low resolution, reproducing the results in~\cite{ISMC}.

Fig.~\ref{fig:gold_dt_convergence} shows the Marshak wave using different time steps and with a spatial resolution of 64 cells. Decreasing the time step in the IMC case, increases the teleportation error, while the runs using ISMC all exhibit the correct behavior. This feature of IMC is disturbing, since you can run with a converged but finite spatial resolution, and when the time-step is decreased, it {\em{does not}} converge to the correct answer. The new teleportation-free ISMC algorithm does not suffer from this feature.

\subsection{Olson 2019}
As a preparation for the multi-group benchmark problem, we consider an optically thin source test, that was studied by~\citet{olson2019}. In this test, the opacity is constant and set to $\sigma_a=0.1\;\mathrm{cm^{-1}}$, and the heat capacity is set to unity. At time $t=0$ a radiation source term is turned on, $Q(x)=\exp{(-700\cdot x^3)}$, and it is turned off at time $ct=2$. 
The initial temperature in the domain $0\leqslant x \leqslant 4$ is set to be $T(t=0)=0.01$ and we run the simulation until a time $ct=3$ using reflective boundary conditions at both ends. For both IMC and the ISMC we use a spatial resolution of 256 equally spaced cells, create $10^4$ new particles each time step, limit the total number of particles to be $10^5$ and set a constant time step of $c\Delta t = 10^{-13}$. For the reference solution, we use the high order $P_N$ solution of~\citet{olson2019}.
\begin{figure}
(a)\includegraphics*[width=7.5cm]{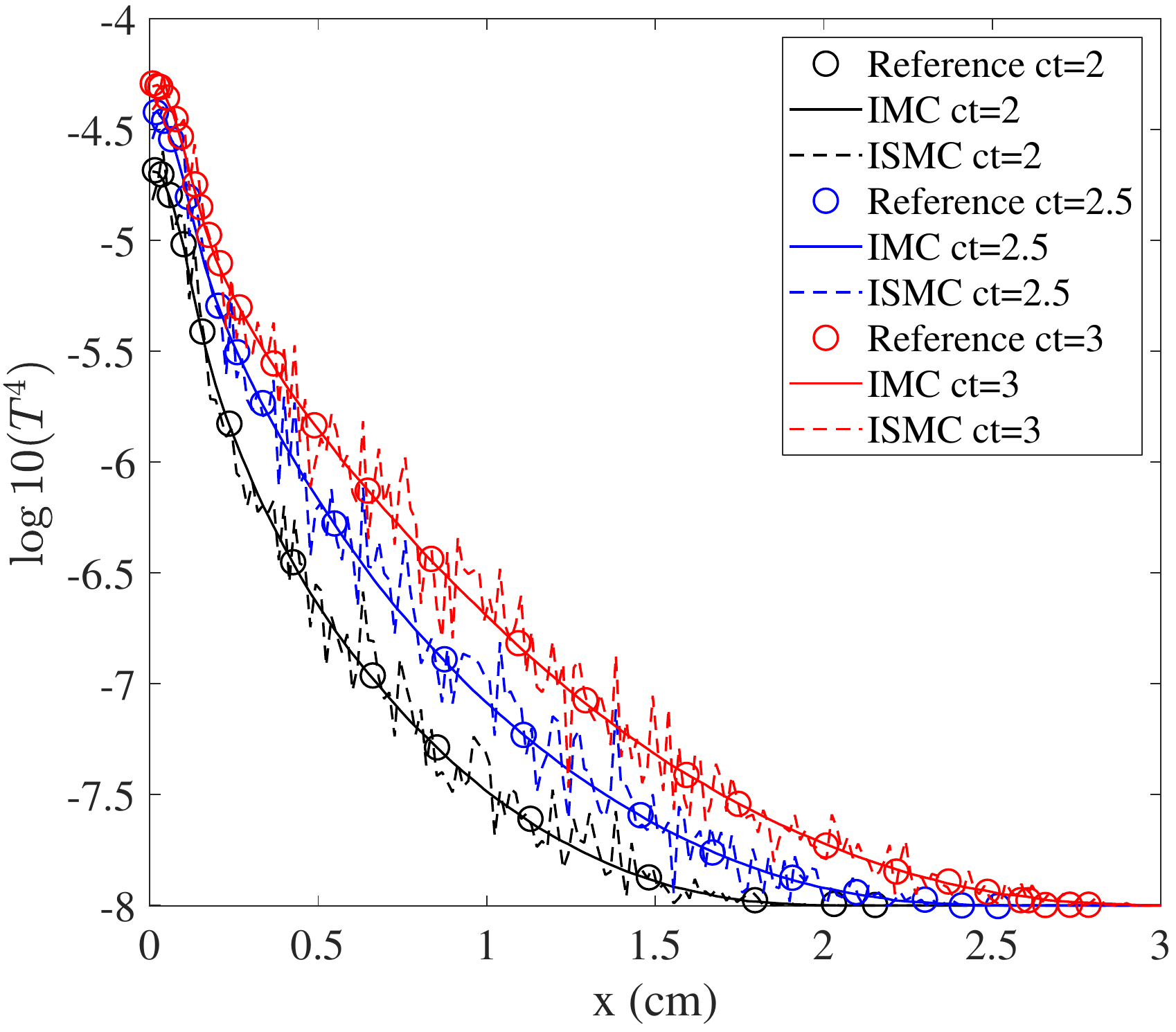}
(b)\includegraphics*[width=7.5cm]{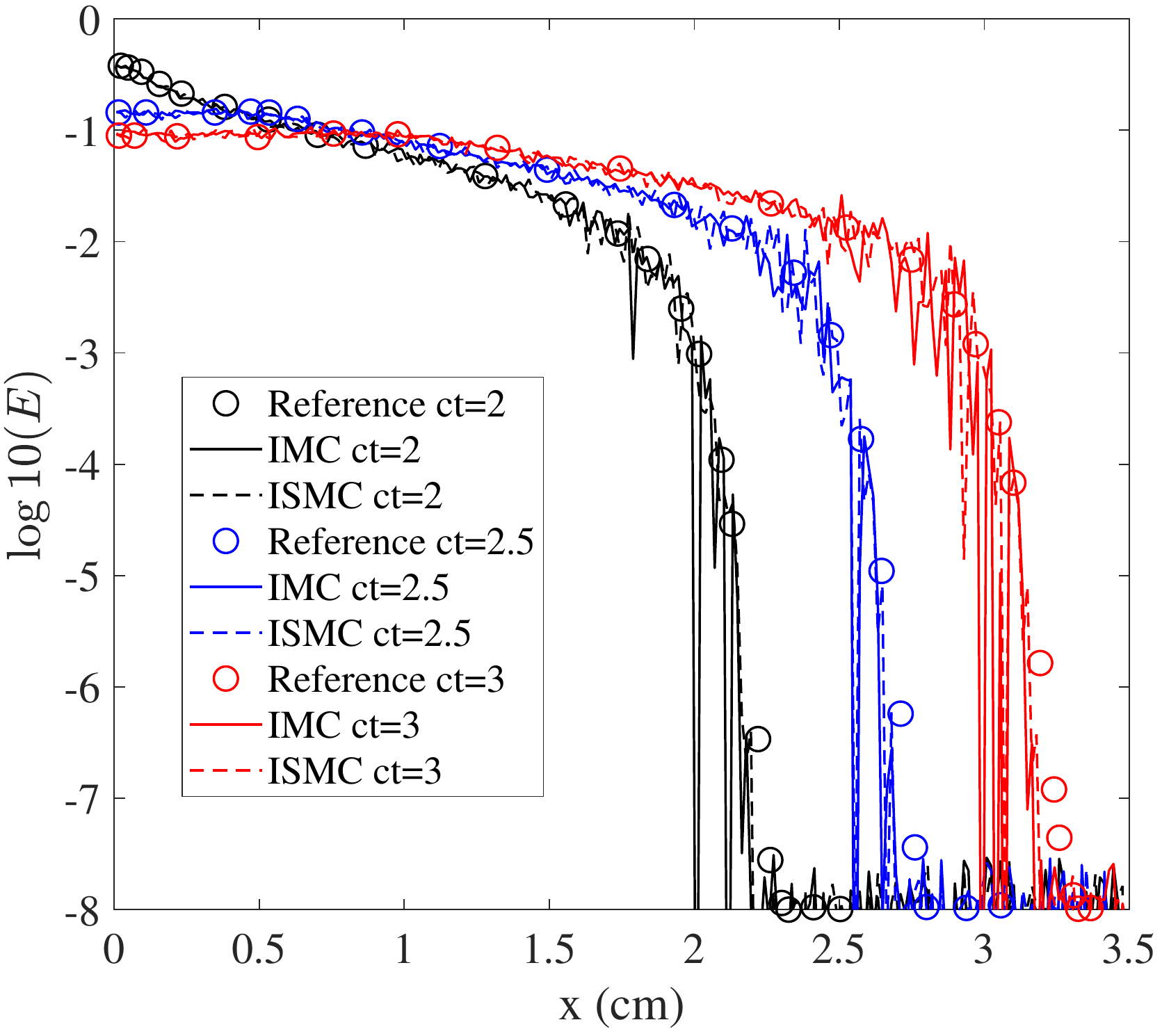}
\caption{(a) The material temperature at different times for the \citet{olson2019} test problem. (b) The radiation energy density at different times for the \citet{olson2019} test problem.}
\label{fig:olson2019}
\end{figure}

In Fig.~\ref{fig:olson2019}(a) we present the material temperature and in Fig.~\ref{fig:olson2019}(b) we show the radiation energy density for different times compared with the reference $P_N$ solution, in logarithmic scale. The large difference between the radiation energy and the material temperature emphasizes how optically-thin this benchmark is. There is a good match between the Monte-Carlo schemes (both IMC and ISMC) and the reference $P_N$. In the radiation energy, both methods exhibit comparable level of noise (as in the~\citet{suol97} test), since both methods have a discretized radiation field and a comparable number of photons.
Concerning the material temperature, once again, the discrete nature of the material field in the ISMC scheme gives rise to a noisier solution compared to the one in the IMC scheme. 

\begin{figure}
(a)\includegraphics*[width=7.5cm]{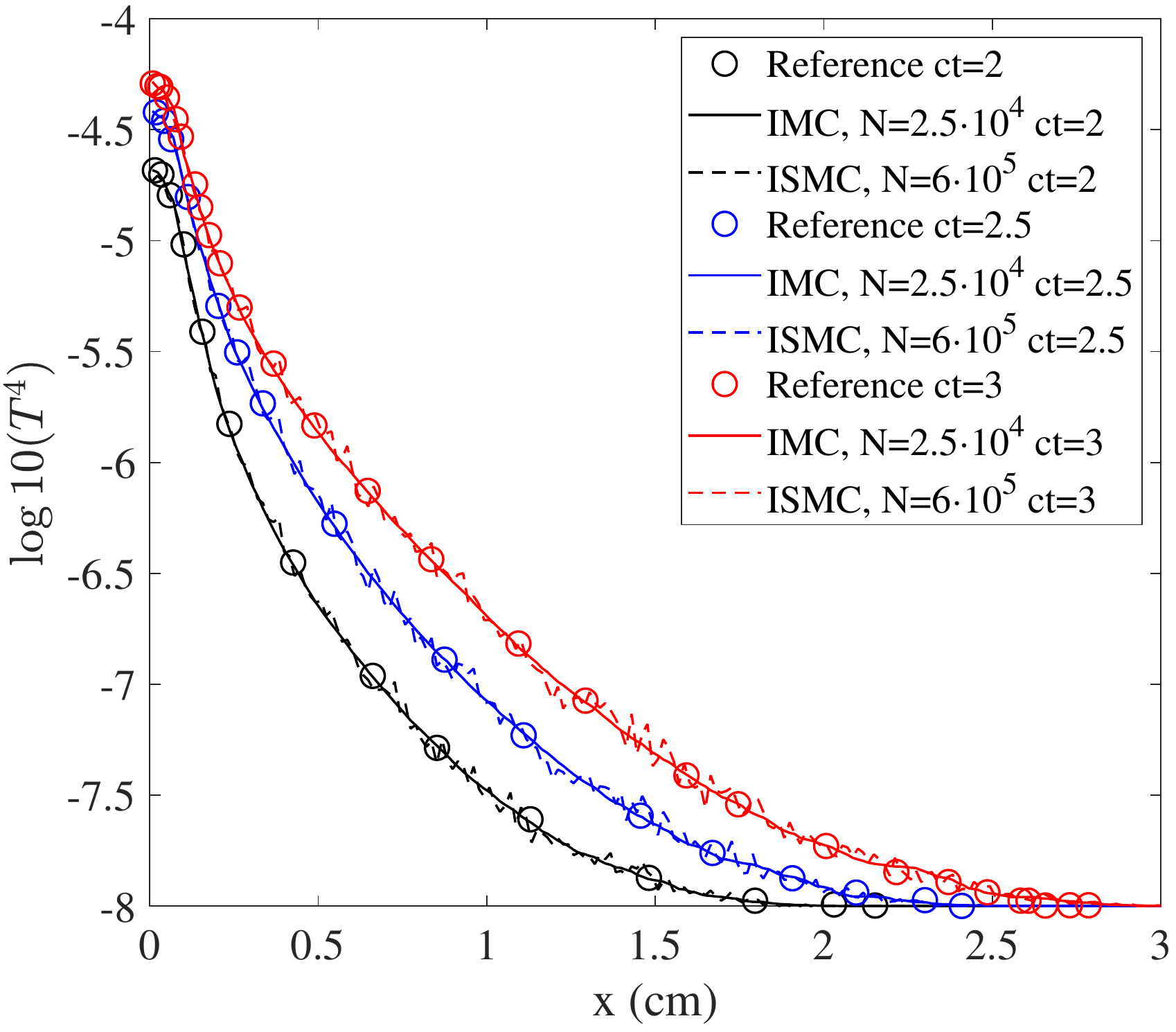}
(b)\includegraphics*[width=7.5cm]{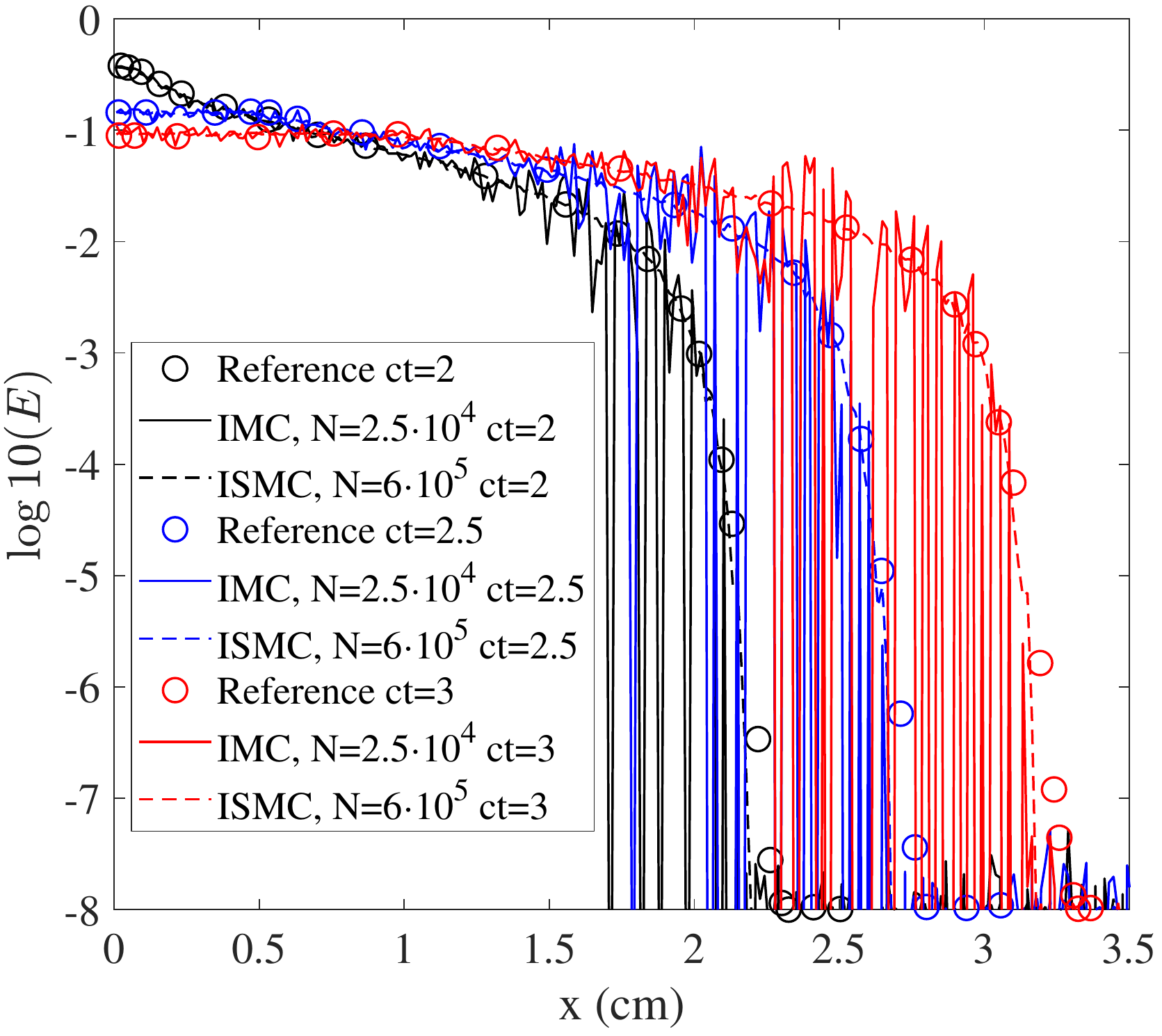}
\caption{Low statistic IMC run vs high statistic ISMC run for (a) The material temperature at different times for the \citet{olson2019} test problem. (b) The radiation energy density at different times for the \citet{olson2019} test problem.}
\label{fig:olson2019_more}
\end{figure}

In order to achieve comparable noise level in the material temperature field for this optically thin test, we reduced the number of photons in IMC and increased them in ISMC. In the IMC case, we create $2\cdot 10^3$ new photons each time step and limit the total number to be $2.5\cdot 10^4$, while in the ISMC case we create $5\cdot 10^3$ new photons each time step and limit the total number of particles to be $6\cdot 10^5$. Figure \ref{fig:olson2019_more} shows the material temperature and radiation temperature for the runs. The noise level in the material temperature is almost comparable while in the radiation field the noise level is much lower in the ISMC run. The overall run time in the ISMC case was a factor of 6.5 longer than the IMC run.

\subsection{Olson 2020 1D}
We now reach to the multi-frequency extension to the ISMC algorithm, comparing it to published test problems results as well as the IMC (classic) algorithm. The multi-frequency tests were published recently in~\citet{olson2020}, both in 1D and 2D, and they are characterized in relatively optically thin media. 
\begin{figure}
(a)\includegraphics*[width=7.5cm]{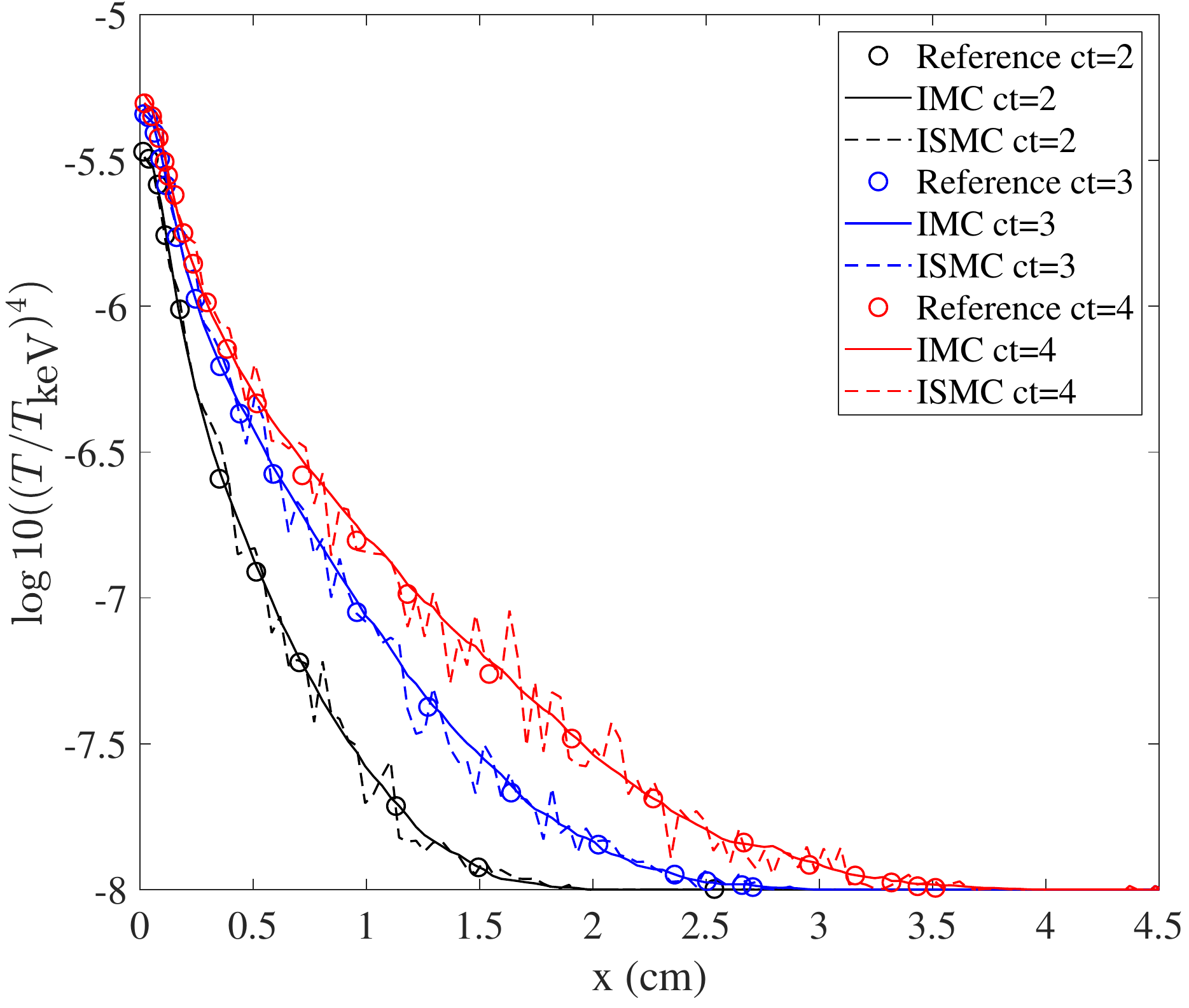}
(b)\includegraphics*[width=7.5cm]{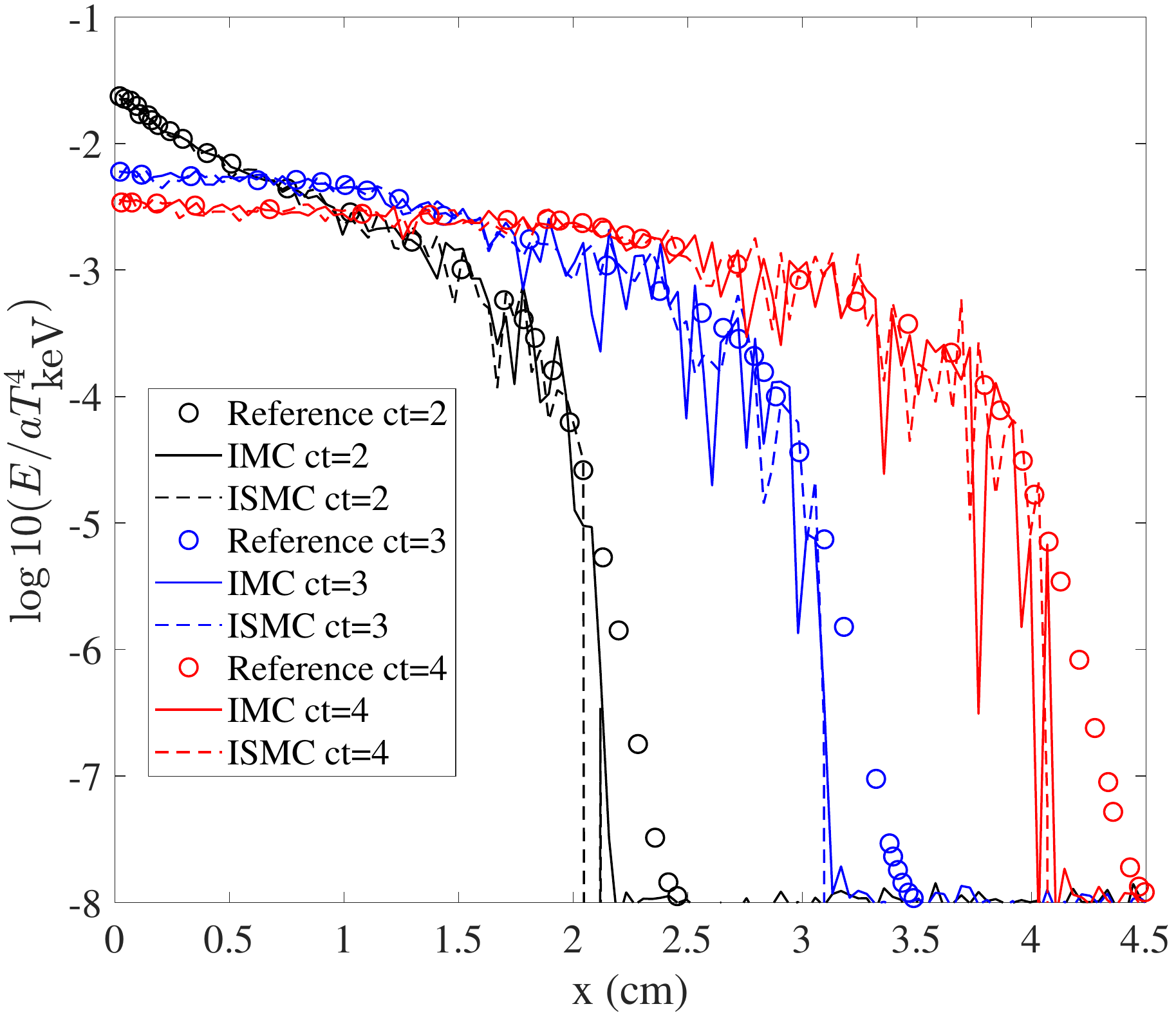}
\caption{(a) The material temperature at different times for the \citet{olson2020} 1D test problem. (b) The radiation energy density at different times for the \citet{olson2020} 1D test problem.}
\label{fig:olson2020}
\end{figure}

In the first test, a similar multi-frequency extension setup to the one used in~\citet{olson2019} is used. The material is composed of carbon-hydrogen foam, whose opacity (in units of $\text{cm}^2\text{g}^{-1}$) is given by:
\begin{equation}
    \kappa_{a,\nu}=\begin{cases}
    \text{min}(10^7,10^9(T/T_\text{ keV})^2)& h\nu<0.008\text{ keV}\\
    \frac{3\cdot 10^6 \left(0.008\text{ keV}/h\nu\right)^2}{(1+200\cdot(T/T_\text{ keV})^{1.5})}& 0.008\text{ keV}<h\nu<0.3\text{ keV}\\
    \frac{3\cdot 10^6 \left(0.008\text{ keV}/h\nu\right)^2\sqrt{0.3\text{ keV}/h\nu}}{(1+200\cdot(T/T_\text{ keV})^{1.5})}+\frac{4\cdot 10^4\left(0.3\text{ keV}/h\nu\right)^{2.5}}{1+8000(T/T_\text{ keV})^2}&
    h\nu >0.3\text{ keV}.
    \end{cases}
\end{equation}
The macroscopic absorption cross-section is given by $\sigma_{a,\nu}=\rho \kappa_{a,\nu}$, where $\rho=0.001$ $\text{g}/\text{cm}^3$. The heat capacity is given by:
\begin{subequations}
\label{heat_cap}
\begin{eqnarray}
    \rho C_V&=&aT_\text{ keV}^3H\left(1+\alpha+\left(T+\chi\right)\frac{\partial\alpha}{\partial T}\right)\\
    \alpha&=&\frac{1}{2}e^{-\chi/T}\left(\sqrt{1+4e^{\chi/T}}-1\right)\\
    \frac{\partial\alpha}{\partial T}&=&\frac{\chi}{T^2}\left(\alpha-1/\sqrt{1+4e^{\chi/T}}\right)
\end{eqnarray}
\end{subequations}
where $\chi =0.1T_{\text{keV}}$ and $H=0.1$.
The source term is given by a black body with a temperature of $0.5$ keV whose spatial extent is given by $Q(x)=B(0.5\text{ keV})\exp{(-693\cdot x^3)}$, and is turned on at time $t=0$ and turned off at time $ct=2$. The initial temperature in the domain $0\leqslant x \leqslant 4.8$ (cm) is set to be $T(t=0)=0.01T_\text{ keV}$ and we run the simulation until a time $ct=4$ using reflective boundary conditions at both ends. For both IMC and the ISMC we use a spatial resolution of 128 equally spaced cells, create $10^3$ new particles each time step, limit the total number of particles to be $2\cdot10^4$ and set a constant time step of $\Delta t = 10^{-13}$. For the reference solution, we use the one calculated using a high order $P_N$ scheme presented in~\citet{olson2020}.

Fig.~\ref{fig:olson2020}(a) shows the material temperature for different times compared with the reference solution, and the radiation energy density is shown in Fig.~\ref{fig:olson2020}(b).
The extension of the ISMC method to include frequency dependent opacity described in \S\ref{sec:ISMC_freq}, gives the correct result and agrees well with both the IMC solution and the reference solution. Near the tails of the distribution, when radiation intensity drops by several orders of magnitudes, there is a difference between the discrete results of MC (as a result of lack of statistics and large statistical noise) and the smooth curves of the $P_N$. Like previous optically thin tests, the IMC method gives a smoother material temperature and comparable noise in the radiation field as the ISMC method.

\begin{figure}
(a)\includegraphics*[width=7.5cm]{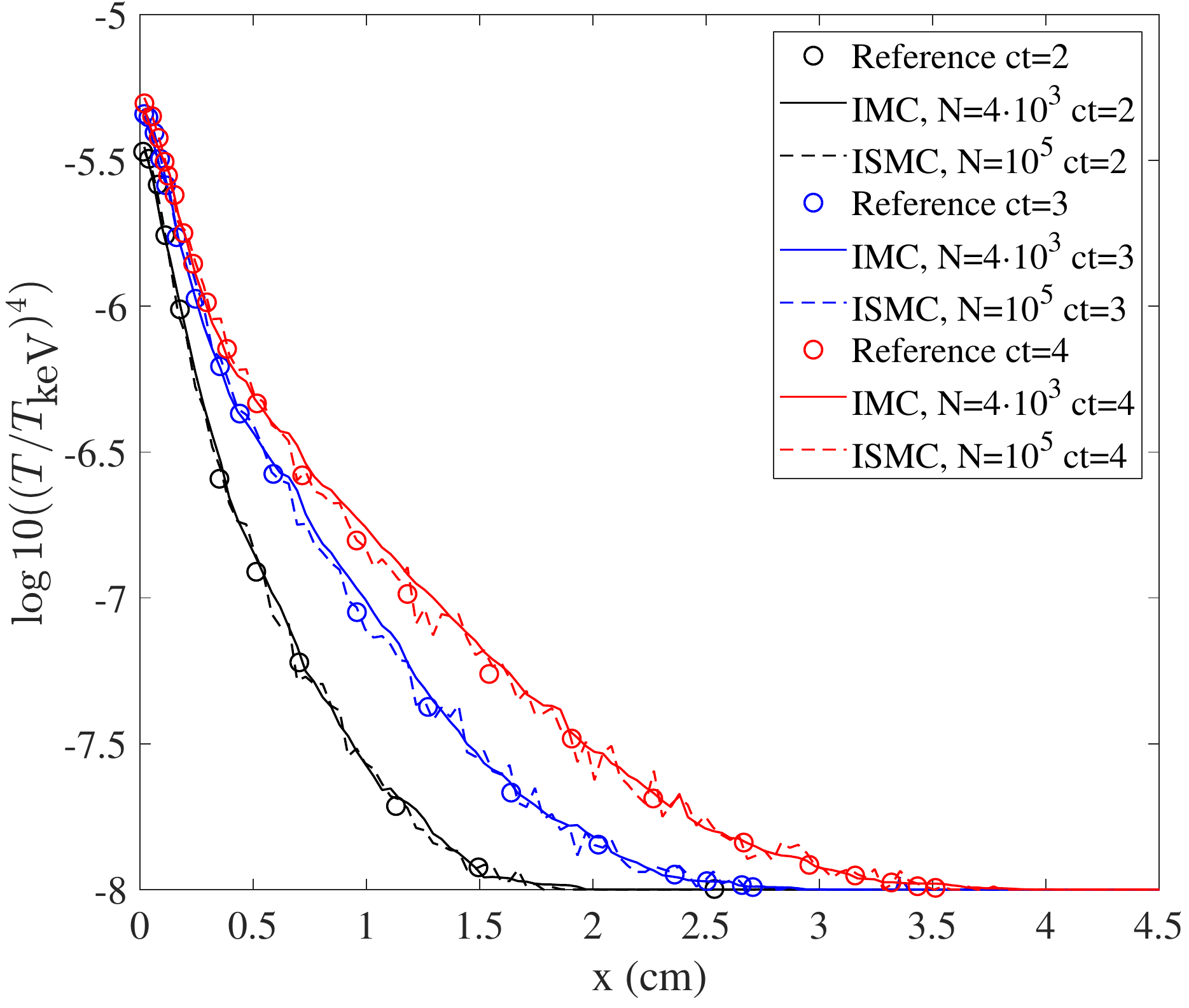}
(b)\includegraphics*[width=7.5cm]{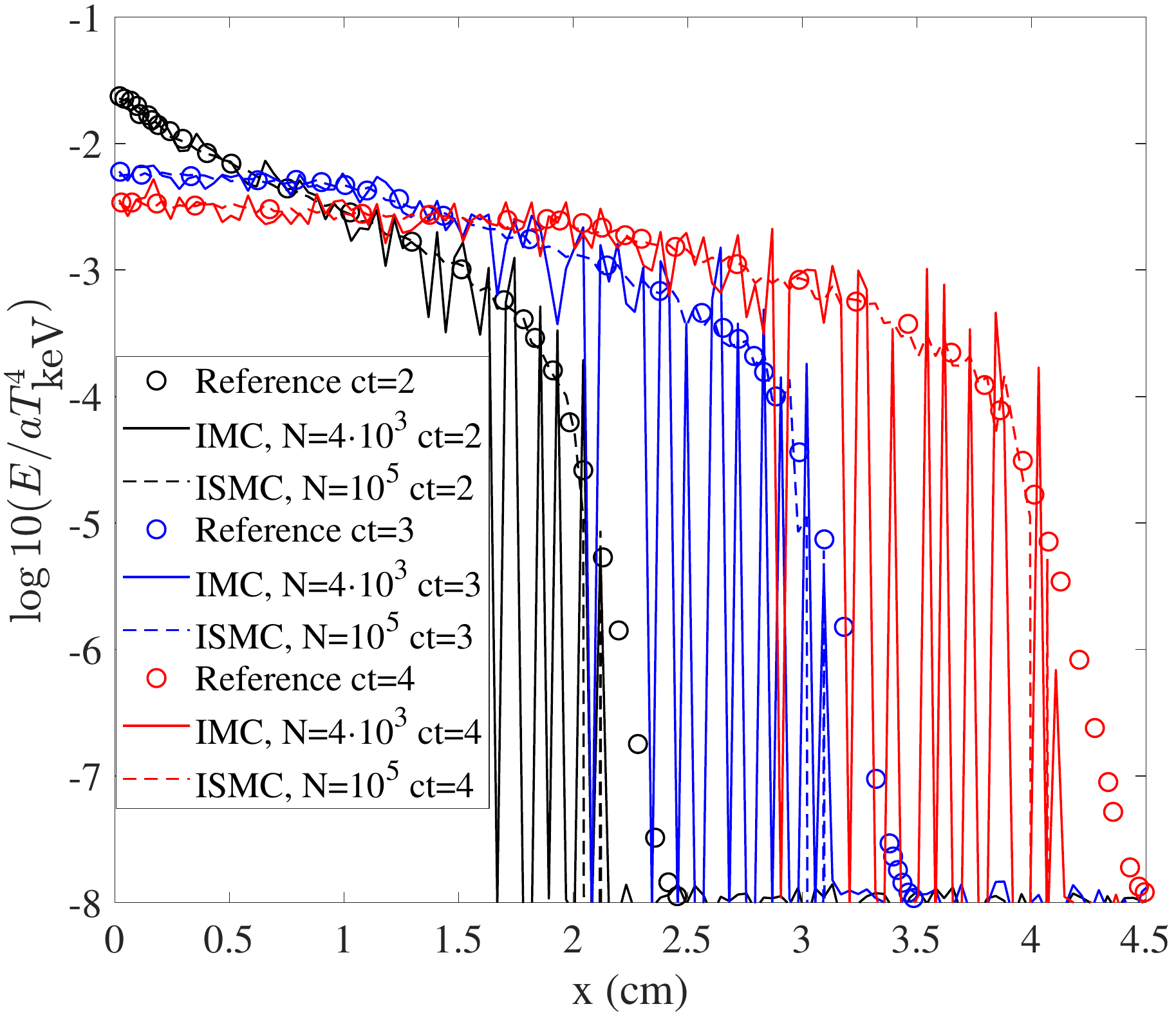}
\caption{Low statistic IMC run vs high statistic ISMC run for (a) The material temperature at different times for the \citet{olson2020} test problem. (b) The radiation energy density at different times for the \citet{olson2020} test problem.}
\label{fig:olson2020_more}
\end{figure}

As in the previous test, we try to achieve comparable noise level in the material temperature field between IMC and ISMC. In the IMC case, we create $500$ new photons each time step and limit the total number to be $4\cdot 10^3$, while in the ISMC case we create $5\cdot 10^3$ new photons each time step and limit the total number of particles to be $10^5$. Figure \ref{fig:olson2020_more} shows the material temperature and radiation temperature for the runs. The noise level in the material temperature is almost comparable while in the radiation field the noise level is much lower in the ISMC run. The overall run time in the ISMC case was a factor of 9.5 longer than the IMC run.

\subsection{Densmore et al. 2012}
\citet{Densmore} (and later \citet{teleportation2}) have presented several interesting one-dimensional frequency-dependent problems that have varying optical depth: optically thin, optically thick, and a combination of the two. The optically-thick problems will emphasis the existence of teleportation using IMC algorithm, and the absence in the ISMC in multi-frequency problems.

In all of the test problems, the opacity has an inverse $\nu^3$ functional form of:
\begin{equation}
    \sigma(x, \nu, T) = \frac{\sigma_0(x)}{\left(h\nu\right)^3\sqrt{k_BT}}
\end{equation}
and the heat capacity is set to be $C_V=10^{15}\text{erg}/T_\text{ keV}/\text{cm}^3$. The initial temperature in the domain $0\leqslant x \leqslant 5$ (cm) is 1 eV, the left boundary is a black body source with a temperature of 1keV and the right boundary is a reflecting wall. The test is run to a time of $t=1$ ns with a constant time step $\Delta t = 0.01$ ns, and for all of the runs we create $10^5$ photons per time step and limit the total number of photons to be $10^6$. The reference solution that we compare to is taken from \citet{Densmore}.

\begin{figure}
(a)\includegraphics*[width=4.75cm]{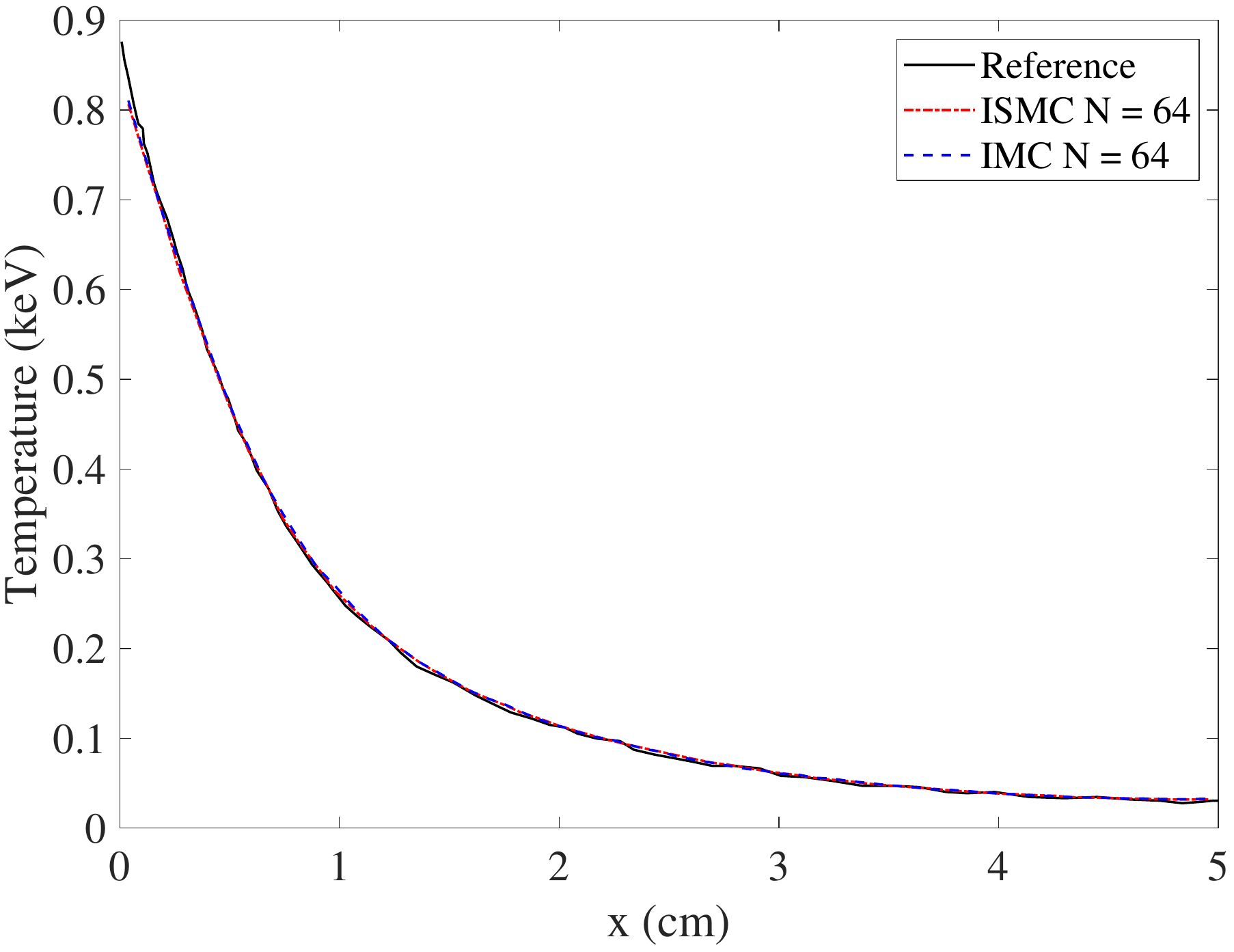}
(b)\includegraphics*[width=4.75cm]{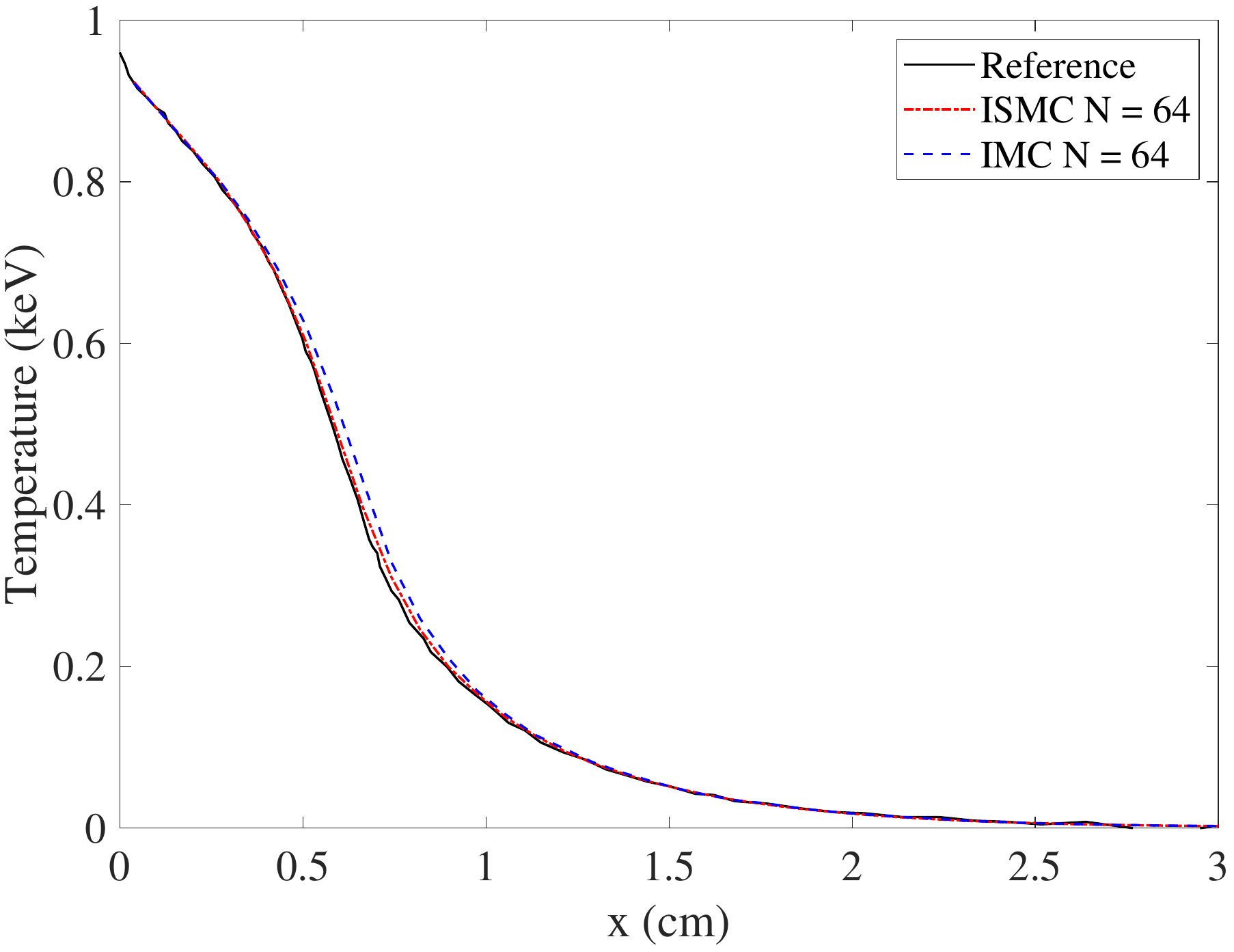}
(c)\includegraphics*[width=4.75cm]{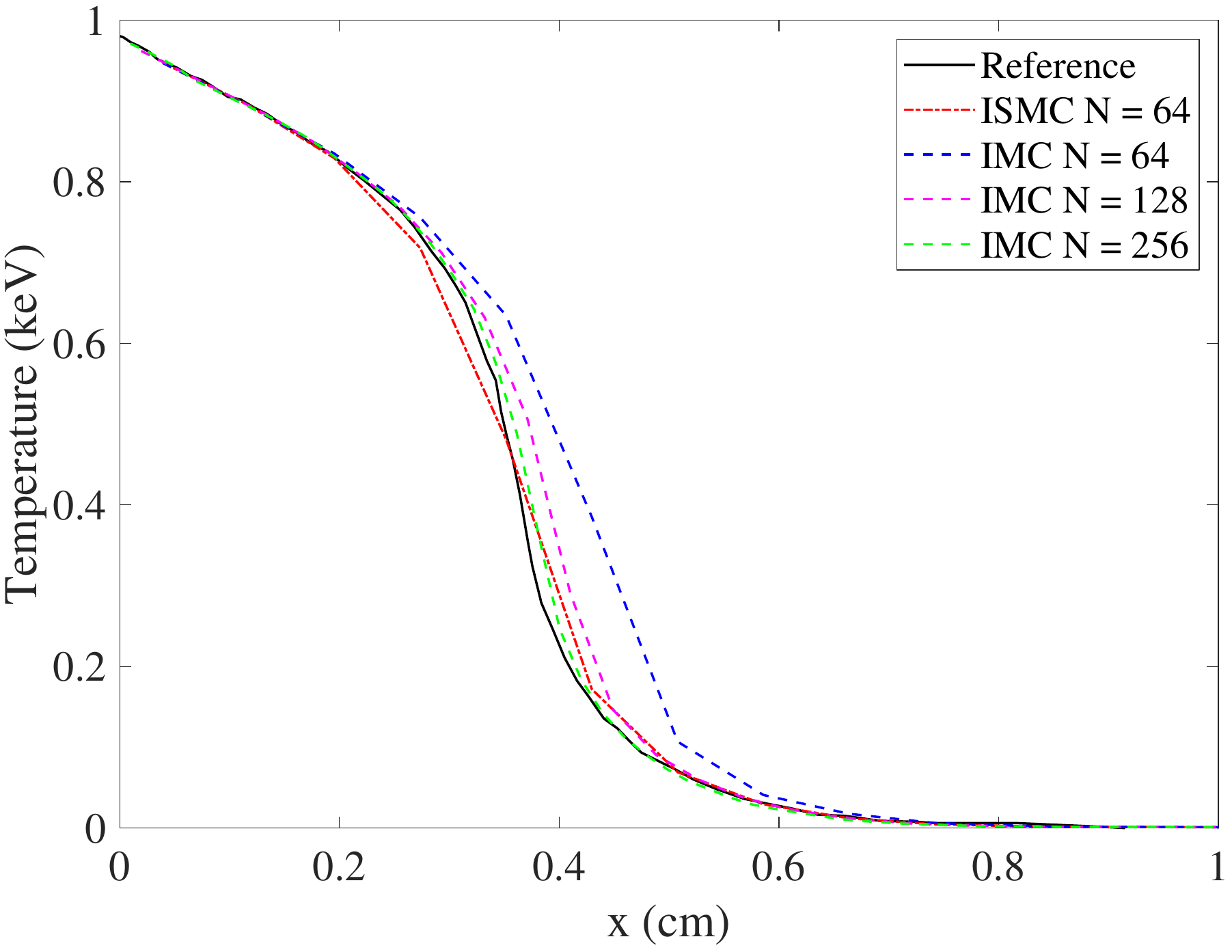}
\caption{The material temperature at time $t=1$ ns for the first three \citet{Densmore} benchmarks. (a) $\sigma_0(x)=10\;\text{keV}^{7/2}$/cm, (b) $\sigma_0(x)=100\;\text{keV}^{7/2}$/cm and (c) $\sigma_0(x)=1000\;\text{keV}^{7/2}$/cm.}
\label{fig:densmore}
\end{figure}

In the first three tests \citet{Densmore} set $\sigma_0(x)=\left[10,\;100,\;1000\right]\;\text{keV}^{7/2}$/cm accordingly, and the reference domain is composed of 64 evenly spaced cells (a small number for presenting the teleportation in opaque problems). The material temperature profiles at $t=1$ ns for both methods along with the reference solution are shown in fig. \ref{fig:densmore}.
There is a good agreement between the two methods and the reference solution for the first two tests (optically thin and medium-opacity problems), while in the third test (optically thick), the IMC requires a spatial resolution of 256 cells in order to avoid teleportation errors, as opposed to the ISMC that gives a good result even with 64 cells.

\begin{figure}
\centering
(a)\includegraphics*[width=4.75cm]{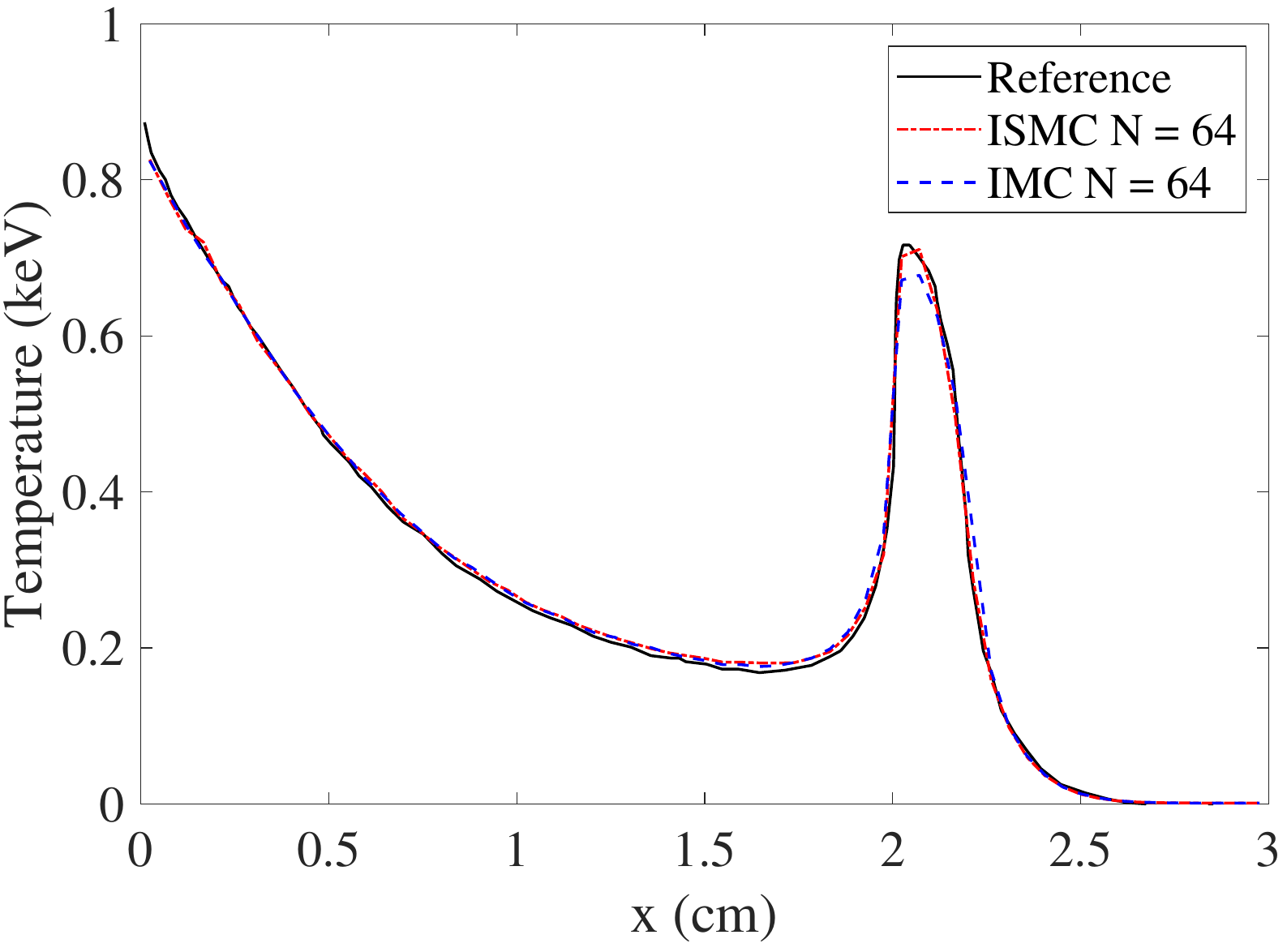}
(b)\includegraphics*[width=4.75cm]{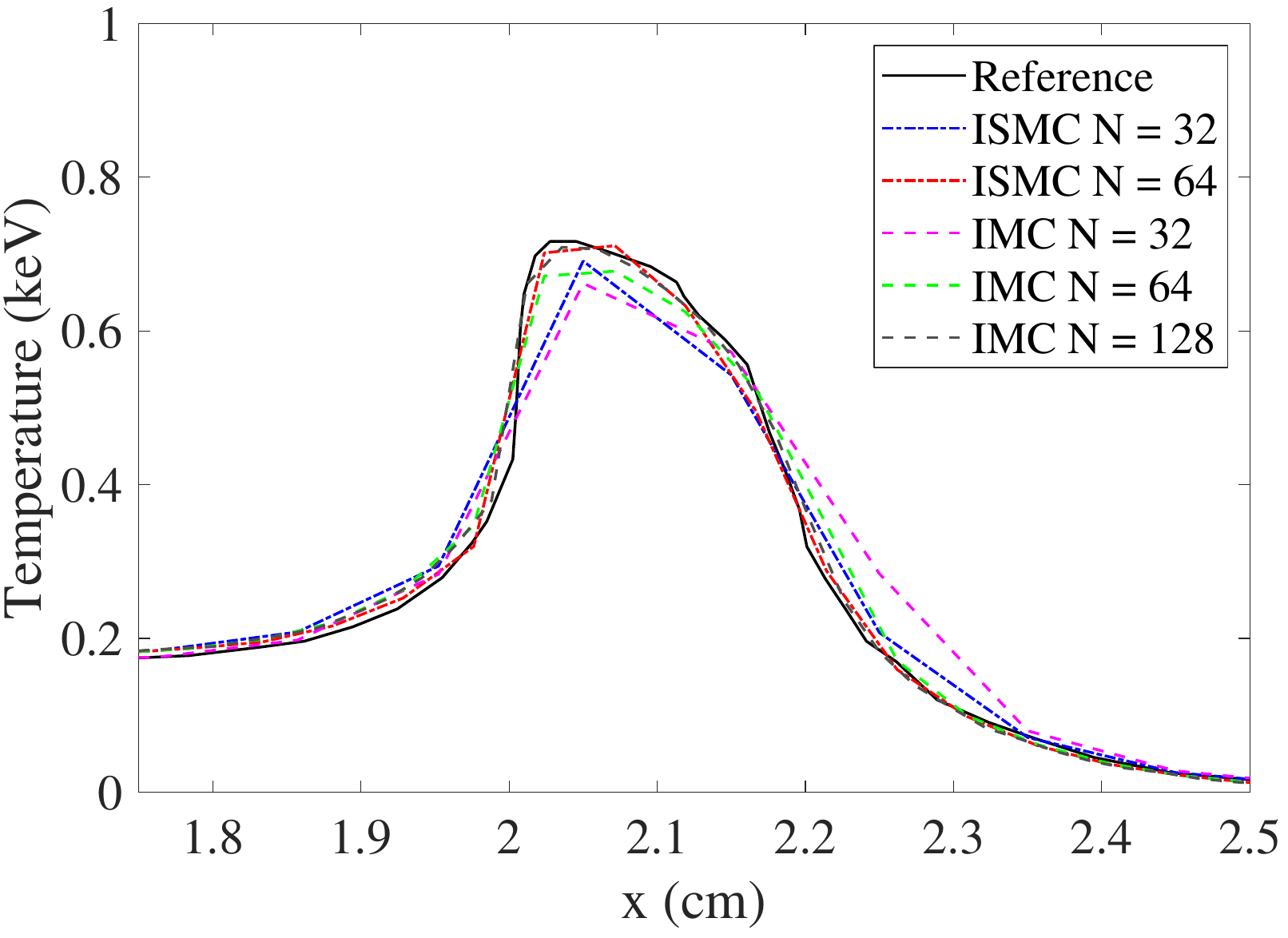}
\caption{The material temperature at time $t=1$ ns for the fourth \citet{Densmore} benchmark. (a) Zoom out showing the entire domain, (b) zoom in showing the interface between the optically thin and optically thick materials located at $x=2$ cm.}
\label{fig:densmore_interface}
\end{figure}

The fourth test measures the codes' ability to handle a sharp transition from an optically thin to an optically thick regime. The domain (with equally spaced cells throughout) is set to $0\leqslant x \leqslant 3$ (cm), and the opacity is
\begin{equation}
    \sigma_0(x)=\begin{cases}
    10\;\text{keV}^{7/2}/\text{cm}& x<2\;\text{cm},\\
    1000\;\text{keV}^{7/2}/\text{cm}& x\ge 2\;\text{cm}.
    \end{cases}
\end{equation}
In Fig. \ref{fig:densmore_interface} we show the material temperature for the fourth test (Fig. \ref{fig:densmore_interface}(b) is zoomed on the interface zone). Once again we see that when optically thick material is present, ISMC requires less spatial resolution than IMC in order to achieve convergence, due to teleportation toward the opaque material in low spatial resolution in the IMC simulation.
\section{2D Tests}
\label{2d_tests}
In this section we present the validity of the ISMC algorithm in various two-dimensional problems, comparing it to IMC. In specific, we will emphasize the teleportation errors that occurs in classic IMC implementation, while the ISMC algorithm yields a converged result in finite spatial resolution and $\Delta t$. The 2D problems show the strength of the ISMC algorithm, where converged IMC results are hard to obtain, in several famous physical problems, containing non-trivial radiation flow. 

\subsection{McClarren \& Hauck 2010}
\begin{figure}
\centering
\includegraphics*[width=5.5cm]{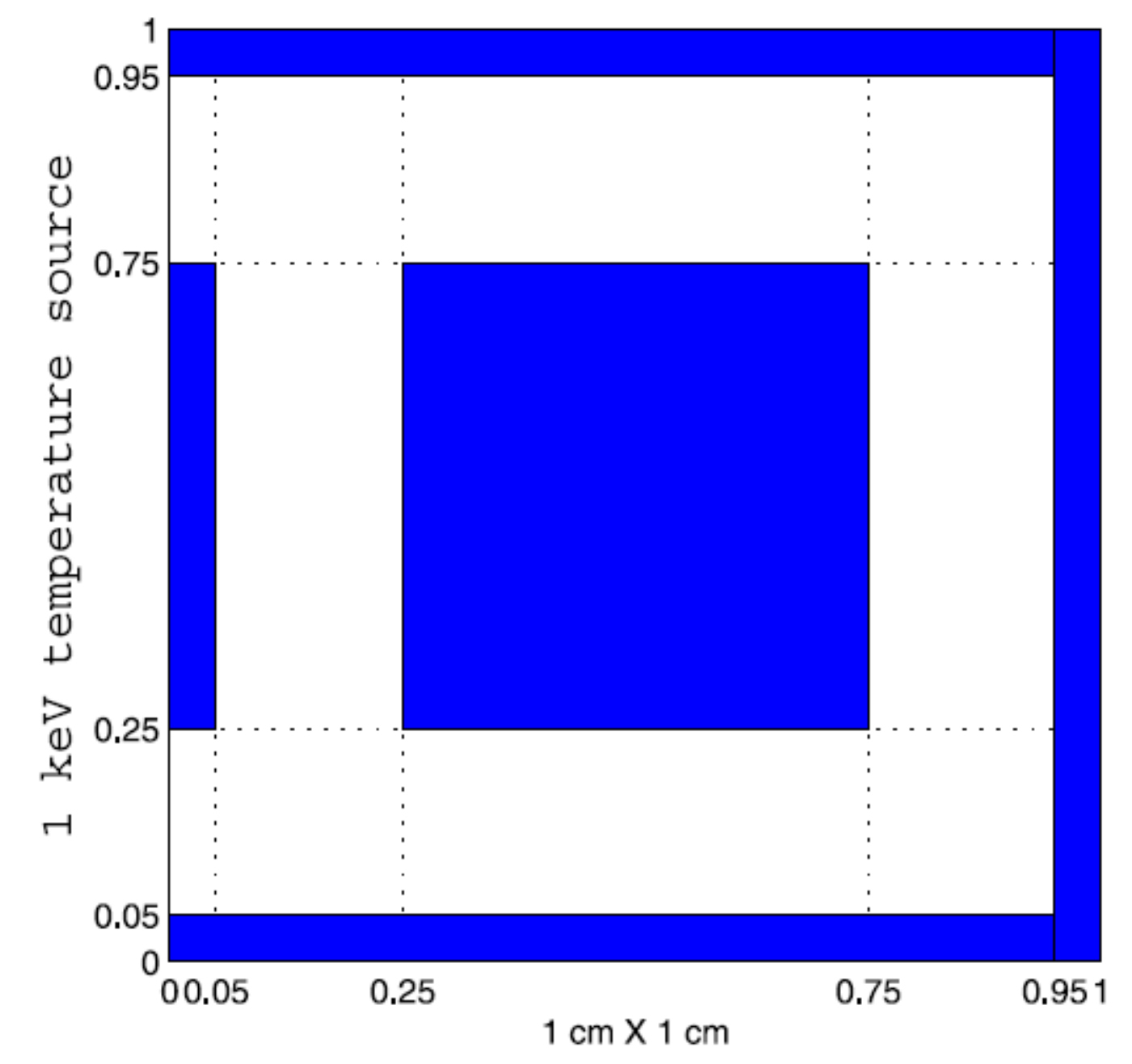}
\caption{The geometry of the McClarren \& Hauck 2010 hohlraum problem. The absorbing material is depicted in blue and material depicted in white is vacuum. The figure is taken from~\cite{MCCLARREN}.}
\label{fig:mcclarren}
\end{figure}

First, we analyse the hohlraum problem presented in~\citet{MCCLARREN,MCCLARREN2}, with a temperature dependent opacity, in XY geometry. The problem consists of a square hohlraum of size 1 cm, whose geometry and dimensions are given in Fig.~\ref{fig:mcclarren}. This benchmark presents quantitative results, that demonstrate the teleportation error in both spatial resolution and time step in a 2D scenario.

The absorbing material, depicted in blue, has an opacity of $\sigma_a=100\left(T/T_\text{ keV}\right)^{-3}\text{cm}^{-1}$ (where $T_{\mathrm{keV}}\approx1.165\cdot 10^7$K), and a constant heat capacity $\rho C_V = 2.5851\cdot 10^8\;\text{erg}/\text{K}/\text{cm}^3$. The material depicted in white is vacuum with zero opacity. The left boundary is a black body source with a constant temperature of 1keV and all the other boundaries are vacuum.

As a nominal resolution, we use a uniform mesh, with a cell size of $\Delta x = \Delta y = 0.005\;\text{cm}$ and a time step of $\Delta t = 1\cdot 10^{-11}\;\text{s}$. The hohlraum is initially cold, and is evolved until time $t=1\;\text{ns}$. For both IMC and ISMC we create $2\cdot 10^6$ new particles each time step, and limit the total number of particles to be $2\cdot10^7$.
\begin{figure}
\centering
(a)\includegraphics*[width=5.5cm]{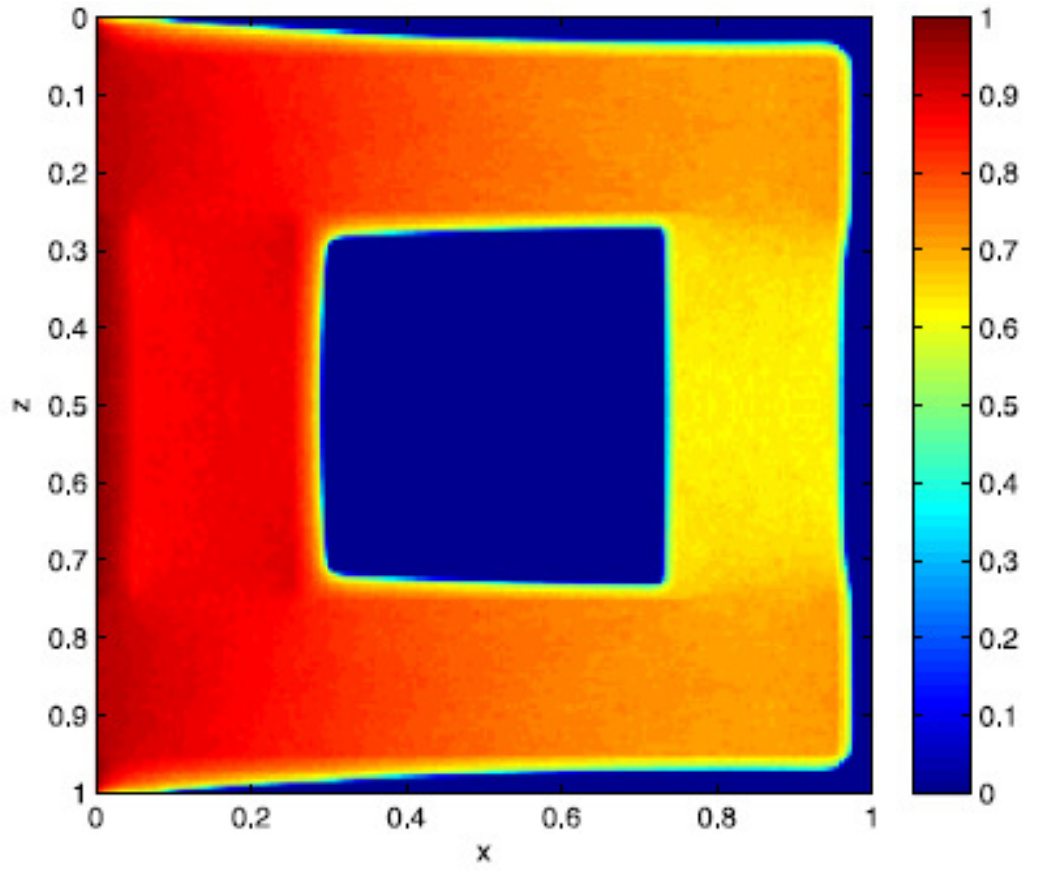}
(b)\includegraphics*[width=5.5cm]{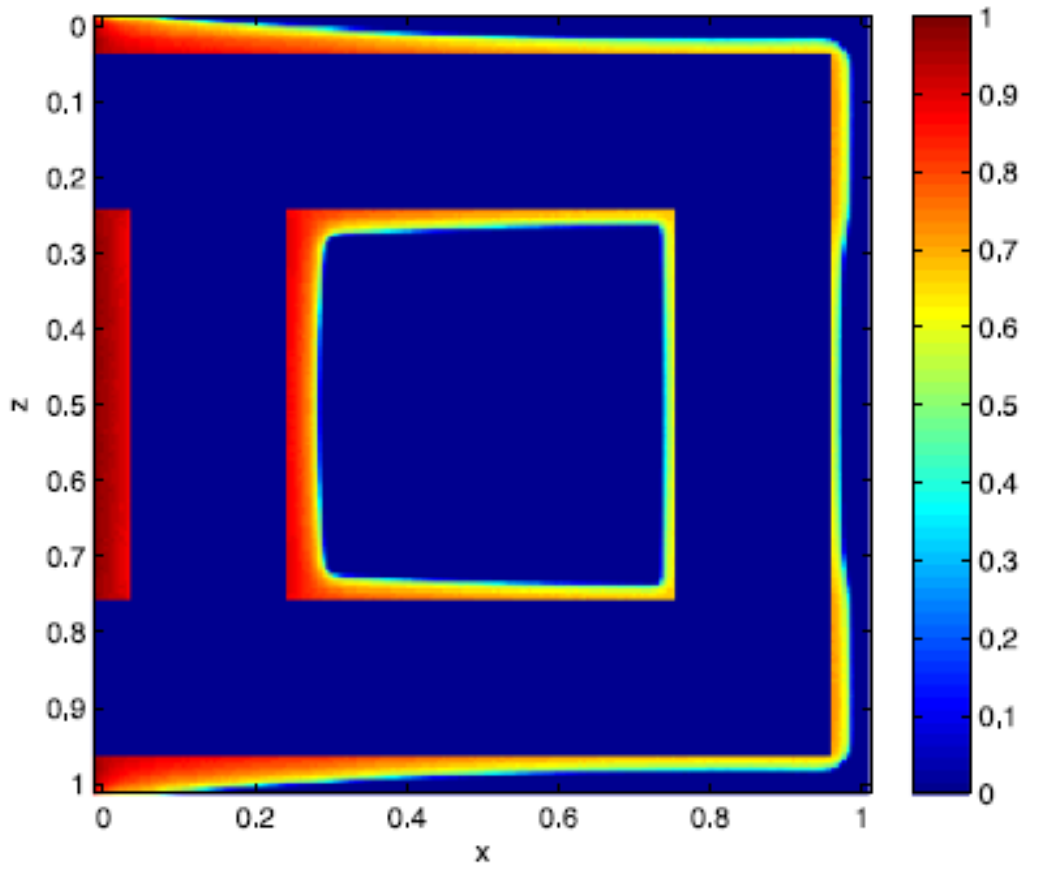}
(c)\includegraphics*[width=5.5cm]{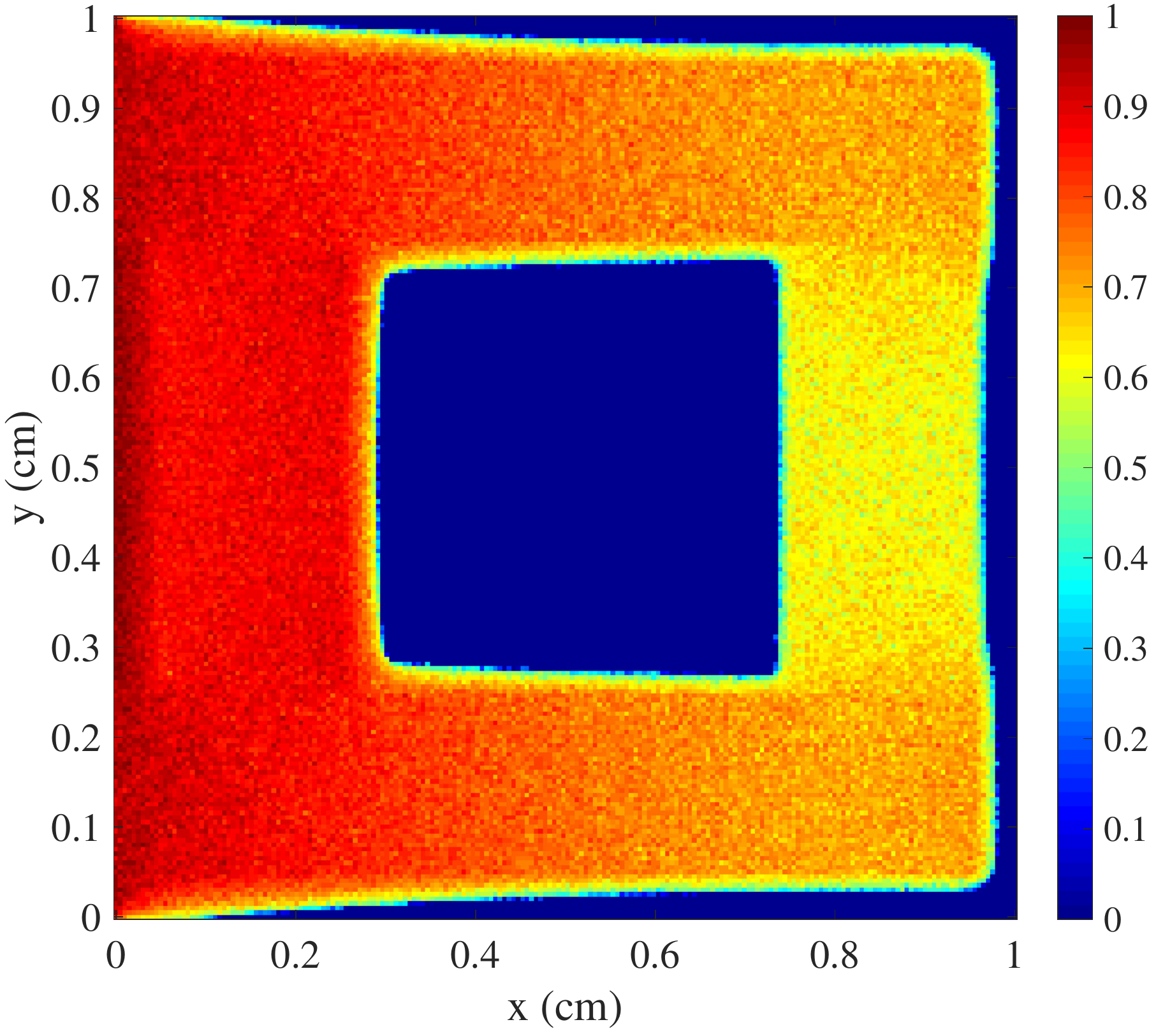}
(d)\includegraphics*[width=5.5cm]{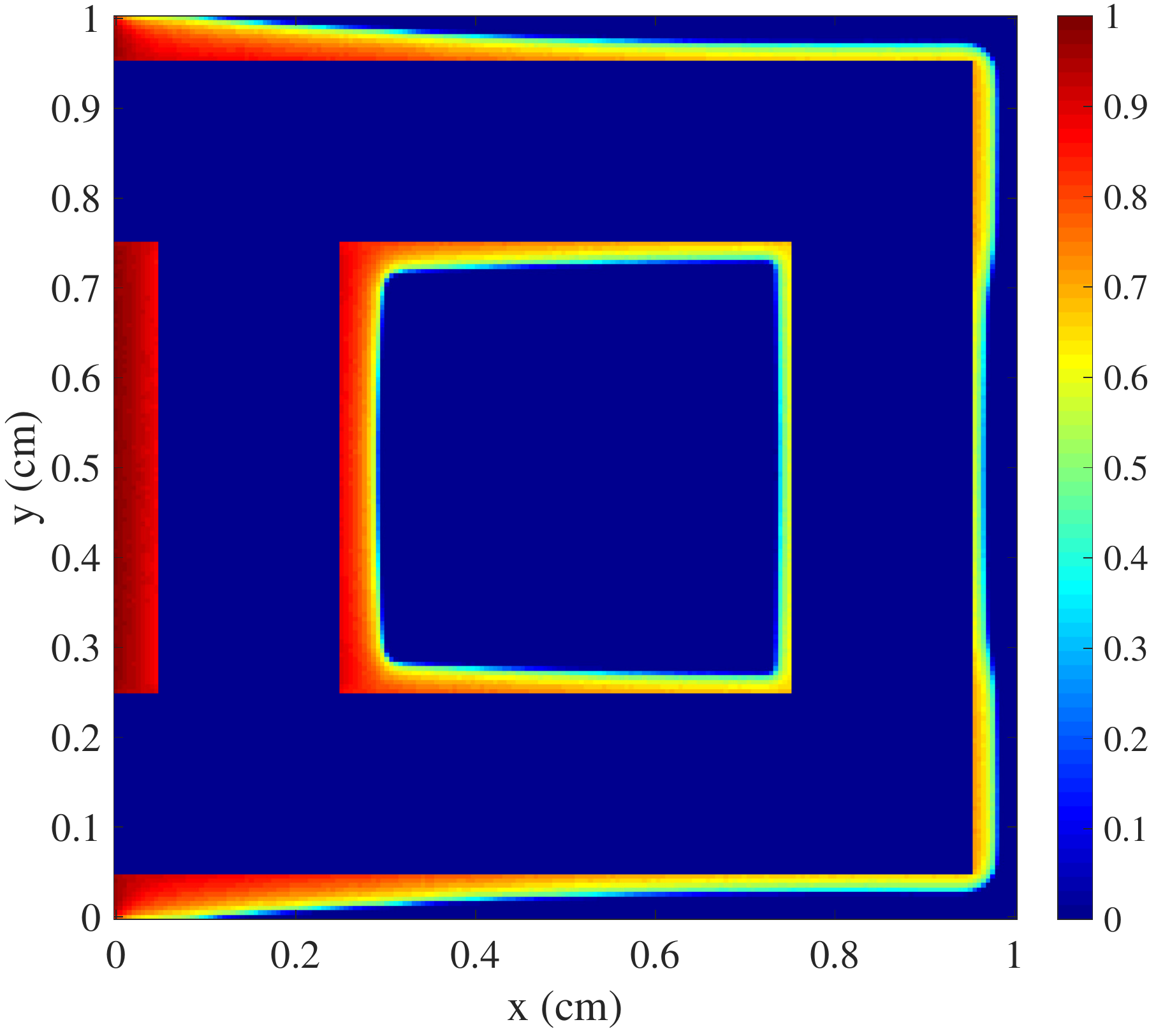}
(e)\includegraphics*[width=5.5cm]{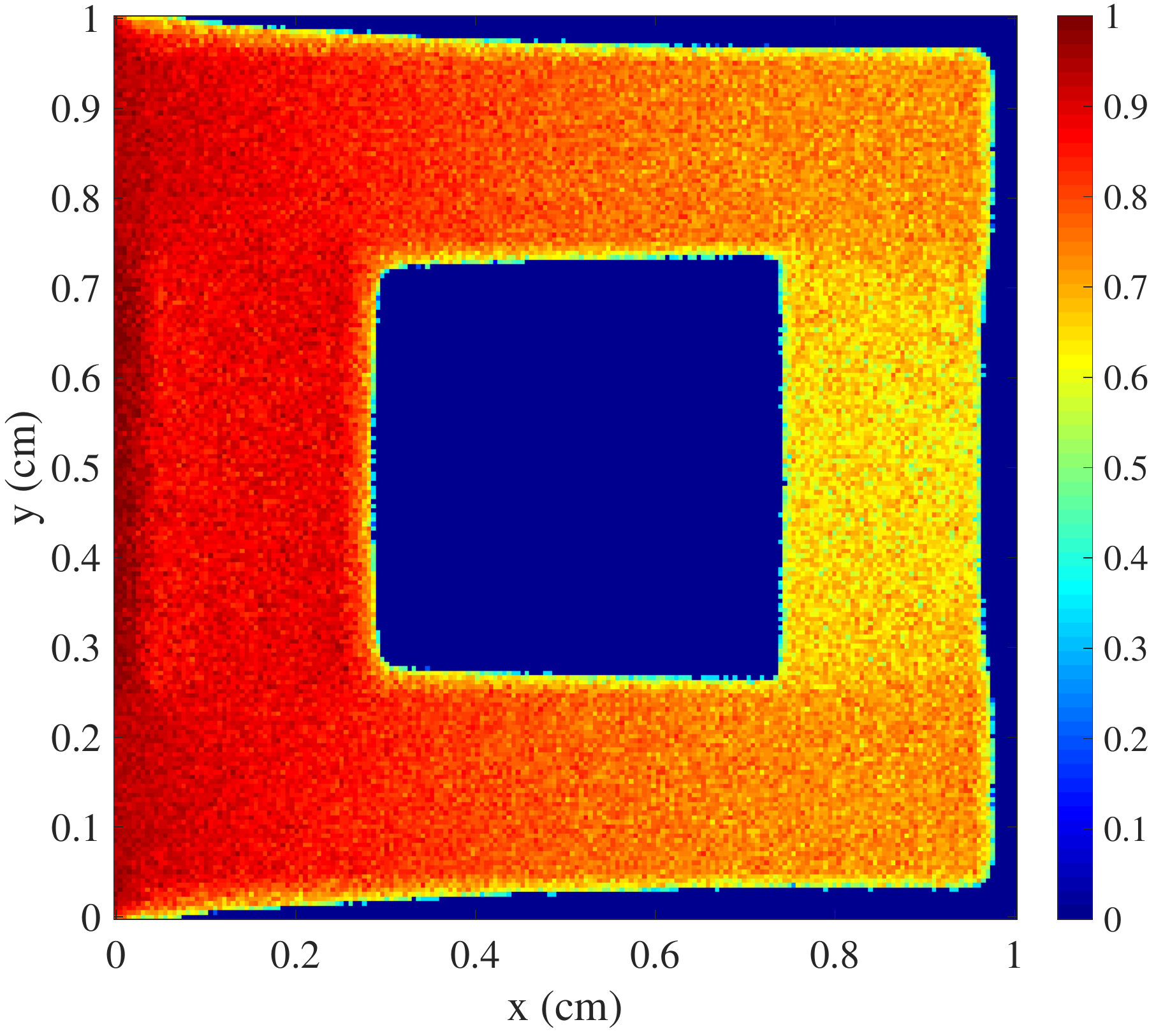}
(f)\includegraphics*[width=5.5cm]{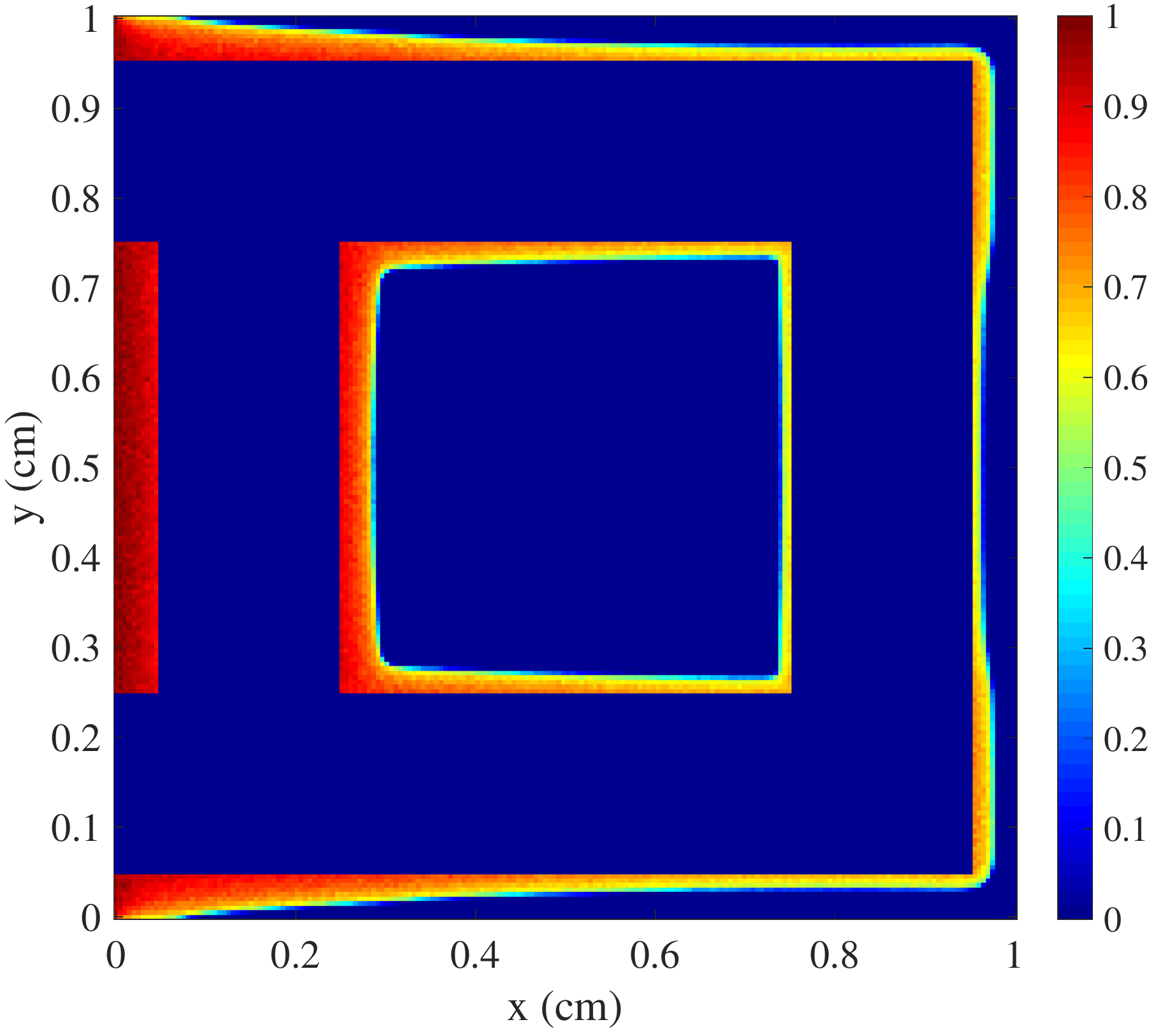}
\caption{The radiation (left) and material (right) temperatures for the McClarren \& Hauck 2010 problem at time $t=1\text{ ns}$. The top row ((a) and (b)) shows the results taken from~\citet{MCCLARREN}, the middle row ((c) and (d)) shows our IMC run and the bottom row ((e) and (f)) shows our ISMC run.}
\label{fig:mcclarren_results}
\end{figure}
\begin{figure}
\centering
(a)\includegraphics*[width=7cm]{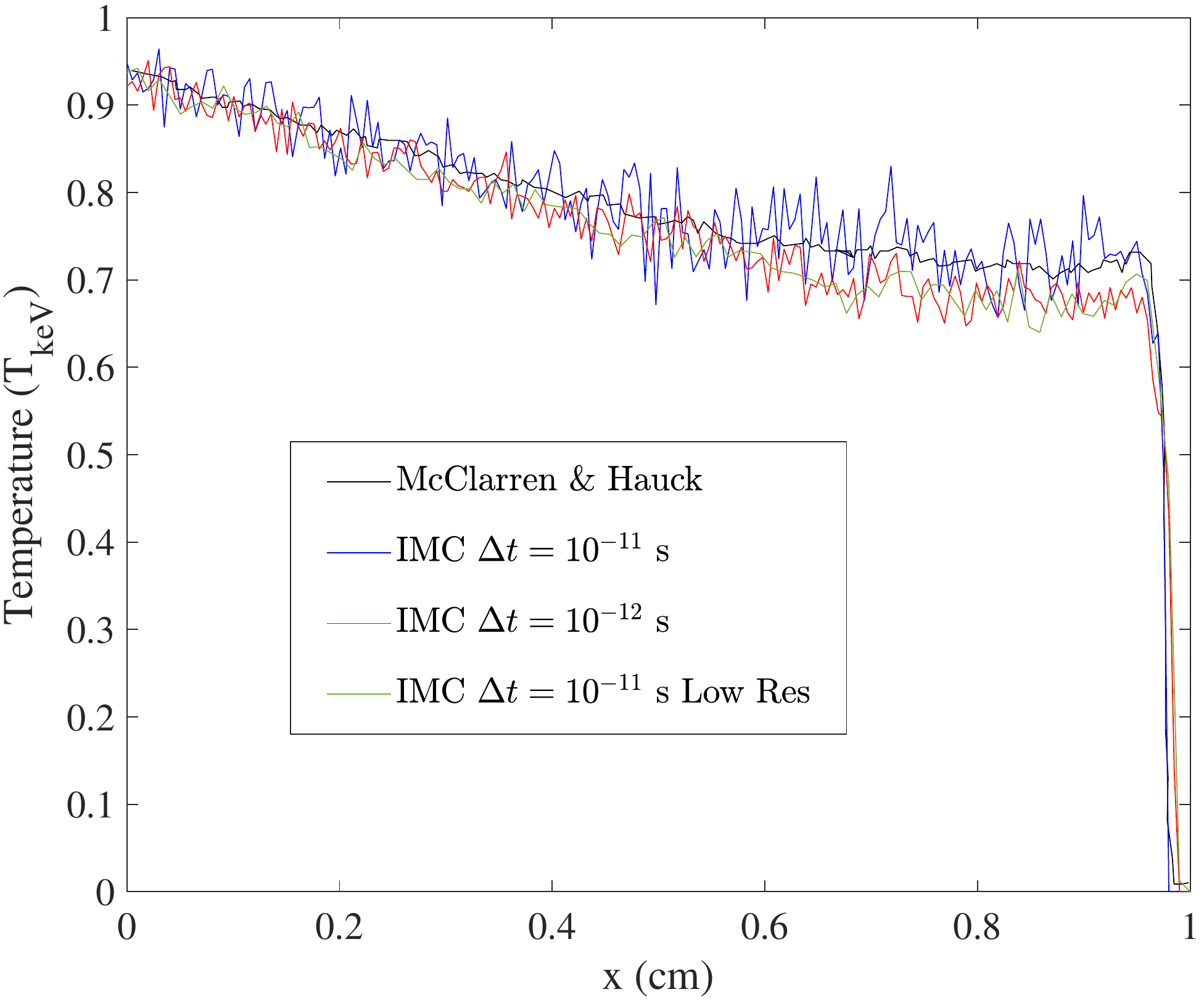}
(b)\includegraphics*[width=7cm]{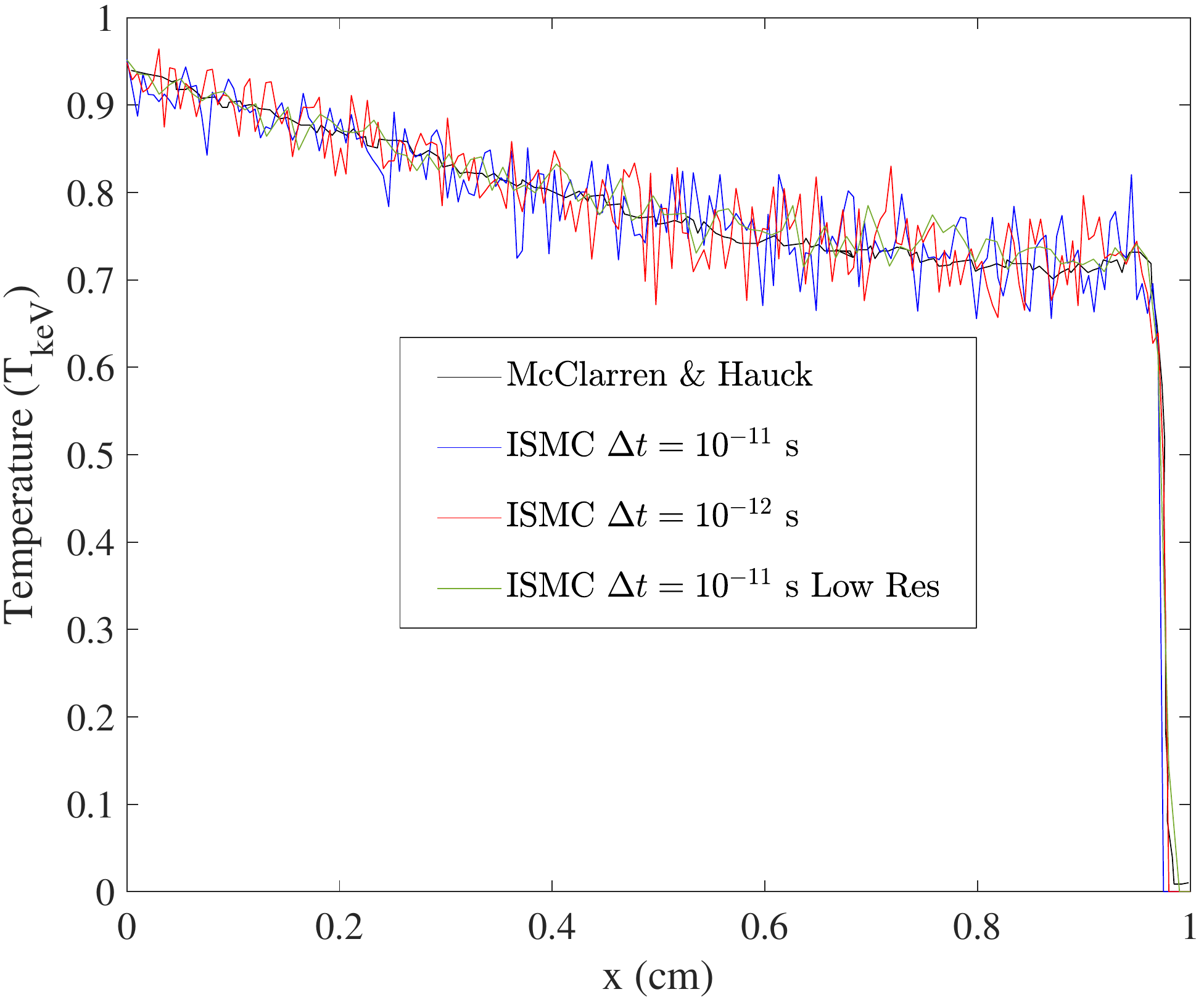}
(c)\includegraphics*[width=7cm]{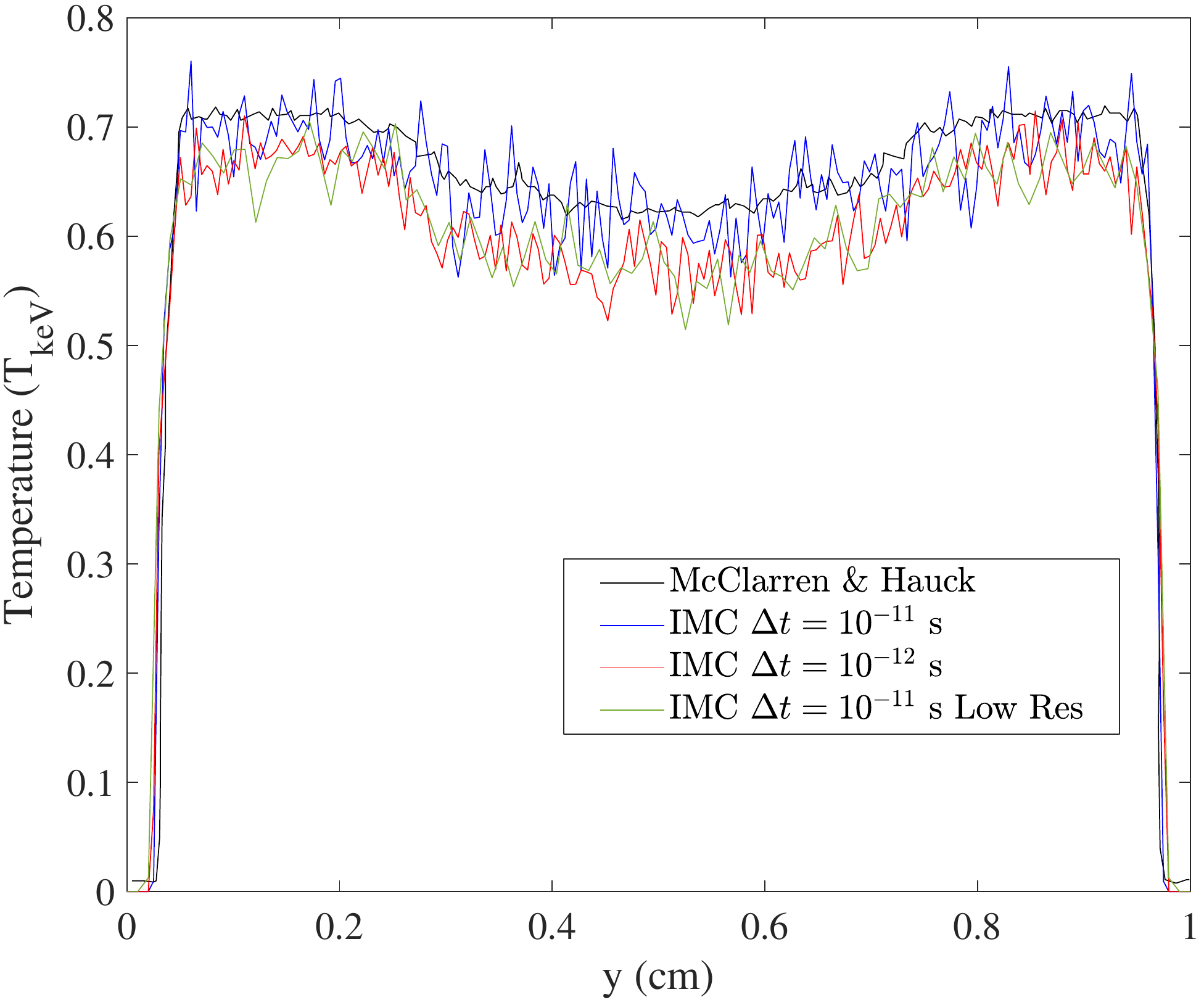}
(d)\includegraphics*[width=7cm]{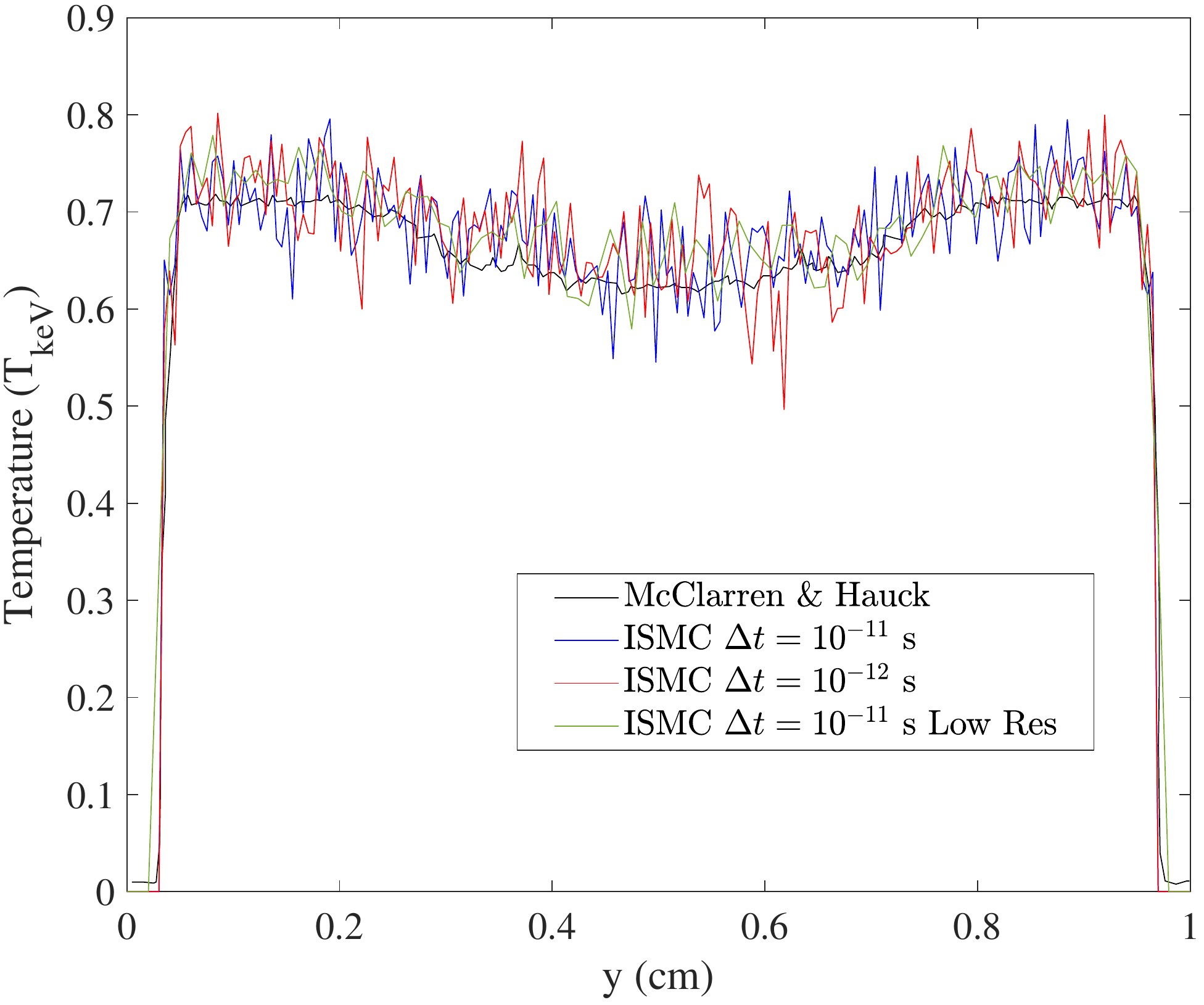}
\caption{The radiation temperature at $y=0.125\text{ (cm)}$ (top) and at $x=0.85\text{ (cm)}$ (bottom), for different time steps and resolutions, at time $t=1\text{ ns}$. IMC results are in the left while ISMC are in the right.}
\label{fig:mcclarren_line}
\end{figure}
In Fig.~\ref{fig:mcclarren_results} we compare our results with the colormaps given in~\citet{MCCLARREN}. Overall, there is good agreement between both methods and the results presented in~\citet{MCCLARREN}. The radiation temperature is slightly higher in the ISMC run due to a slight teleportation error in the IMC run. Again, the radiation energy density is noisier in the ISMC run than the IMC run, due to its discrete nature of material particles.

In Fig.~\ref{fig:mcclarren_line} we present the radiation temperature for the different runs at two slices, a horizontal $y=0.125\text{ (cm)}$ slice, and a vertical $x=0.85\text{ (cm)}$ slice. This enables us to quantify the different schemes, and in specific, the teleportation errors of IMC that vanishes in ISMC. In our nominal spatial resolution and $\Delta t$, both IMC and ISMC yield results that are comparable with~\citet{MCCLARREN}. However, reducing the time step to $\Delta t = 10^{-12}\text{ s}$, or using half the spatial resolution in IMC, yield a too low radiation temperature (see especially in Fig.~\ref{fig:mcclarren_line}(c)). This is because too much energy is teleported into the opaque walls and internal square. In contrast, all of the ISMC runs give similar, and thus, converged results, without regard to the time step or lower resolution.

\subsection{McClarren \& Urbatsch 2009}
The solution to the teleportation error in the ISMC algorithm, is demonstrated well in the following benchmark problem, the RZ~\citet{MCCLARREN_URBATSCH} hohlraum. In addition, this benchmark enables to test our MC implementation in a cylindrical geometry, as opposed to the previous XY hohlraums. This problem is investigated lately in several works~\cite{shi_2020_1,shi_2020_3,shi_2019}, that we compare to in order to benchmark our codes.
\begin{figure}
\centering
\includegraphics*[width=7.5cm]{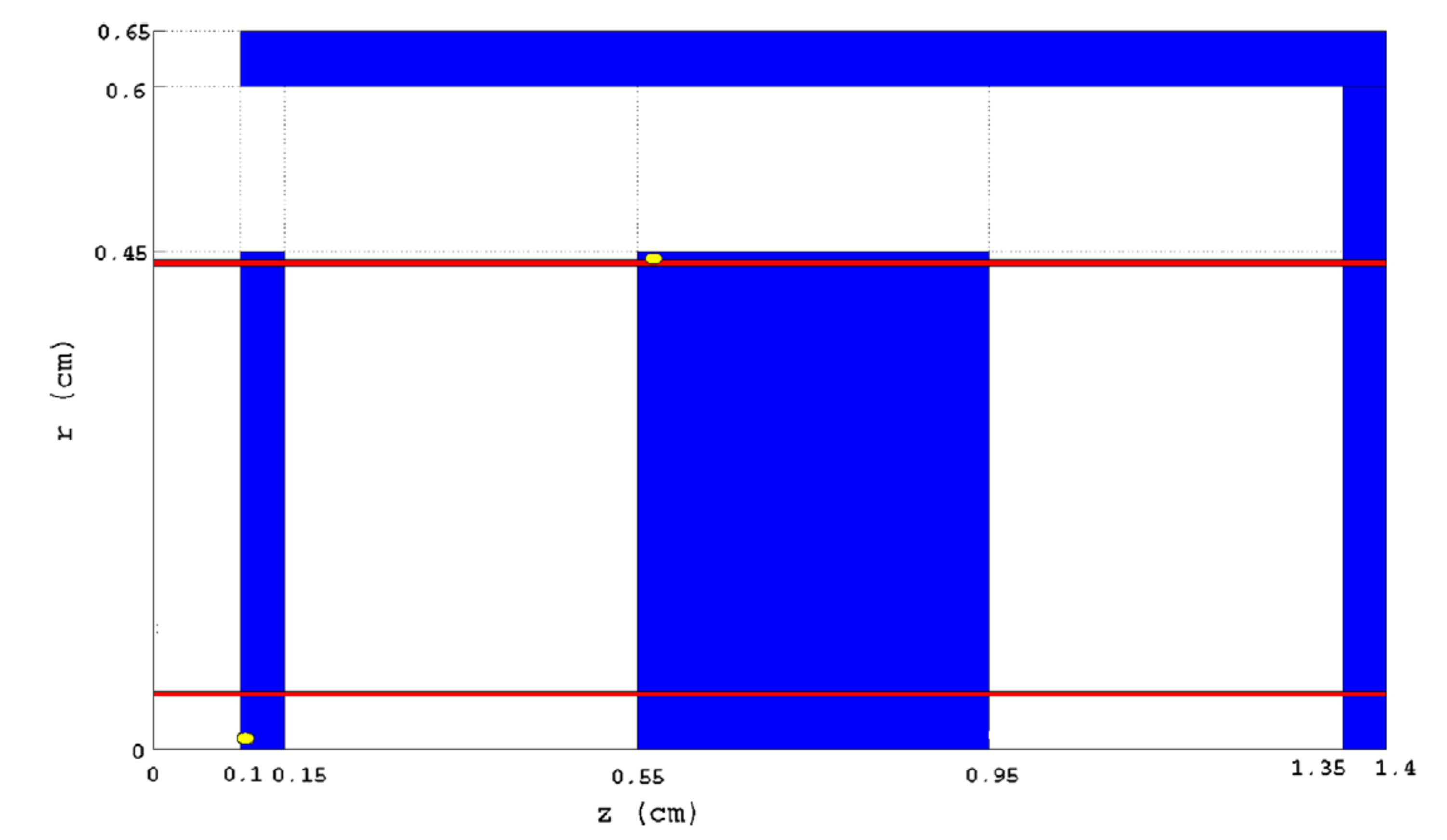}
\caption{The geometry of the~\citet{MCCLARREN_URBATSCH} cylindrical hohlraum problem. The absorbing material is depicted in blue and material depicted in white is vacuum. The figure is taken from~\cite{MCCLARREN_URBATSCH}.}
\label{fig:RZ_geom}
\end{figure}

The problem consists of a cylindrical hohlraum with a radius of 0.65 cm and a length of 1.4 cm, whose geometry and dimensions are given in Fig.~\ref{fig:RZ_geom}. The absorbing material, depicted in blue, has an opacity of $\sigma_a=300\left(T/T_\text{keV}\right)^{-3}\text{cm}^{-1}$, and a constant heat capacity $\rho C_V = 3\cdot 10^{15}\;\text{erg}/T_\text{keV}/\text{cm}^3$. The material depicted in white is vacuum with zero opacity. The $z=0$ boundary is a black body source with a constant temperature of 1keV and all the other boundaries are vacuum. As a nominal resolution, we use a uniform mesh, with a cell size of $\Delta r = \Delta z = 0.01\;\text{cm}$ and a time step of $\Delta t = 10^{-11}\;\text{s}$. The hohlraum is initially cold, and is evolved until time $t=10\;\text{ns}$. For both the IMC and the ISMC runs, we create $10^6$ new particles each time step, and limit the total number of particles to be $5\cdot10^6$. 
\begin{figure}
\centering
(a)\includegraphics*[width=5cm]{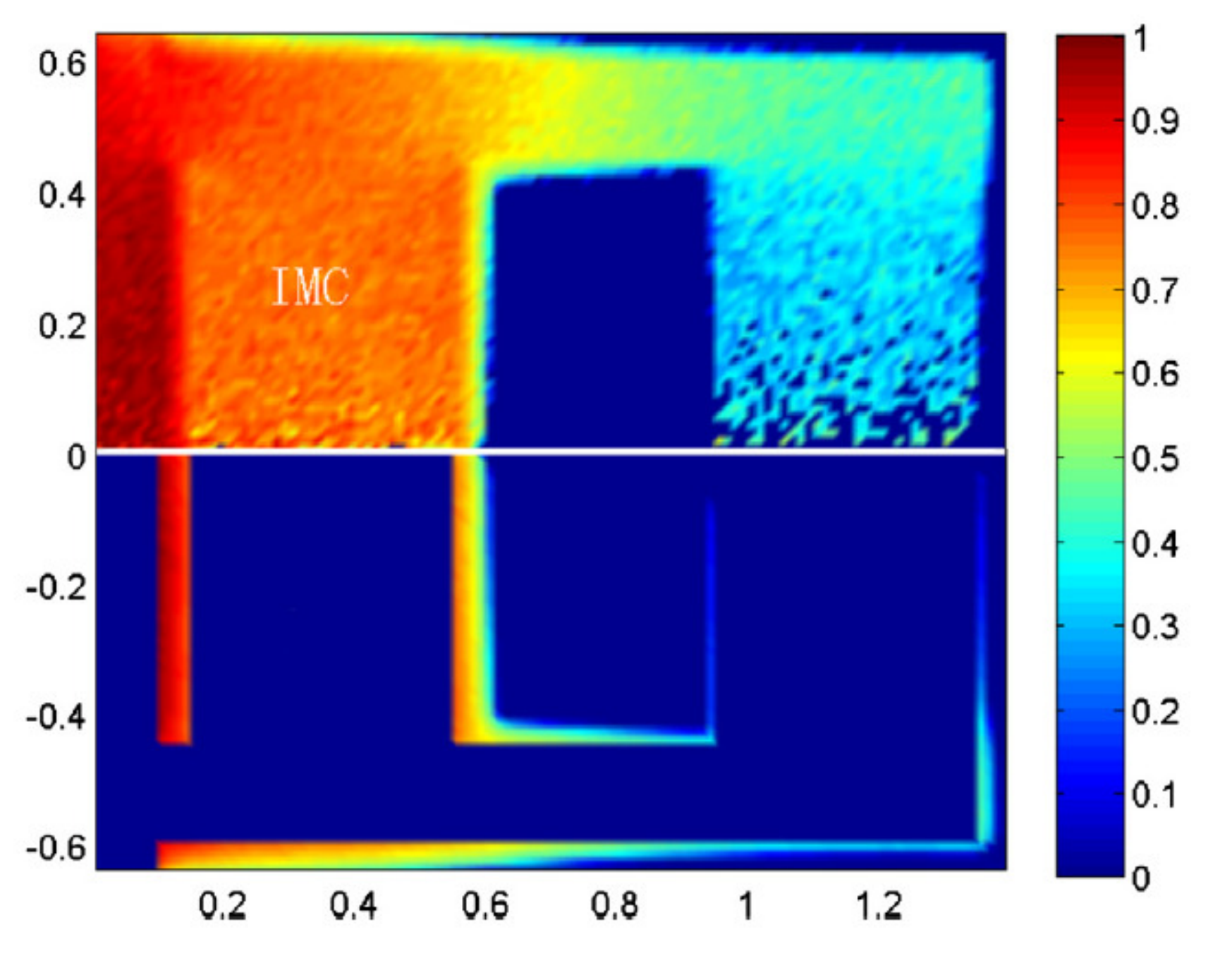}
(b)\includegraphics*[width=5cm]{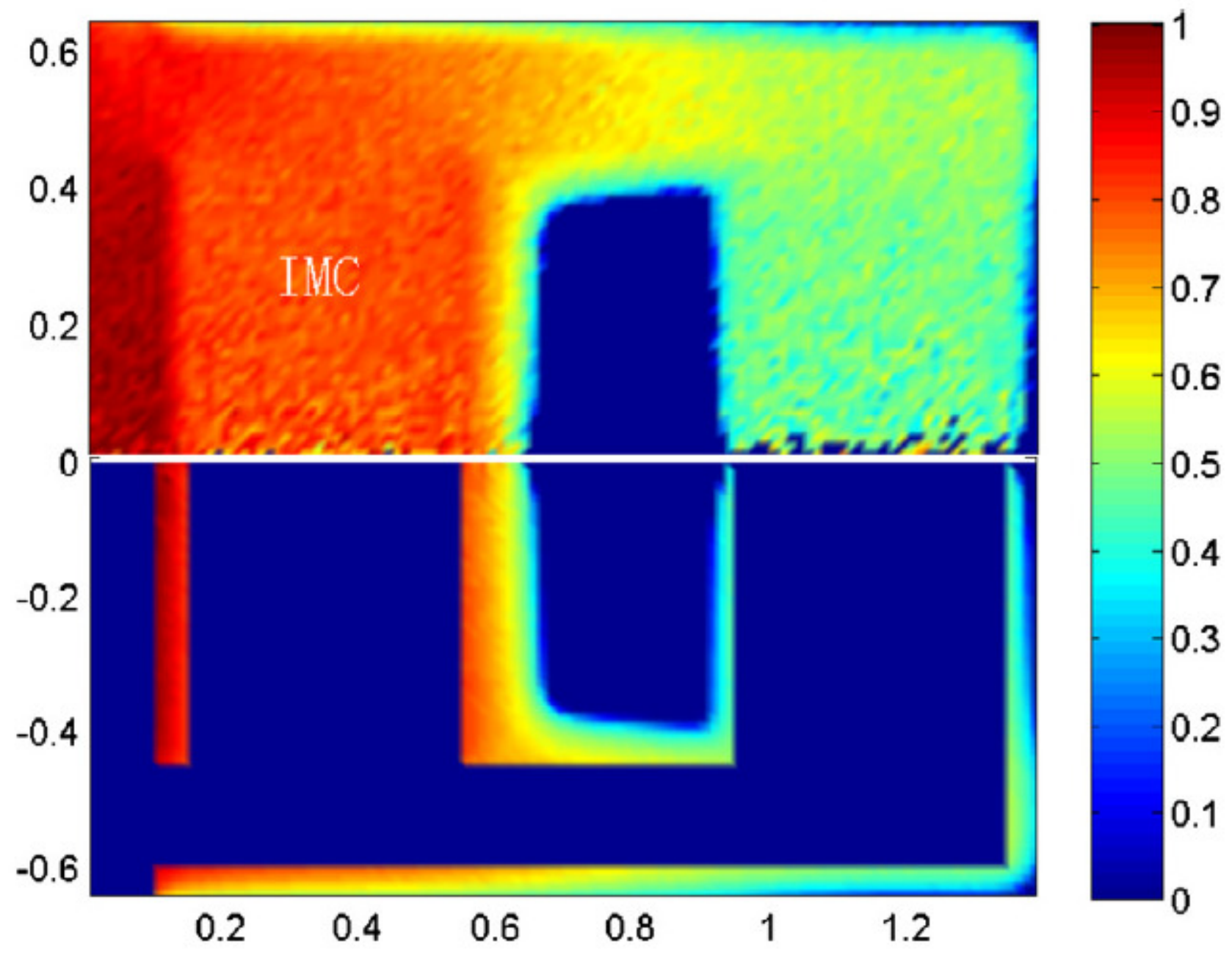}
(c)\includegraphics*[width=5cm]{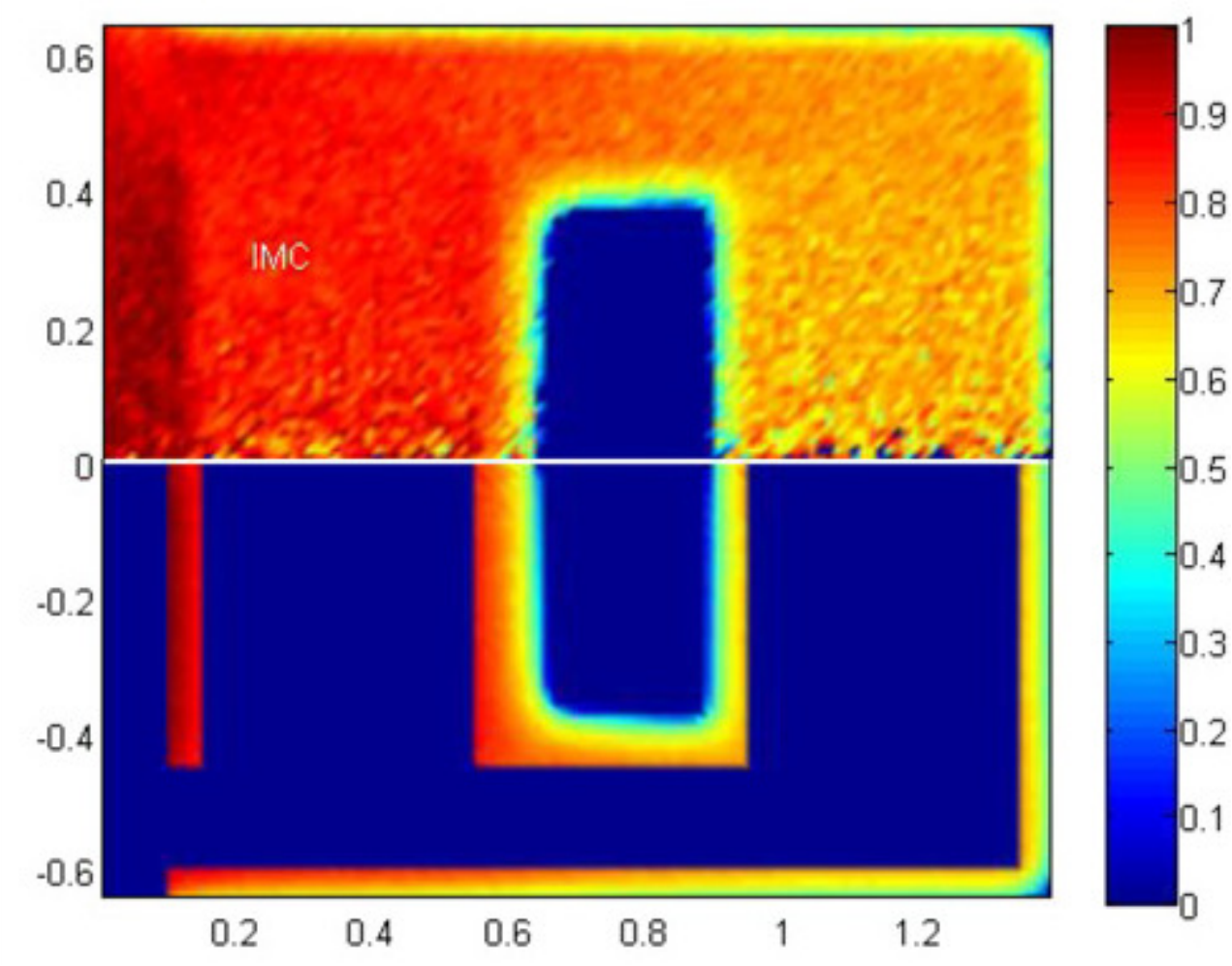}
(d)\includegraphics*[width=5cm]{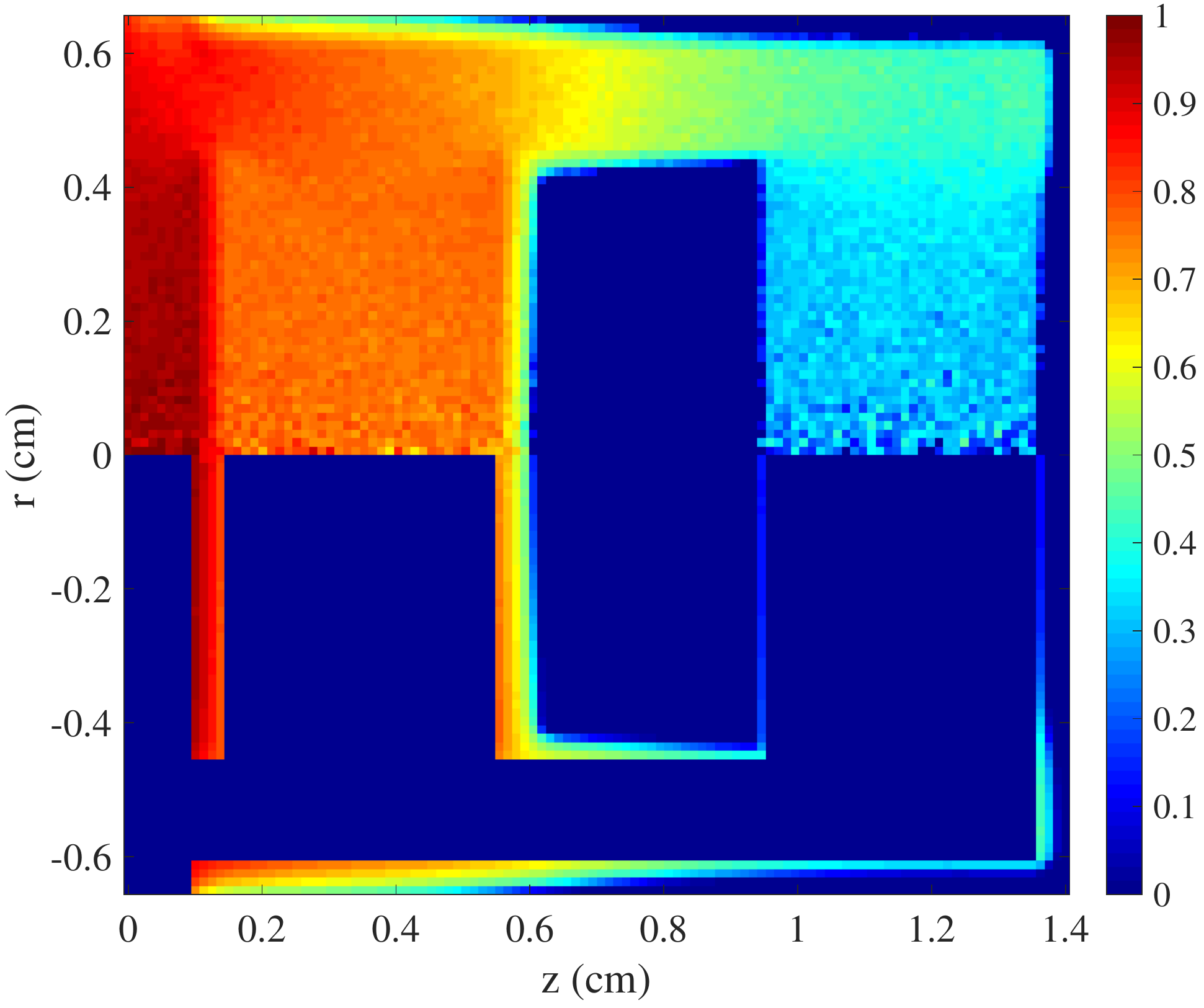}
(e)\includegraphics*[width=5cm]{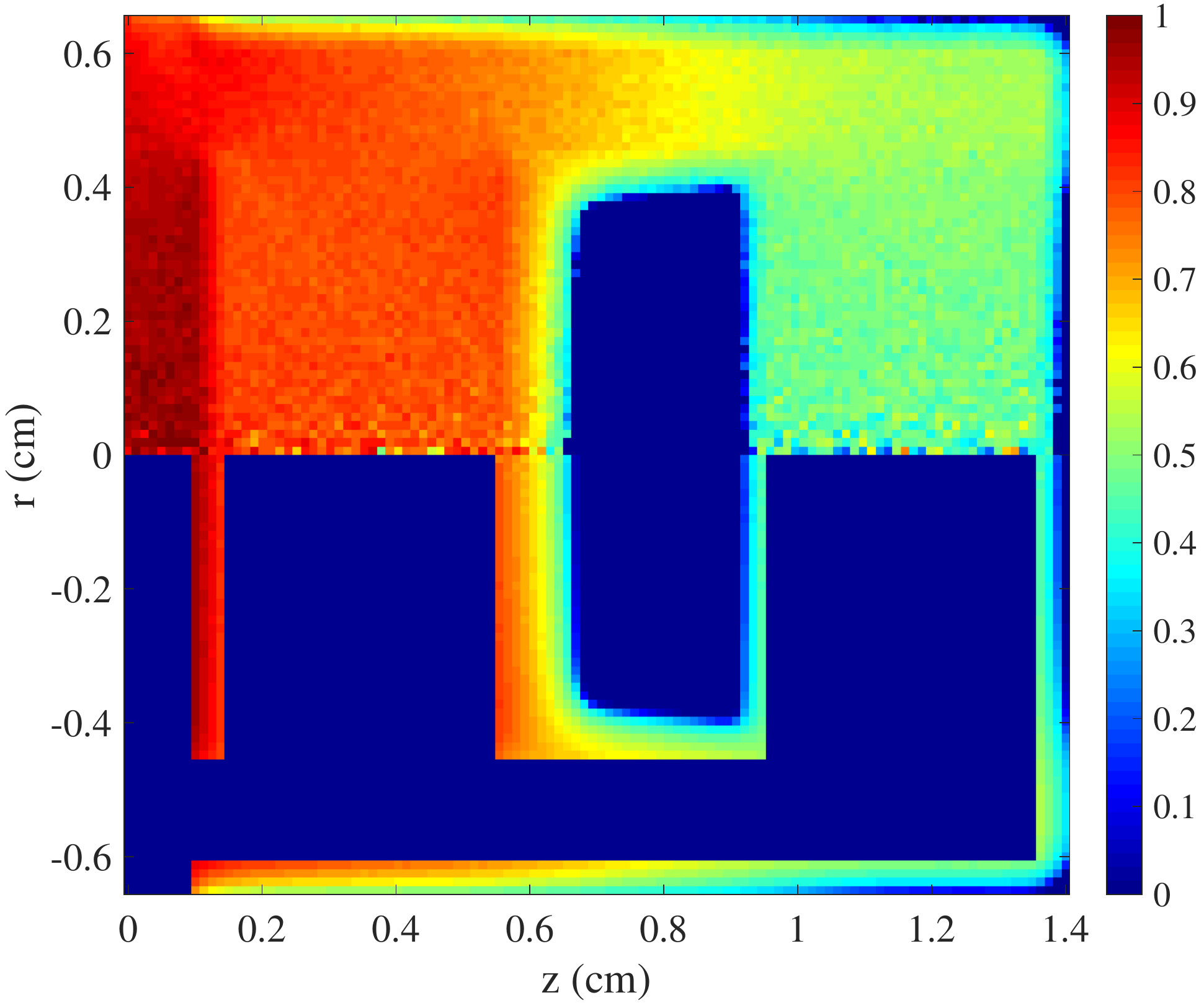}
(f)\includegraphics*[width=5cm]{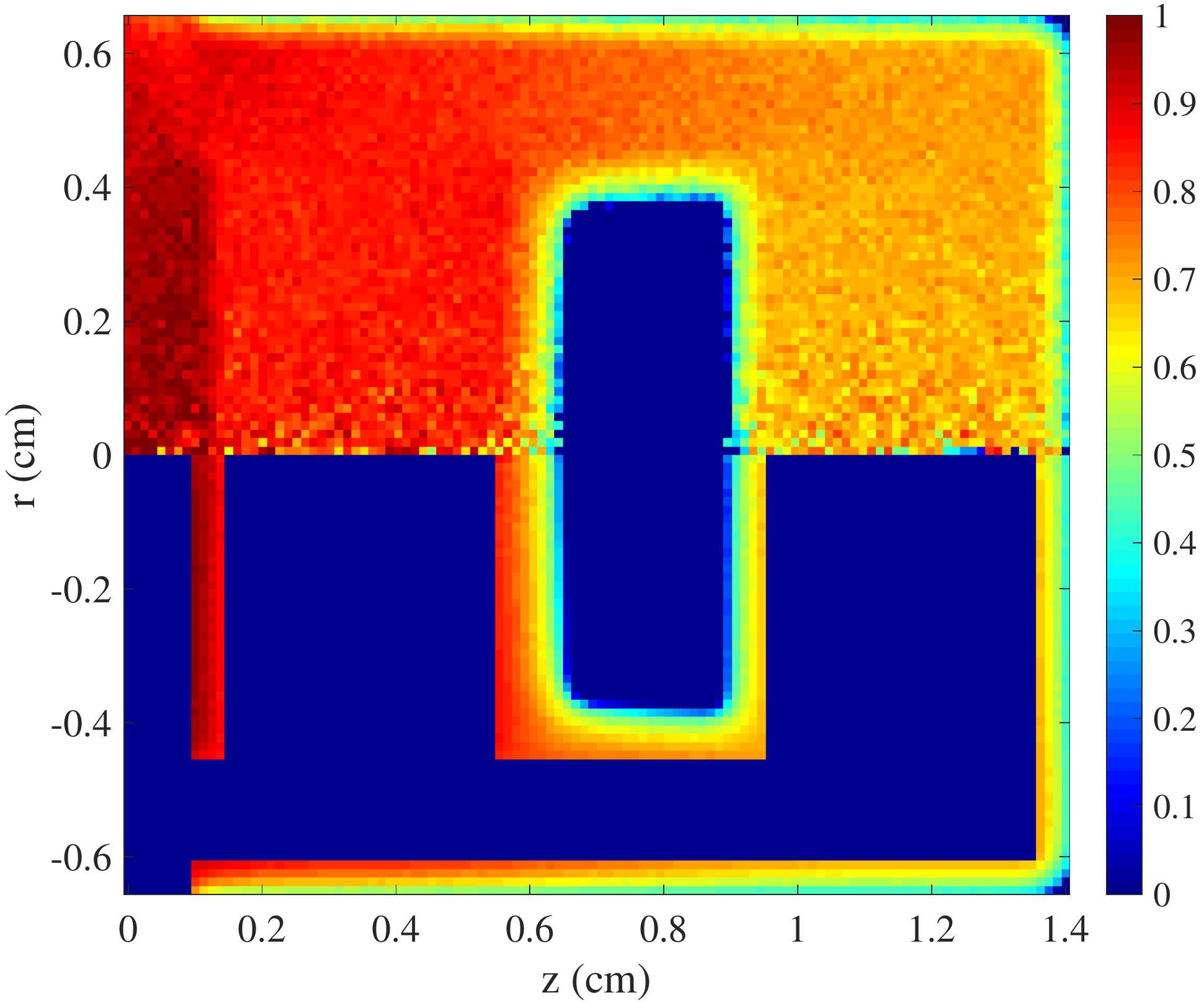}
(g)\includegraphics*[width=5cm]{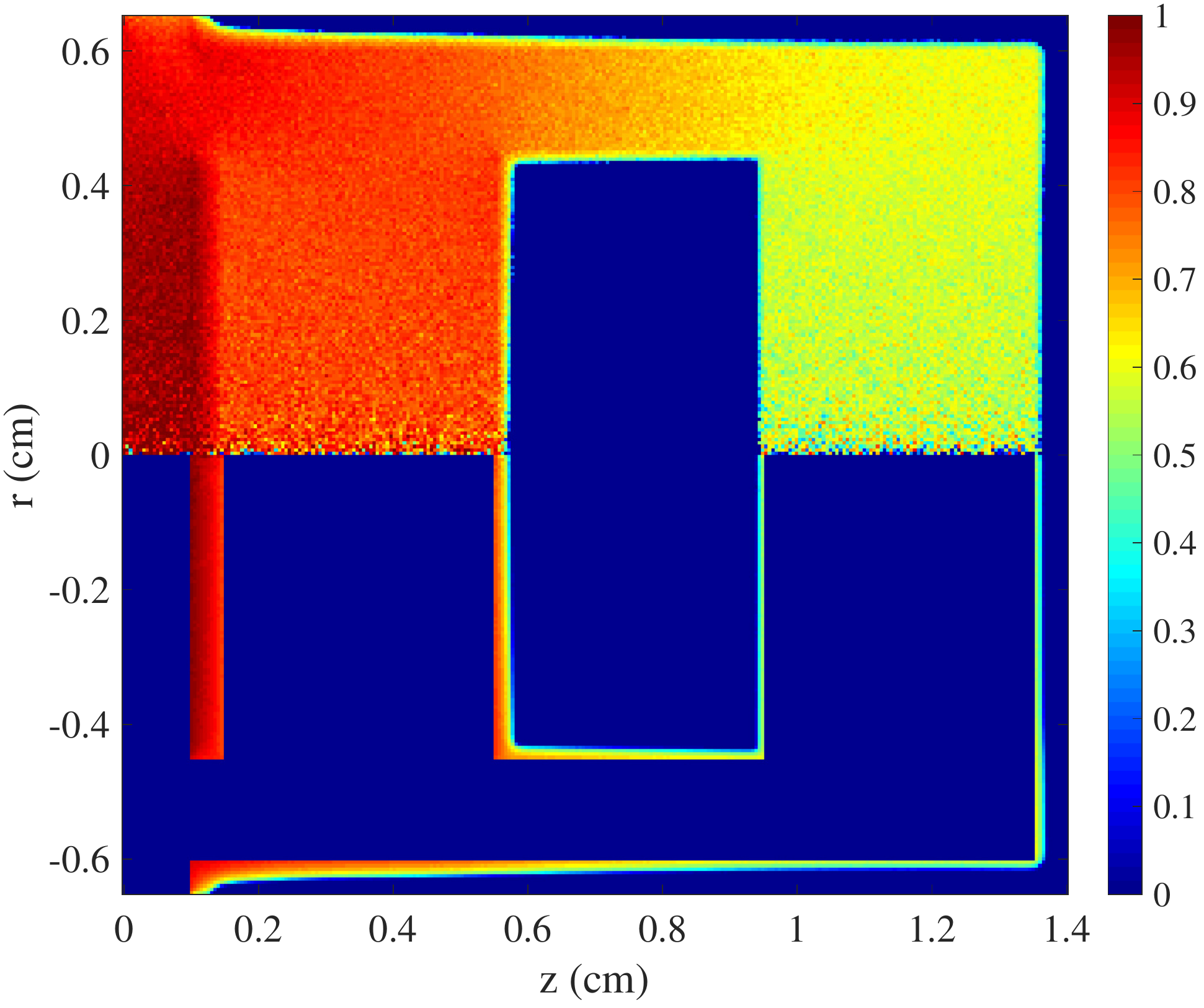}
(h)\includegraphics*[width=5cm]{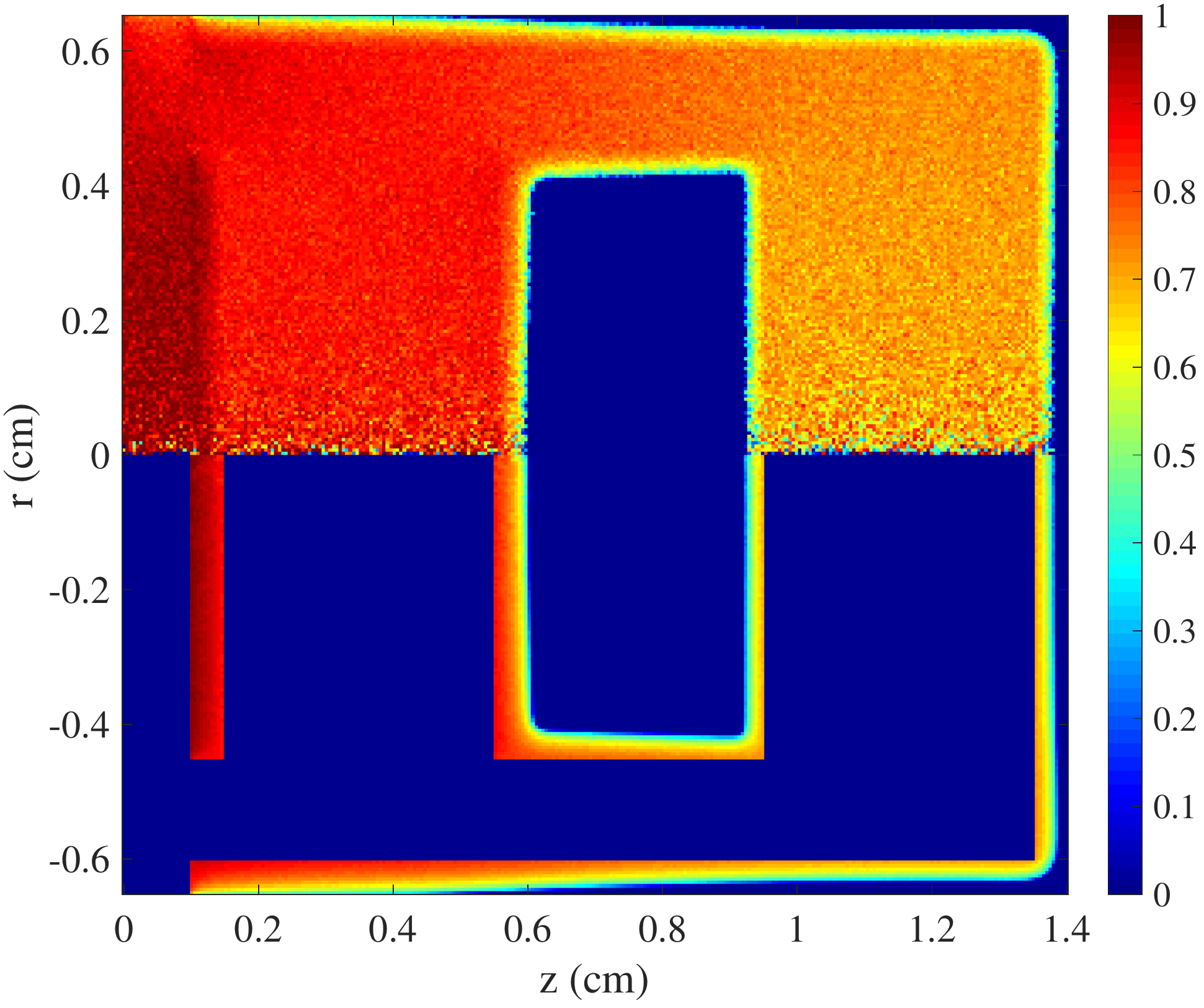}
(i)\includegraphics*[width=5cm]{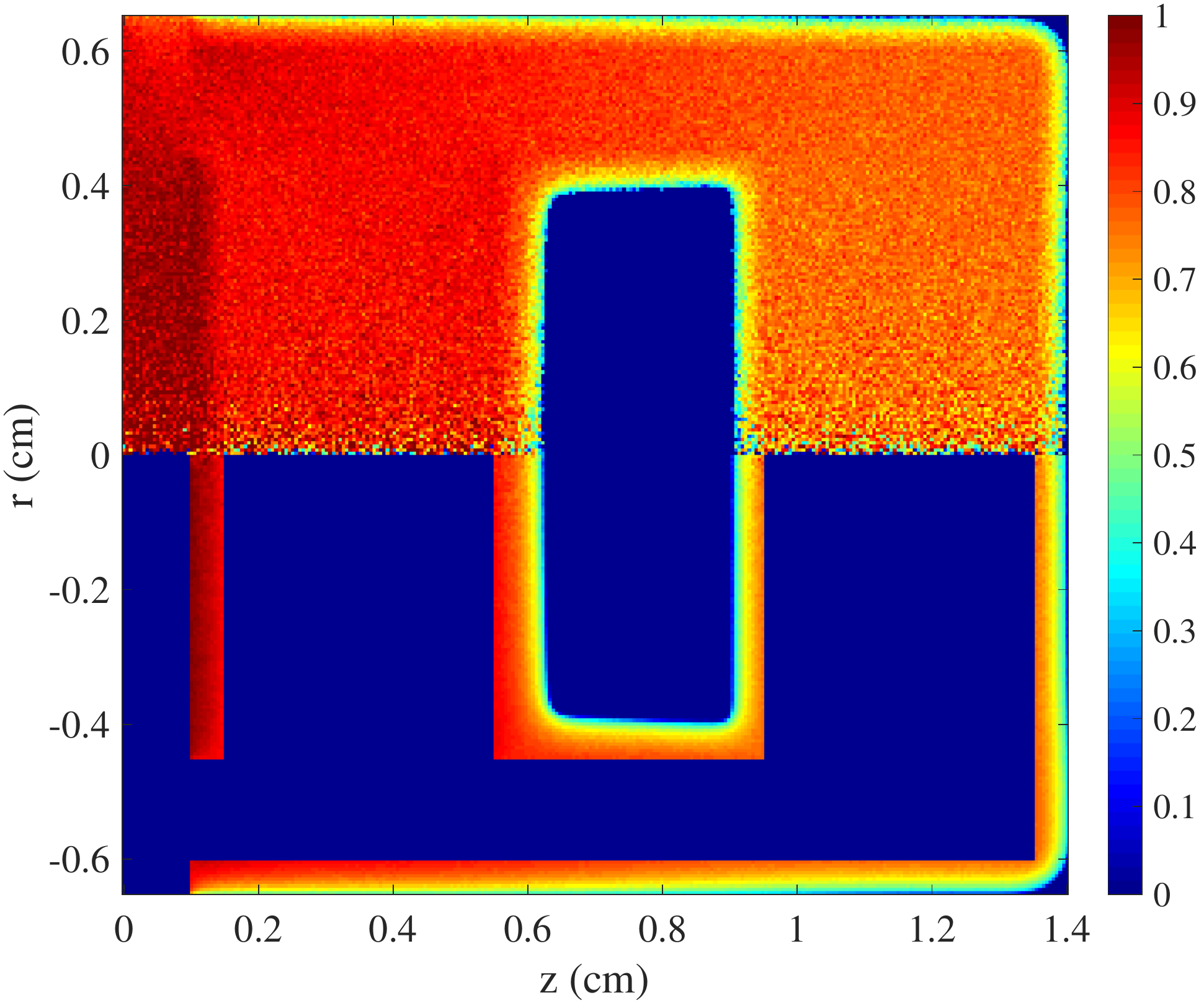}
(j)\includegraphics*[width=5cm]{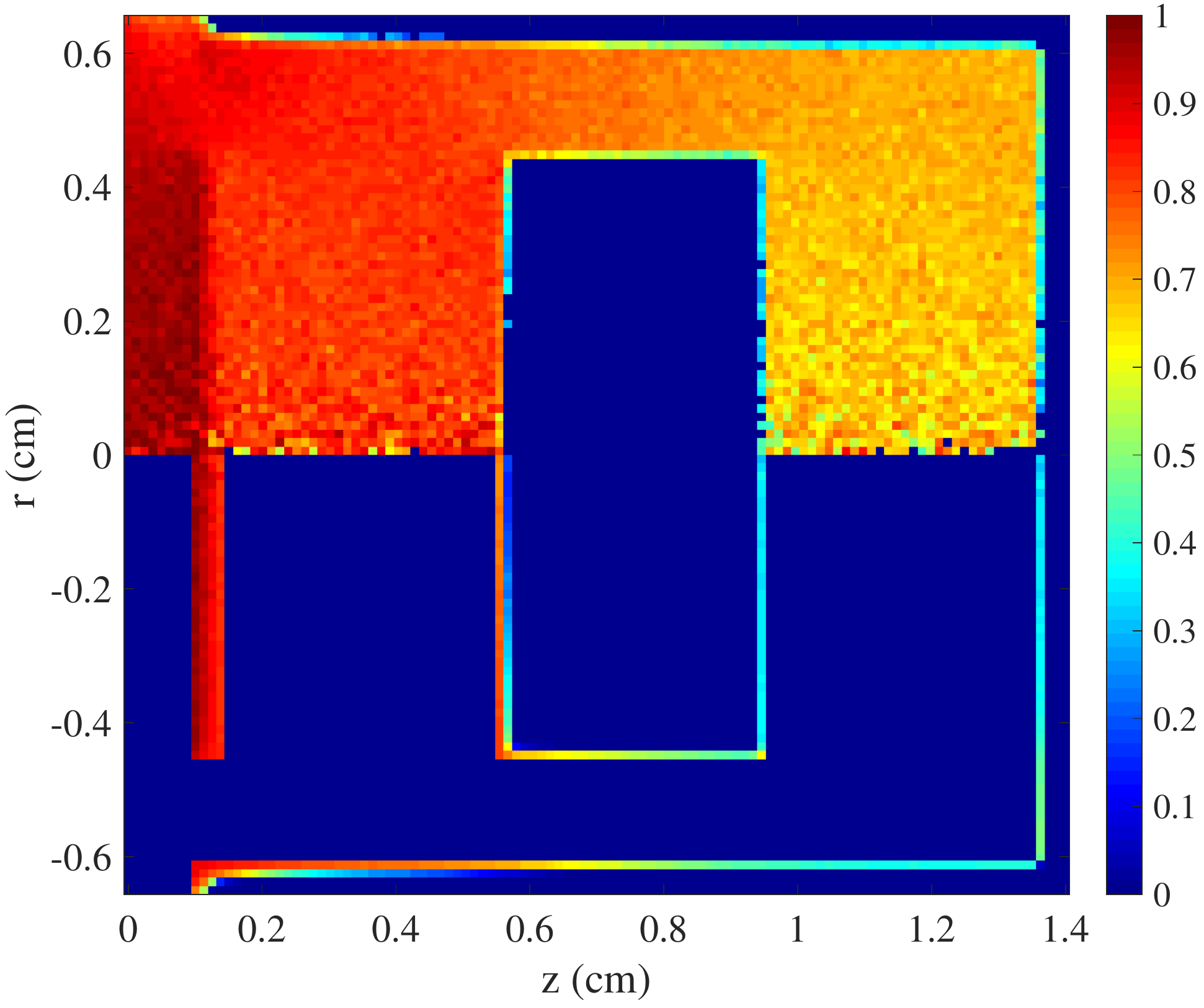}
(k)\includegraphics*[width=5cm]{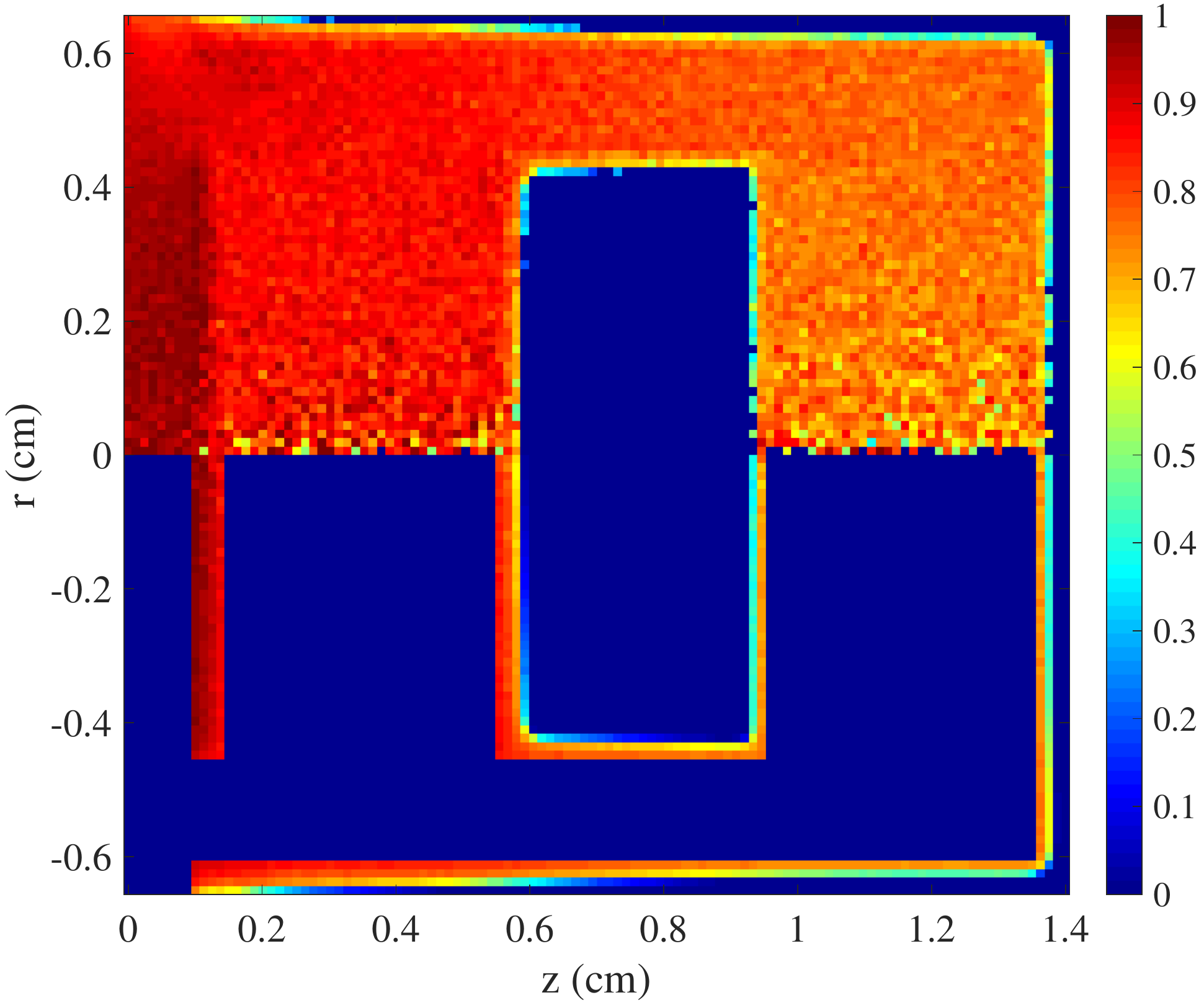}
(l)\includegraphics*[width=5cm]{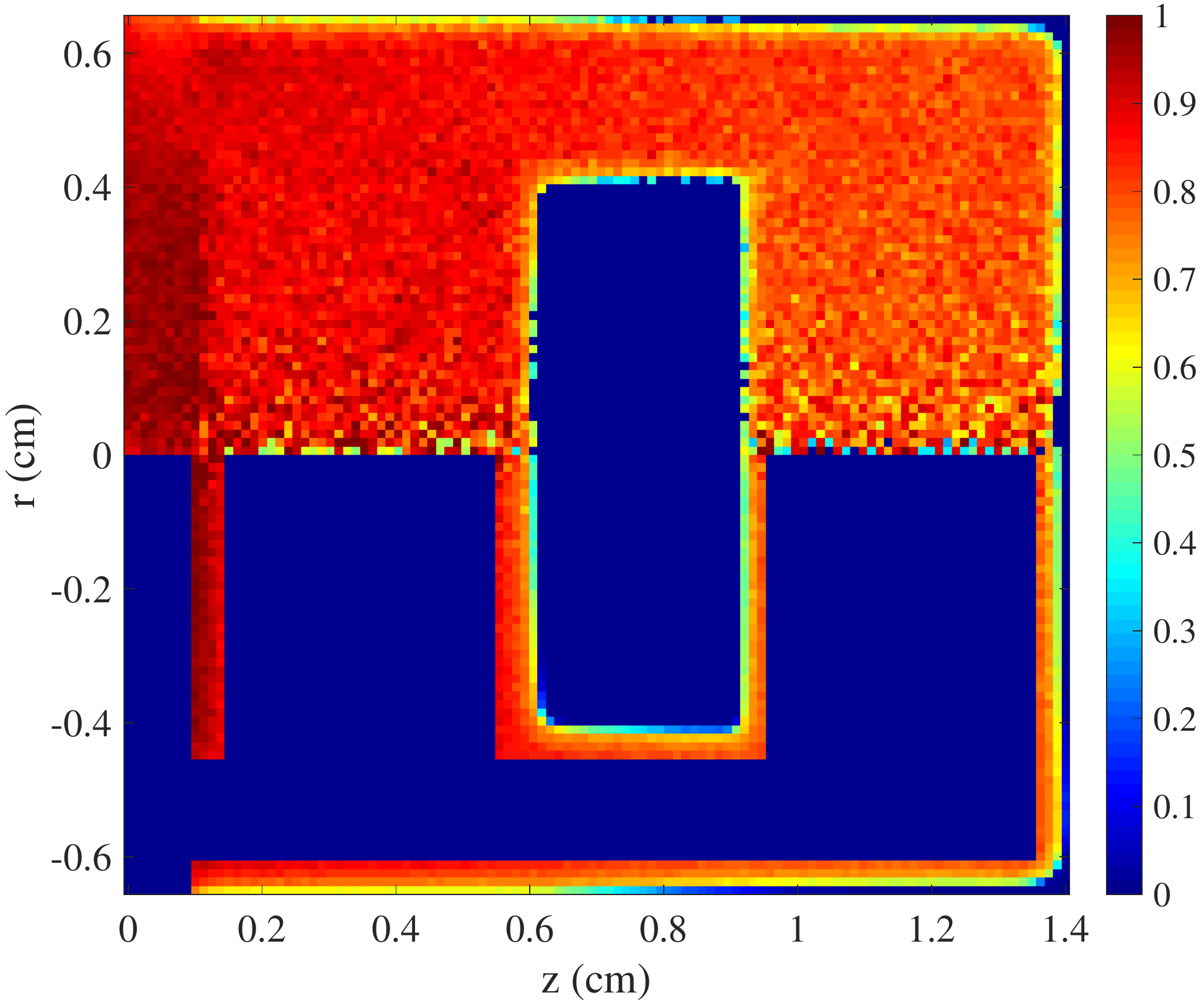}
\caption{The radiation temperature (top half of figure) and material temperature (bottom half of figure) for the cylindrical hohlraum test problem presented in \citet{MCCLARREN_URBATSCH}. In the first row, the temperature is shown for reference from \citep{shi_2020_1}, \citep{shi_2020_3} and \citep{shi_2019}, at times $t=1.2\text{ ns}$ (a), $t=3.2\text{ ns}$ (b) and $t=5.6\text{ ns}$ (c) respectively. For completeness, we state that the time steps used were $\Delta t = 4\cdot10^{-4}$ ns (a), $\Delta t =2\cdot 10^{-3}$ ns (b) and $\Delta t = 0.01$ ns (c). The middle row ((d), (e) and (f)) shows the temperature using our IMC for the same run parameters as \citep{shi_2019, shi_2020_1, shi_2020_3}, while the third row ((g), (h) and (i)) shows the results for our high resolution IMC run with $\Delta t = 10^{-11}\text{ s}$ for comparison. In the bottom row ((j), (k) and (l)) we show the results for our ISMC runs using the parameters of \citep{shi_2019, shi_2020_1, shi_2020_3}.}
\label{fig:RZ_hoh_results}
\end{figure}
\begin{figure}
\centering
(a)\includegraphics*[width=7cm]{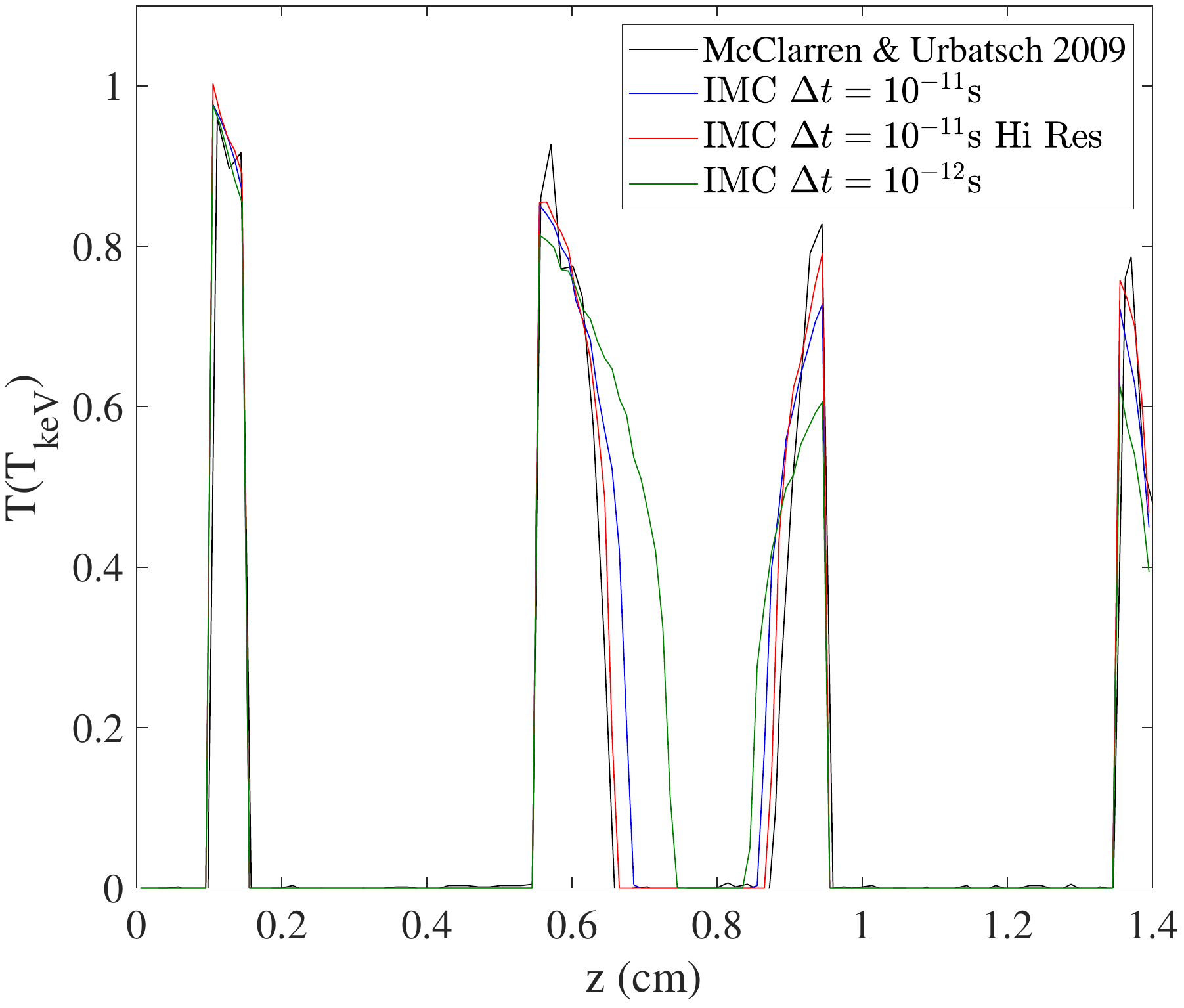}
(b)\includegraphics*[width=7cm]{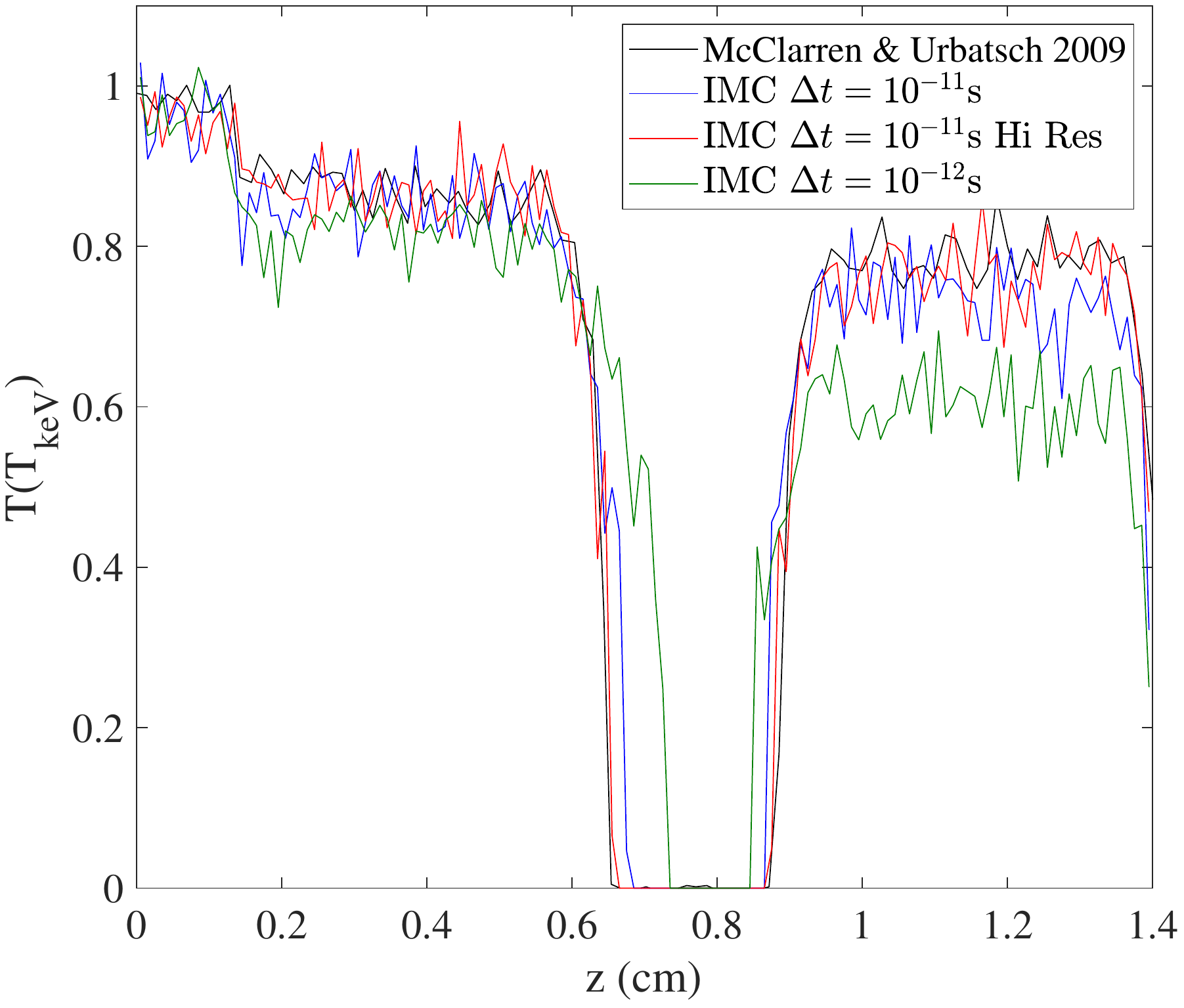}
(c)\includegraphics*[width=7cm]{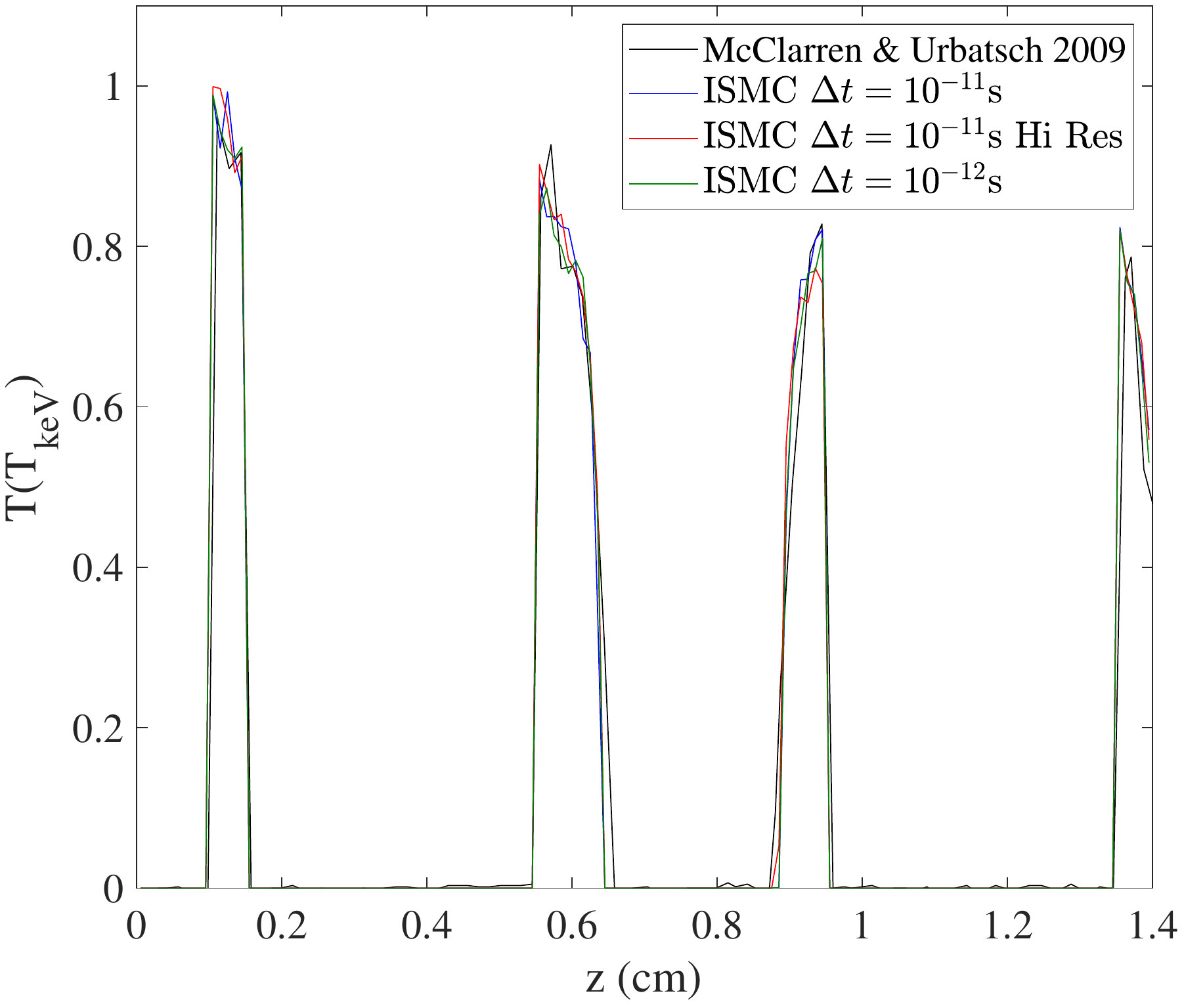}
(d)\includegraphics*[width=7cm]{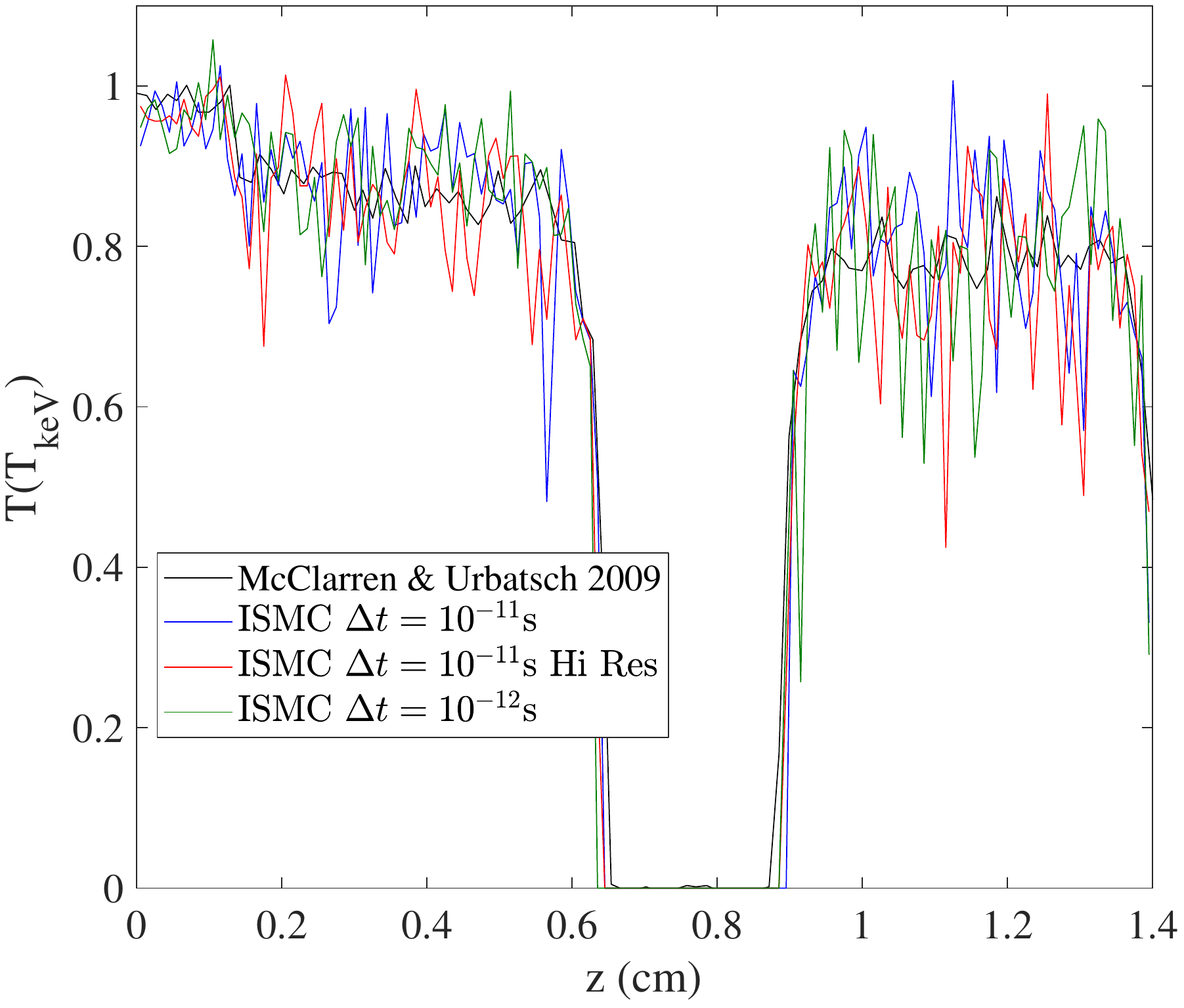}
\caption{The material temperature (left column) and radiation temperature (right column) at time $t=10\text{ ns}$ along the line $r=0.05\text{ cm}$, for the cylindrical hohlraum test problem presented in \citet{MCCLARREN_URBATSCH}. The results for the IMC runs are shown in top row and the results for the ISMC runs in the bottom row.}
\label{fig:RZ_hoh_lines}
\end{figure}

Fig.~\ref{fig:RZ_hoh_results} shows the radiation (top) and material (bottom) temperatures at times $t=1.2\text{ns}$, $t=3.2\text{ns}$ and $t=5.6\text{ns}$ for our IMC and ISMC runs, as well as reference IMC runs taken from \citep{shi_2020_1, shi_2020_3, shi_2019}. First, it can be seen that IMC (Fig.~\ref{fig:RZ_hoh_results}(g-i)) gives results close to the ISMC (Fig.~\ref{fig:RZ_hoh_results}(j-l)), only using the larger time step, $\Delta t = 10^{-11}\;\text{s}$ (and even then, it is not fully converged). This is due to teleportation errors; it can be seen that in the IMC runs, the heat wave penetrates deeper than in the ISMC runs, toward the opaque material, and as a result, the radiation temperature inside the vacuum drops. Moreover, the teleportation error increases dramatically using smaller time-steps (Fig.~\ref{fig:RZ_hoh_results}(d-f)). In these simulations we use the time-steps that were used in the benchmark papers~\cite{shi_2020_1, shi_2020_3, shi_2019}, as a sanity check. We see that the benchmark results (Fig.~\ref{fig:RZ_hoh_results}(a-c)), are very close to our IMC results (Fig.~\ref{fig:RZ_hoh_results}(d-f)), however, both of them contain large teleportation errors, and the heat wave penetrates deeply inside the opaque material. This means that the benchmark IMC results are not converged.
\begin{figure}
\centering
\includegraphics*[width=7.5cm]{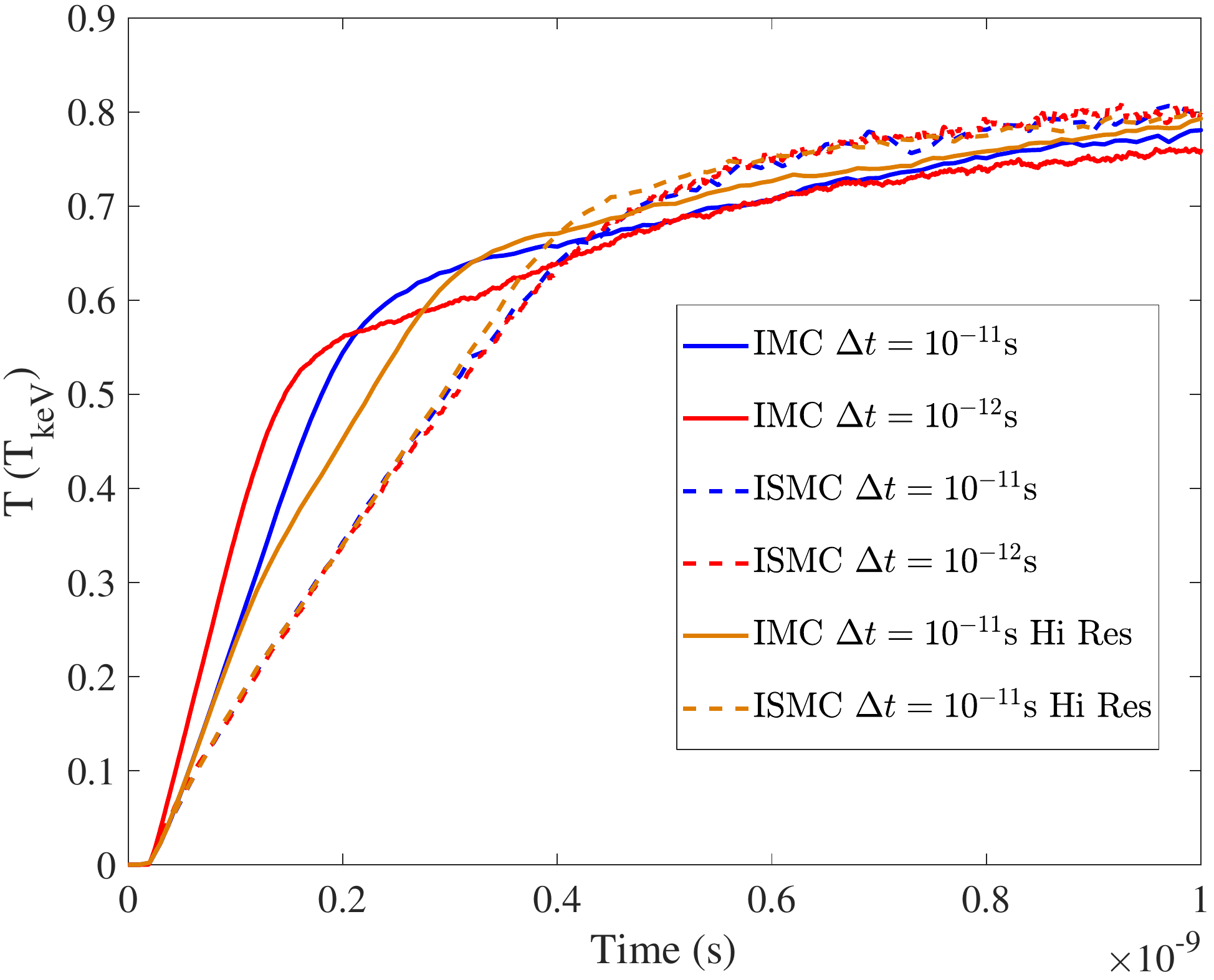}
\caption{The material temperature as a function of time at $(r,z)=(0.44, 0.56)$ for the different runs in the test problem from \citet{MCCLARREN_URBATSCH}.}
\label{fig:cyl_holo_p1}
\end{figure}

To better show the effect of resolution and time step, we perform additional runs, changing the time step to be $\Delta t = 10^{-12}$ s and a run with twice the spatial resolution. Fig.~\ref{fig:RZ_hoh_lines} shows the radiation and material temperature for the different runs at a horizontal slice, $r=0.05\text{ cm}$. The main result is that while the IMC runs show considerable teleportation errors once the time step is decreased (green curves), and a slight teleportation with the spatial resolution (blue curves), all of the ISMC runs give comparable results, without regard to the time step or higher resolution. ISMC is completely converged both in spatial resolution and in time step. In addition, we plot in Fig.~\ref{fig:cyl_holo_p1} the temperature as a function of time at the point $(r,z)=(0.44, 0.56)$ for the different runs. We see once more that all of the ISMC runs are converged and agree with each other, while the IMC runs all differ from each other, and convergence is questionable.

\subsection{Graziani's ``crooked-pipe'' problem (2000)}
\begin{figure}
\centering
\includegraphics*[width=10cm]{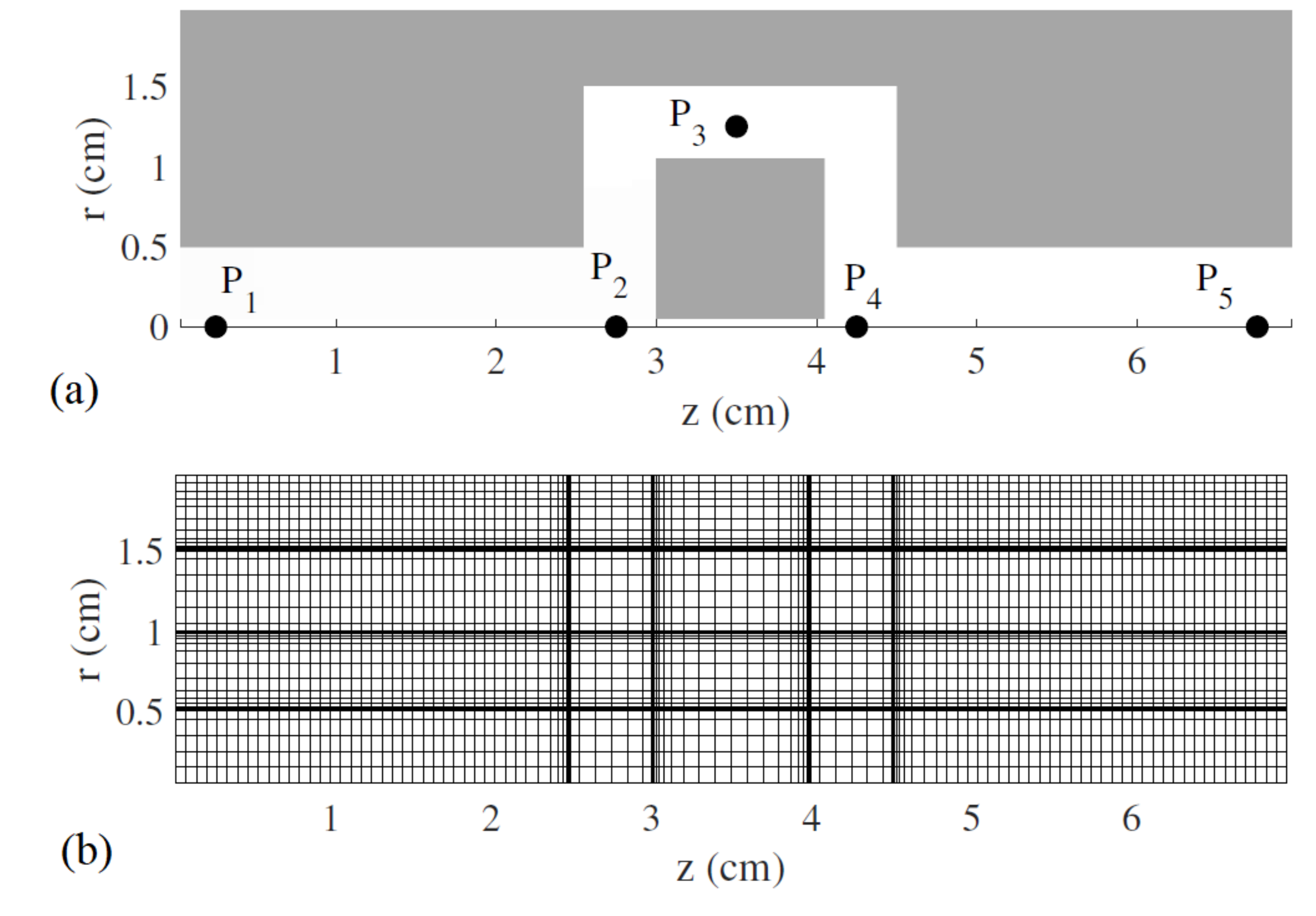}
\caption{(a) The geometry for the Gentile 2001 crooked pipe test problem. The high opacity material is in gray and the low opacity material is in white. (b) The grid used in the problem.}
\label{fig:pipe_geo}
\end{figure}

A challenging test that measures the ability of a scheme to deal with teleportation errors in a complicated radiation-flow regime is the Graziani crooked pipe problem, as presented by \citet{Gentile2001,teleportation}, using RZ geometry. The geometrical setup of the problem is presented in Fig.~\ref{fig:pipe_geo}(a), showing the high opacity material in gray and low opacity material in white.

At $z=0$ boundary there is a black-body source with a temperature of 0.5keV at the edge of the low opacity material. The high opacity material has an absorption opacity of $\sigma_a=2000\text{ cm}^{-1}$, and a constant heat capacity of $ C_V = 10^{16}\;\text{erg}/T_\text{ keV}/\text{cm}^3$, while the low opacity material has an absorption opacity of $\sigma_a=0.2\text{ cm}^{-1}$, and a constant heat capacity of $ C_V = 10^{13}\;\text{erg}/T_\text{ keV}/\text{cm}^3$. At time $t=0$ the domain has a temperature of 0.05keV and the initial time step is set to be $10^{-11}\text{ s}$. As given in~\citet{teleportation}, the time step is increased by a factor of $1.1$ every time step, until it reaches its maximal value of $10^{-8}\text{ s}$.
The spatial grid has a resolution of $0.1\text{ cm}$, except in the interfaces between the high and low opacity regions, where a logarithmically spaced grid is used, as shown in Fig. \ref{fig:pipe_geo}(b). In each interface, the logarithmically spaced grid is composed of 10 cells, where we vary the size of the smallest cell to be $1.5\cdot 10^{-3}\text{ cm},\;2.5\cdot 10^{-3}\text{ cm}$ or $4\cdot 10^{-3}\text{ cm}$ (this choice enables to check the spatial resolution convergence, due to the size of the first opaque cell, since the heat wave does not propagate too much inside the opaque material).

A total of $10^6$ photons are created each time step in the IMC scheme, and we limit the total number of photons to $10^7$. In contrast, in the ISMC scheme we create $10^7$ new photons each time step and limit the total number of particles to be $2\cdot 10^8$. We note that this problem is very challenging for the ISMC scheme, since the probability that a photon will be absorbed in the cells where we measure the temperature (the optically-thin material) is very small. Thus, an extremely large number of particles is needed to get reasonable noise. This is in contrast with the IMC scheme, where each photon that travels through a cell deposits a fraction of its energy, due to the continuous absorption mechanism~\cite{IMC}. 
\begin{figure}
\centering
(a)\includegraphics*[width=7.5cm]{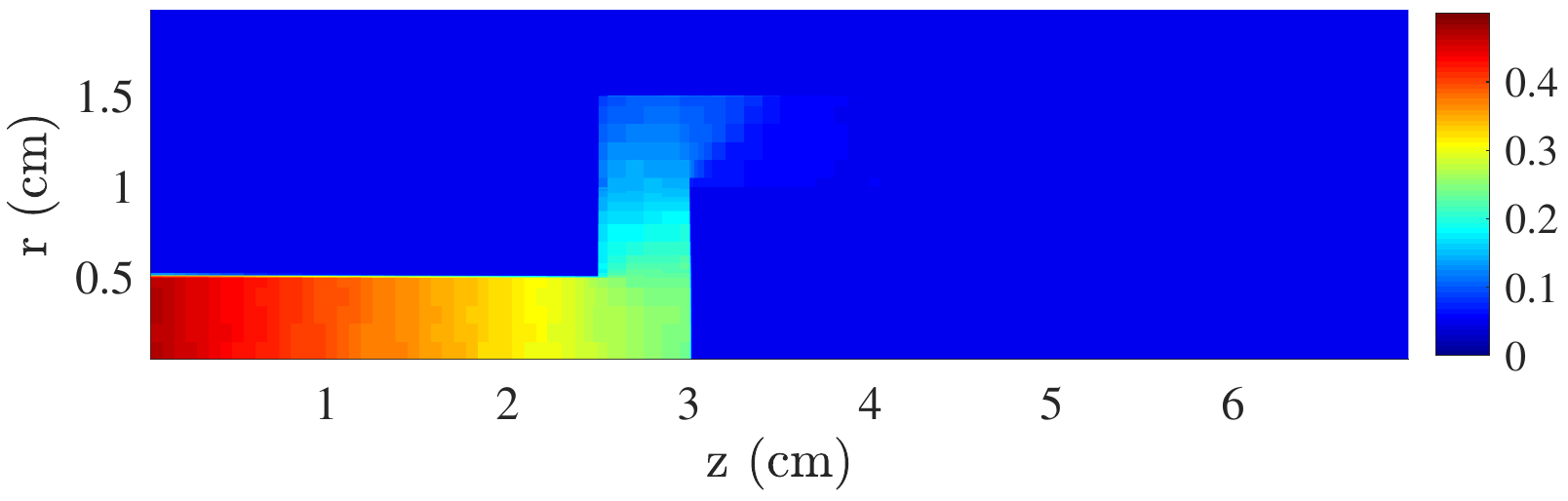}
(b)\includegraphics*[width=7.5cm]{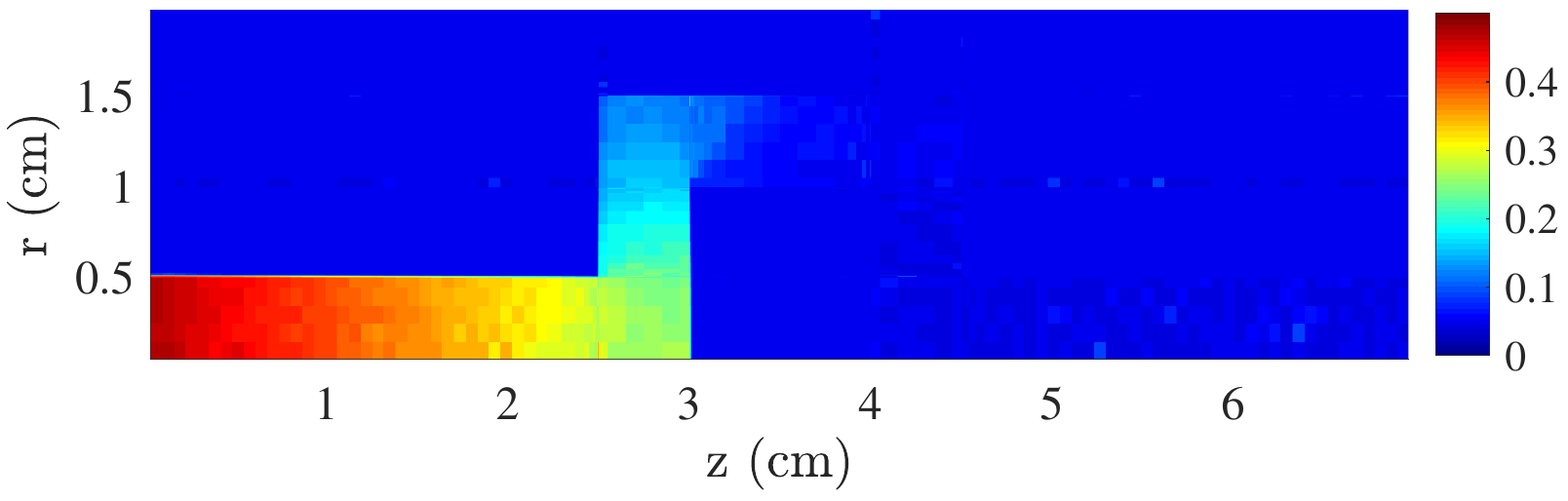}
(c)\includegraphics*[width=7.5cm]{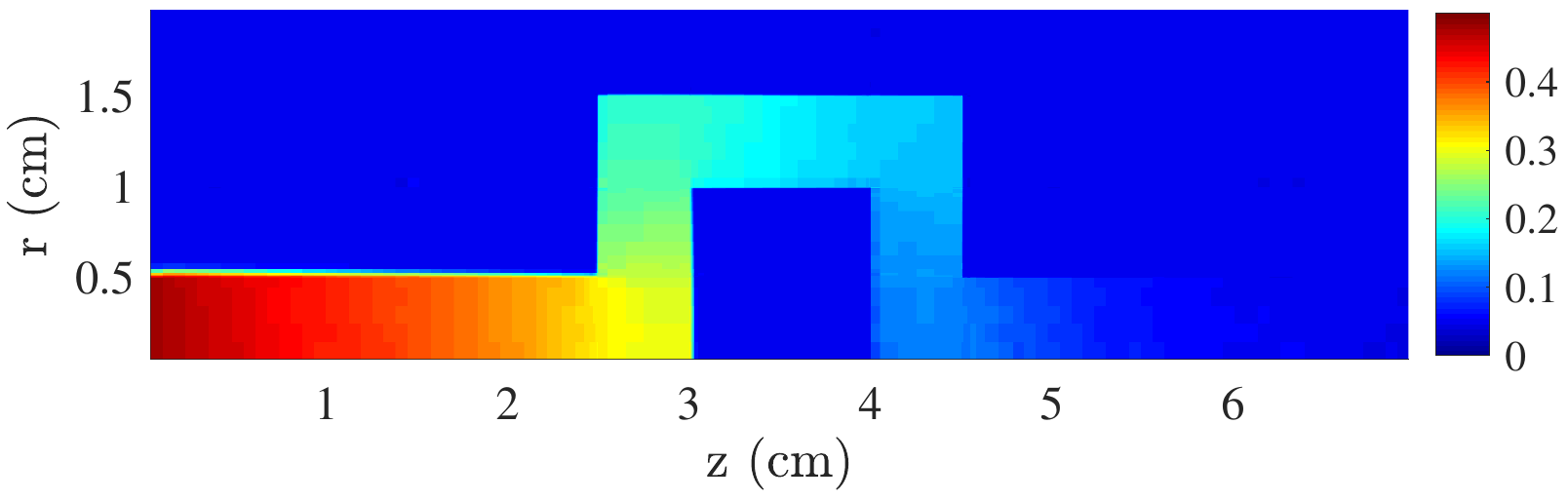}
(d)\includegraphics*[width=7.5cm]{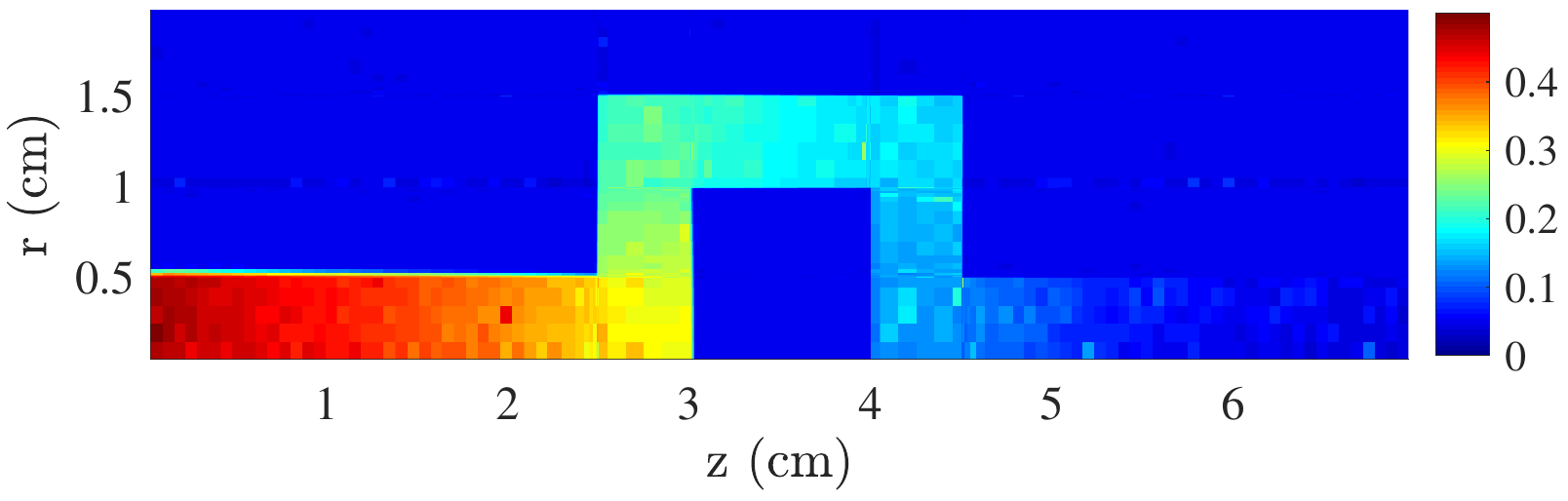}
(e)\includegraphics*[width=7.5cm]{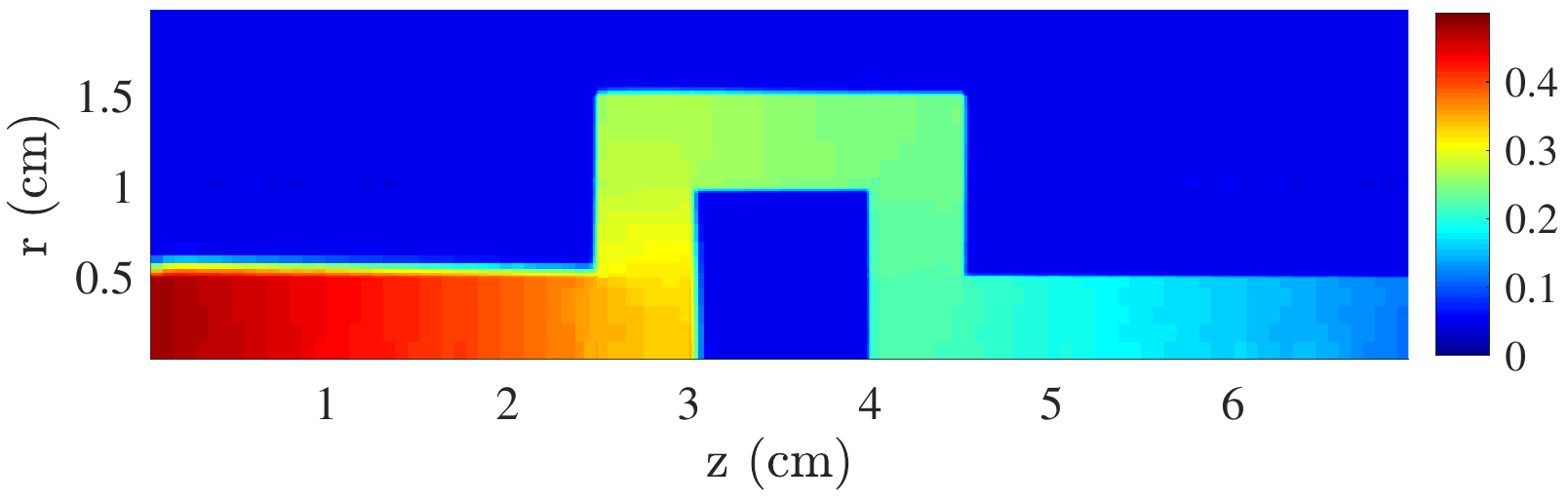}
(f)\includegraphics*[width=7.5cm]{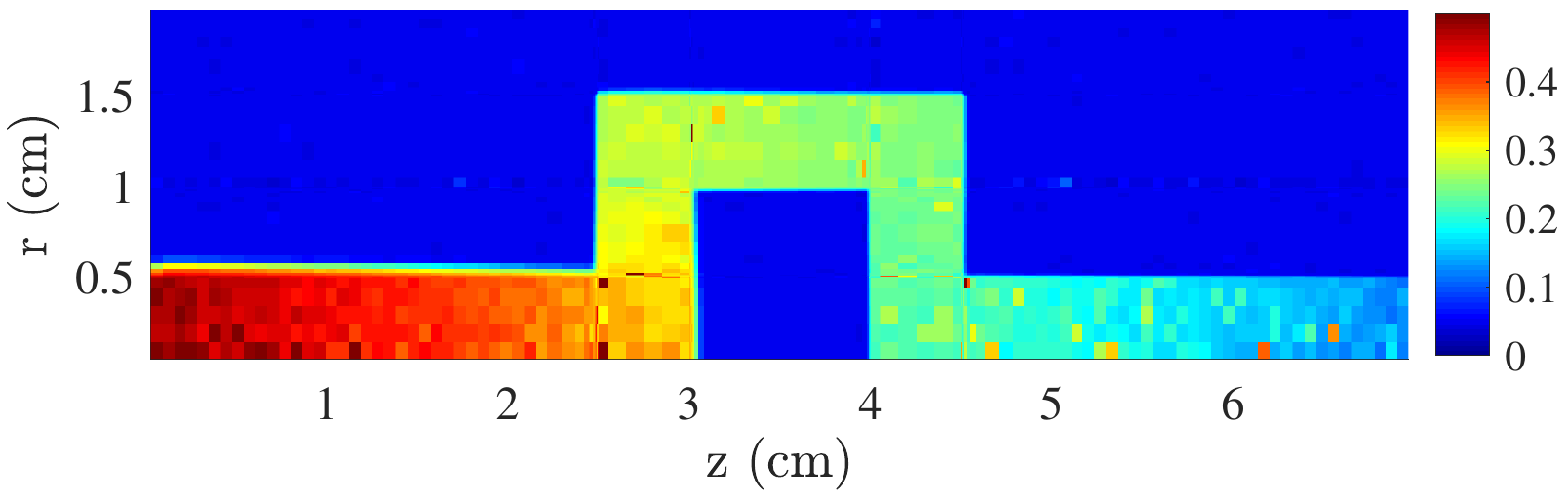}
\caption{The material temperature for the Graziani ``crooked-pipe'' problem using our high spatial resolution for IMC ((a), (c) and (e)) and ISMC ((b), (d) and (f)) at times $t=10^{-8}\text{ s}$, $t=5\cdot 10^{-8}\text{ s}$ and $t=2\cdot 10^{-7}\text{ s}$ respectively.}
\label{fig:pipe_colormap}
\end{figure}

\begin{figure}
\centering
(a)\includegraphics*[width=7.5cm]{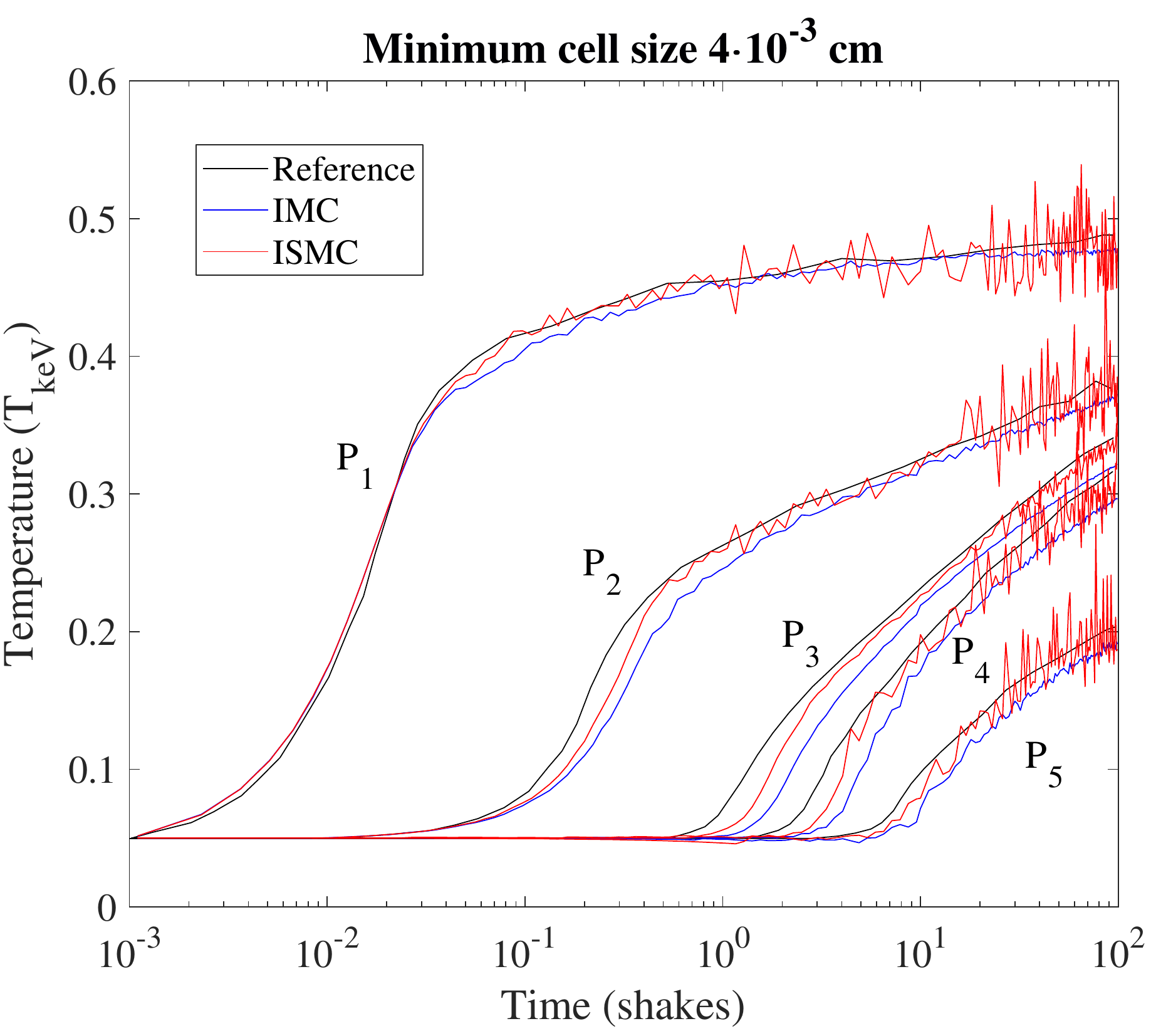}
(b)\includegraphics*[width=7.5cm]{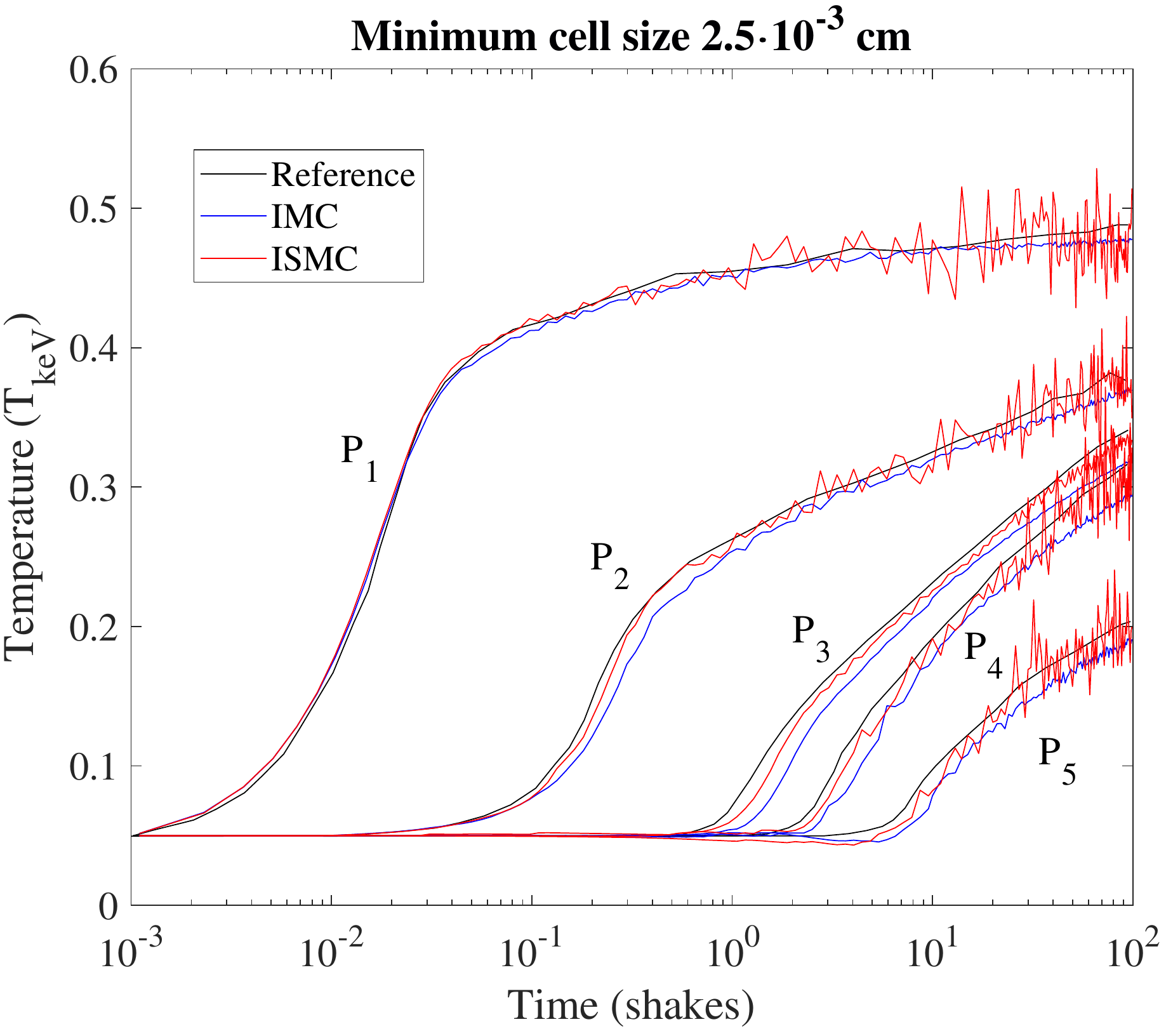}
(c)\includegraphics*[width=7.5cm]{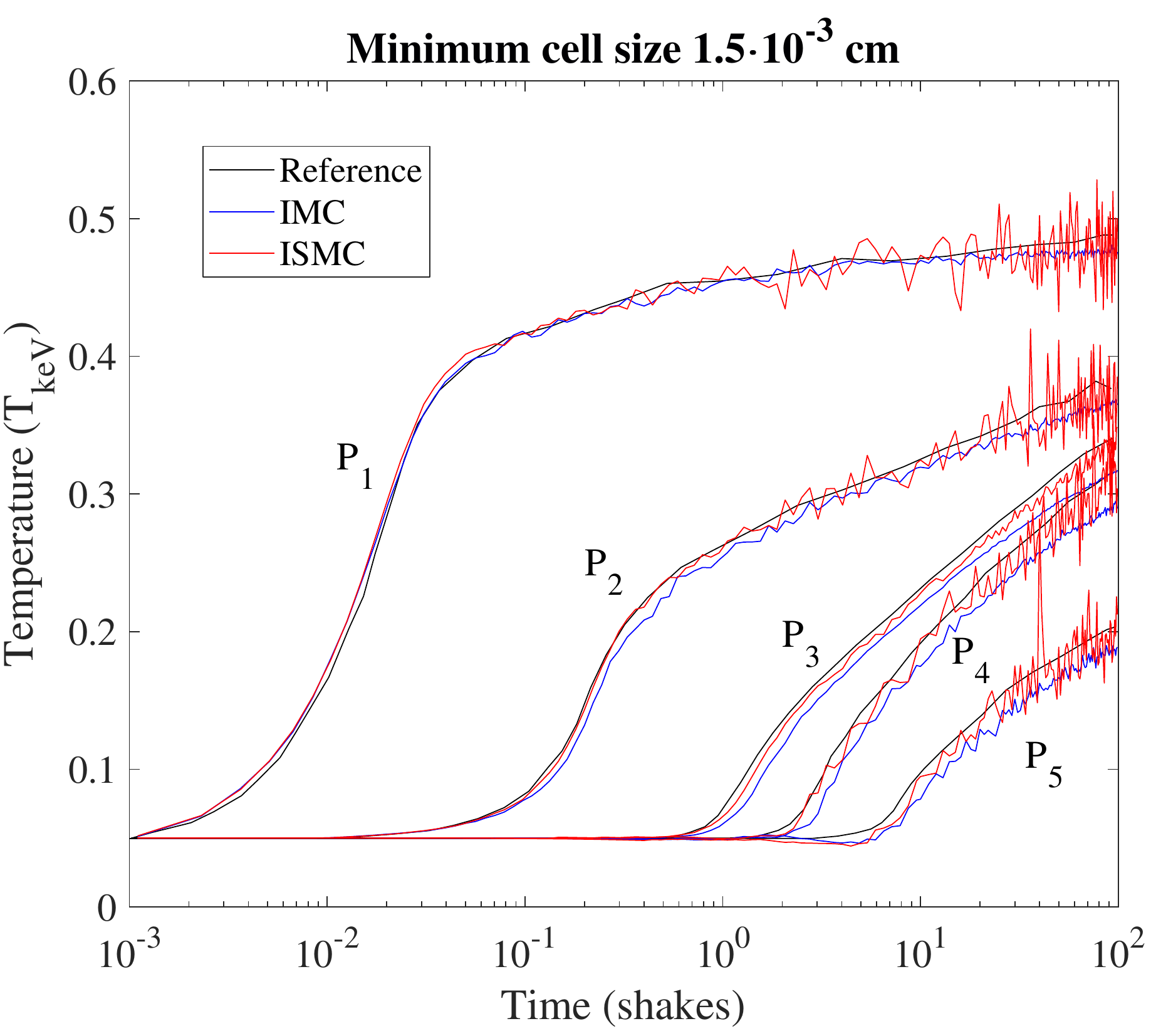}
\caption{The material temperature as a function of time for the five reference points, as described in the text. The results for IMC are shown in blue, ISMC in red and the reference solution taken from \citet{Gentile2001} in black. The temperature is shown for grids in which the smallest cell in the boundary layer between the high opacity and low opacity materials, is $1.5\cdot 10^{-3}\text{ cm},\;2.5\cdot 10^{-3}\text{ cm}$ or $4\cdot 10^{-3}\text{ cm}$ ((a), (b) and (c)).}
\label{fig:pipe_result}
\end{figure}

In Fig.~\ref{fig:pipe_colormap} we can see several snapshots of the radiation heat flow through the crooked pipe. We can see that the ISMC result is noisier than the IMC result. However, IMC requires a finer spatial resolution in order to converge.
The material temperature as a function of time is measured at five different points: $(r=0,\;z=0.25),\;(r=0,\;z=2.75),\;(r=1.25,\;z=3.5),\;(r=0,\;z=4.25)$ and $(r=0,\;z=6.75)$. In Fig.~\ref{fig:pipe_result} the material temperature profiles are presented for the different points, in the two edges of the pipe and around the optically-thick inner bypassed square, for the IMC and ISMC schemes and for different minimum cell sizes. When we reduce the spatial resolution (larger cell sizes), the IMC scheme suffers from both teleportation errors and finite discretization errors, while the ISMC does not suffer from any teleportation errors, and the temperatures profiles are converged even the low resolution (larger cell sizes), especially in points P2-P4 (around the inner square). Nevertheless, even when the ISMC scheme utilizes more than an order of magnitude more particles than IMC, it is still very noisy compared to IMC.

\subsection{Olson 2020 2D}
\begin{figure} 
\centering
(a)\includegraphics*[width=7.5cm]{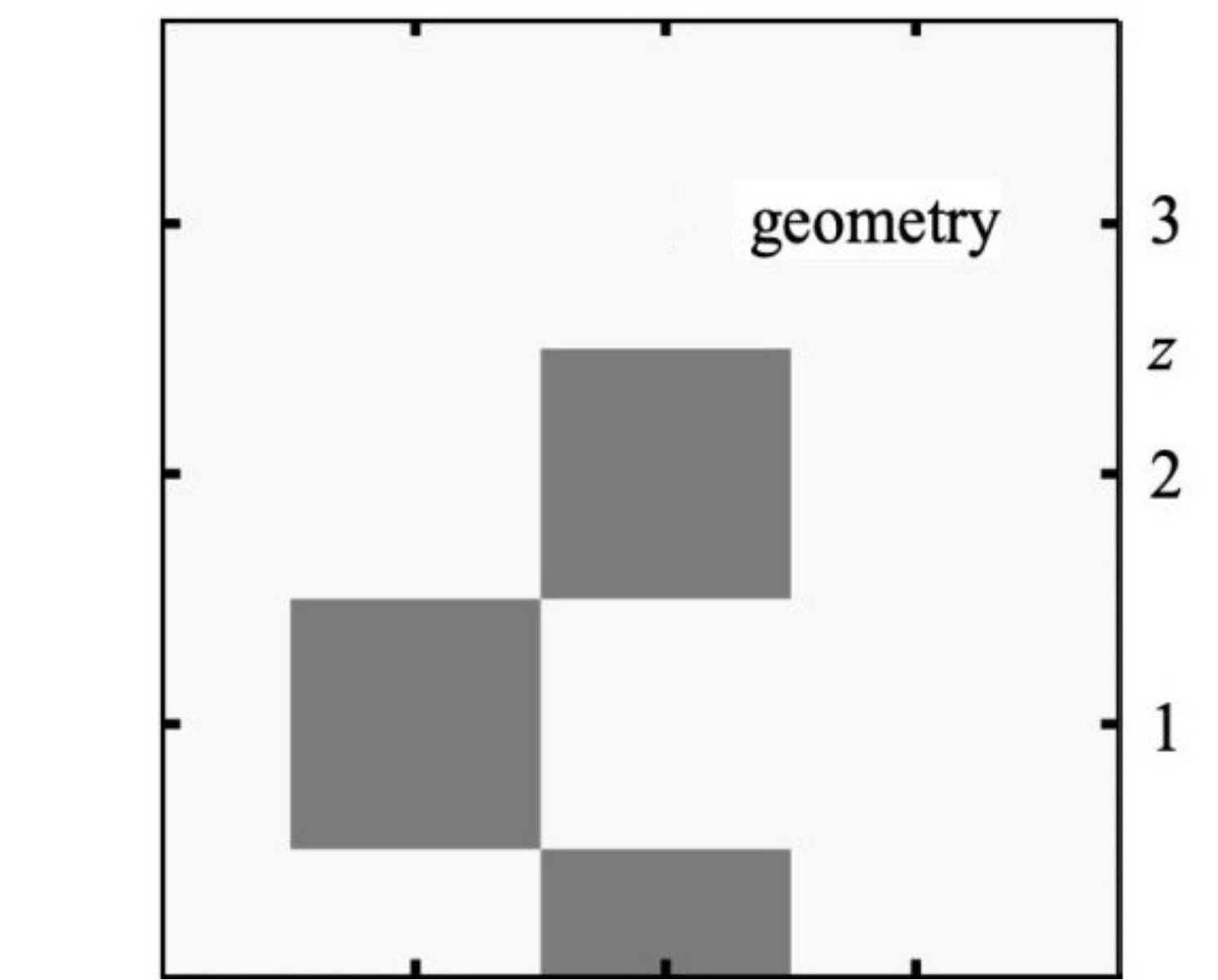}
(b)\includegraphics*[width=7.4cm]{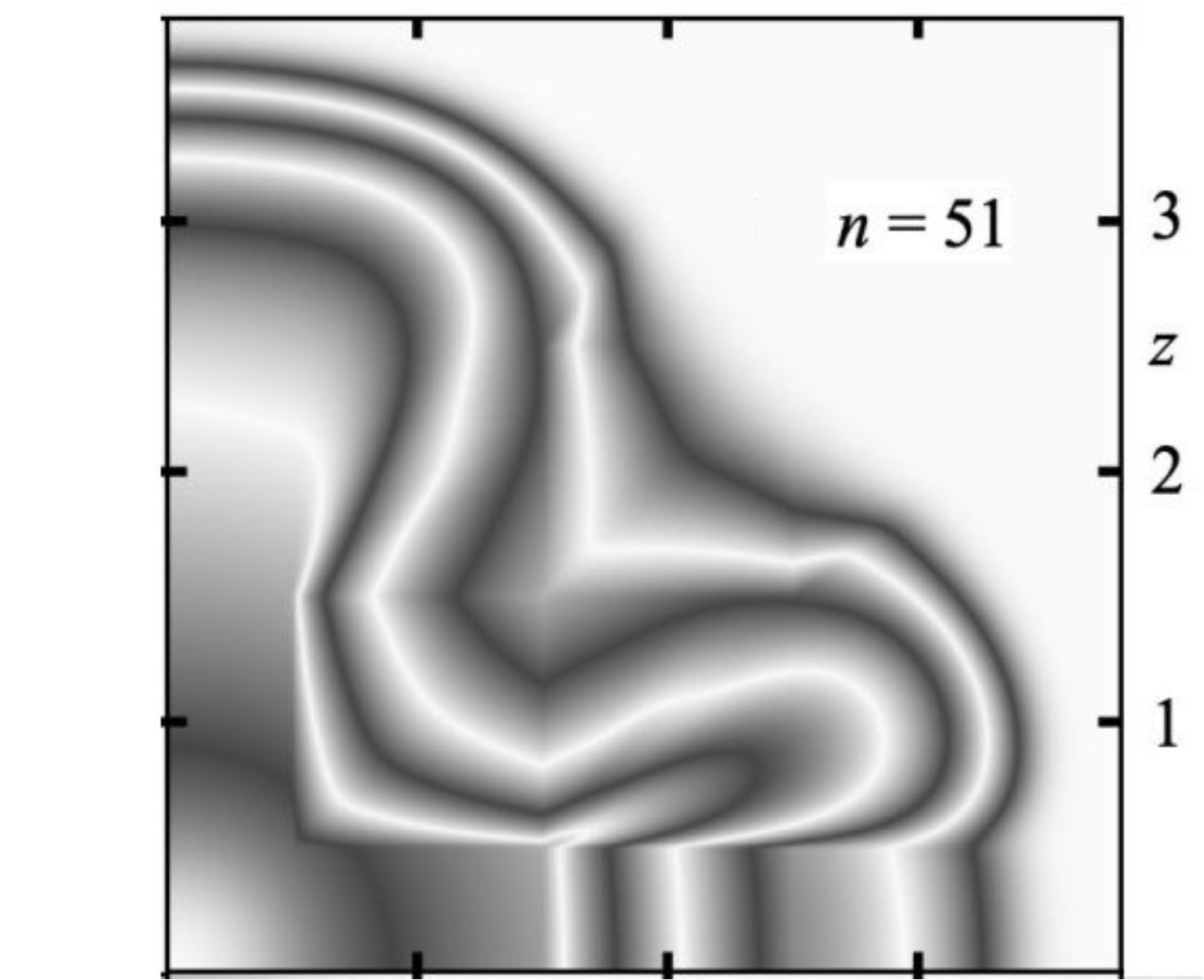}
(c)\includegraphics*[width=7.5cm]{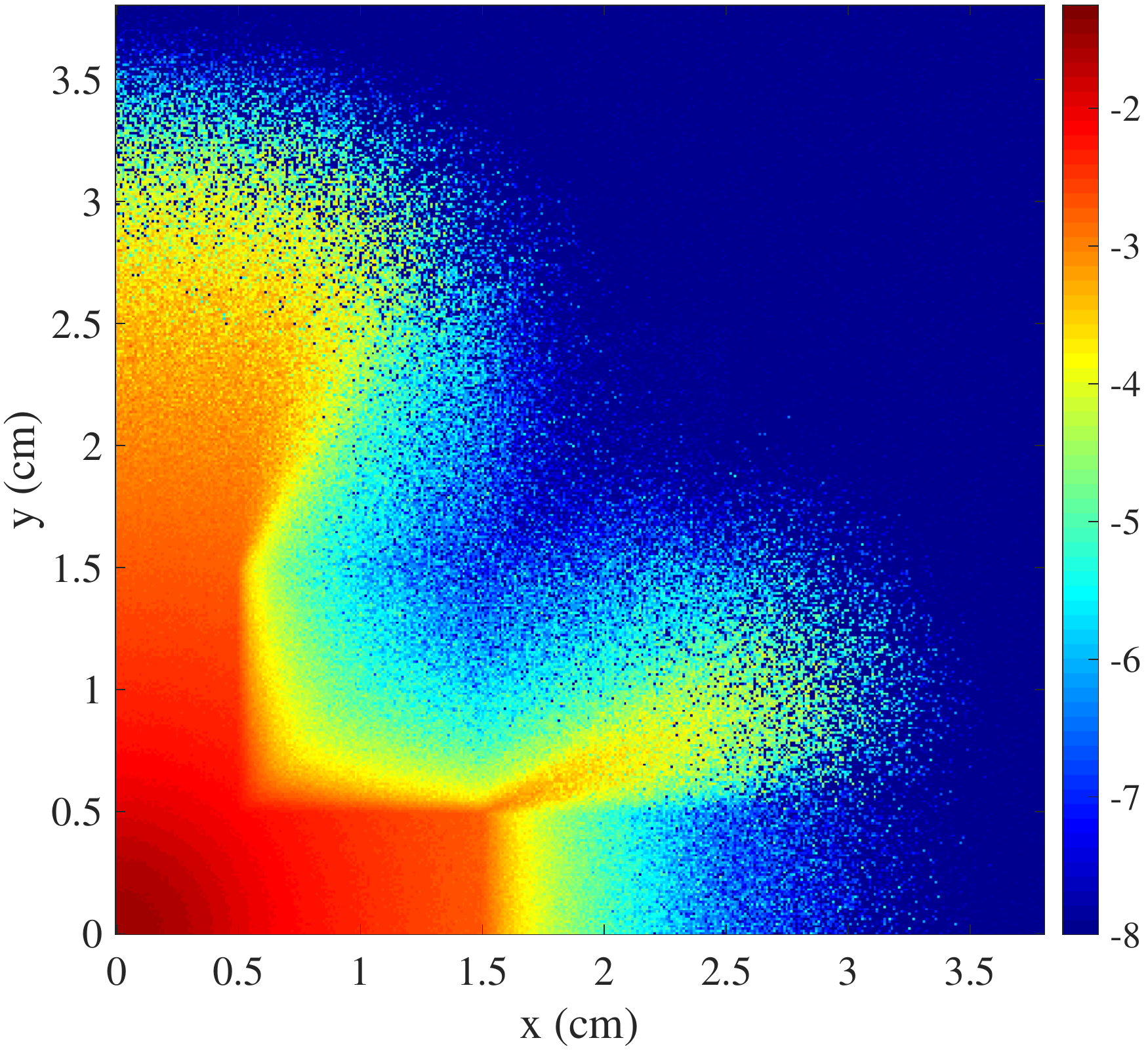}
(d)\includegraphics*[width=7.5cm]{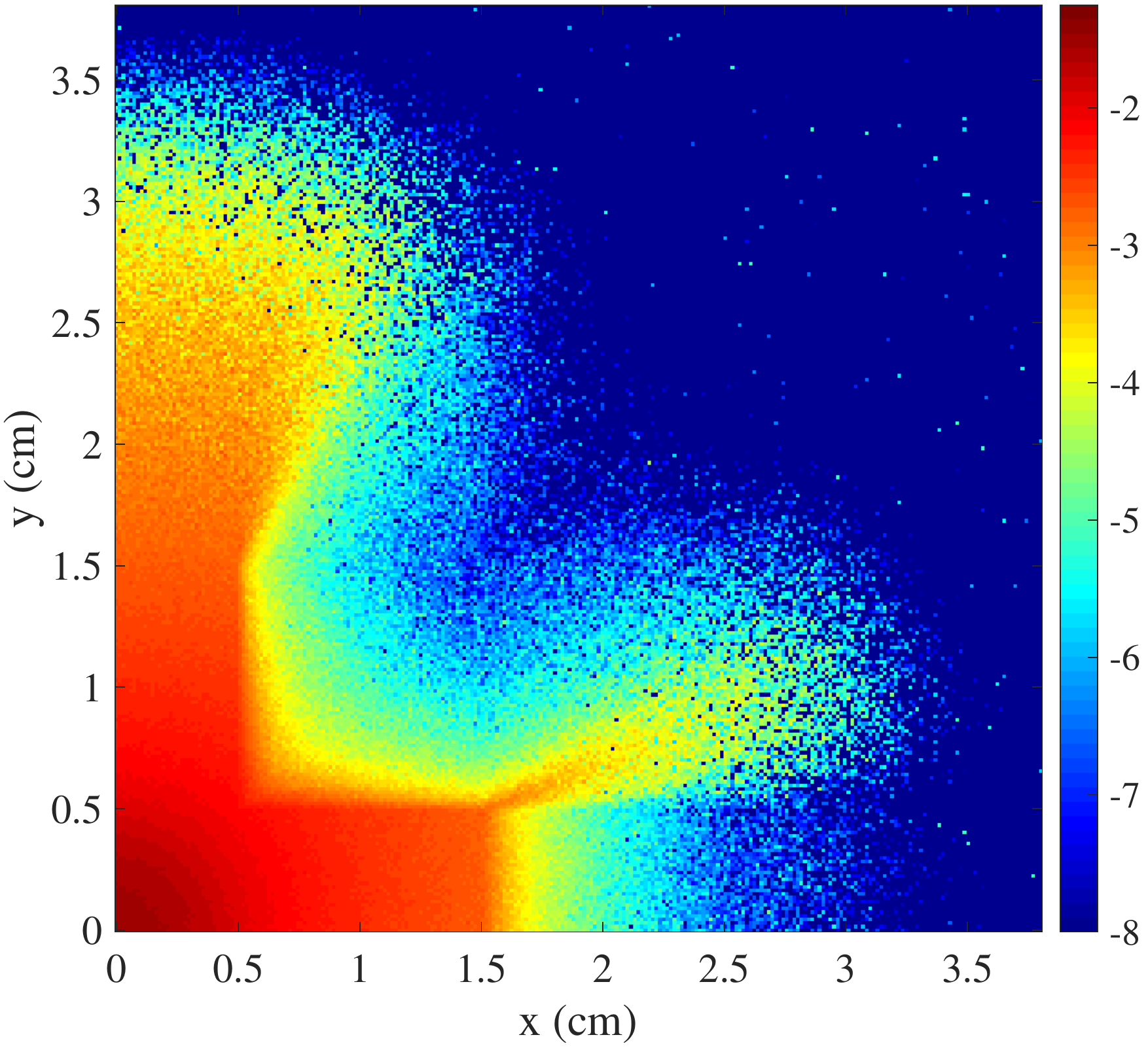}
\caption{(a) The geometrical setup to the 2D multi-frequency benchmark of~\citet{olson2020}. The gray squares represent opaque regions. (b) Contours of the logarithm of the radiation energy density at $ct=3$ using $P_{51}$ approximation. Both (a) and (b) are taken from~\cite{olson2020}. (c) The radiation energy density at $ct=3$ using IMC method. (d) Same for ISMC method.}
\label{fig:olson20202D_rad}
\end{figure}

We close with a multi-frequency implementation in a full two-dimensional problem, using the most complex 2D problem presented in \citet{olson2020}. In this problem, a lattice is composed of (relatively-) opaque aluminum blocks surrounded by the same foam from the previous 1D section; the geometry is shown in Fig.~\ref{fig:olson20202D_rad}(a). The aluminum blocks have the same density as the foam and their heat capacity has the same functional form as the foam (Eq.~\ref{heat_cap}), but with $H=0.5$ and $\chi=0.3T_\text{ keV}$. The opacity for the aluminum blocks is given by:
\begin{equation}
    \kappa_{a,\nu}=\begin{cases}
    \text{min}(10^7,10^8T/T_\text{ keV})& h\nu<0.01\text{ keV}\\
    \frac{10^7 \left(0.01\text{ keV}/h\nu\right)^2}{(1+20\cdot(T/T_\text{ keV})^{1.5})}& 0.01\text{ keV}<h\nu<0.1\text{ keV}\\
     \frac{10^7 \left(0.01\text{ keV}/h\nu\right)^2}{(1+20\cdot(T/T_\text{ keV})^{1.5})} + \frac{10^6\left(0.1\text{ keV}/h\nu\right)^2}{1+200\cdot(T/T_\text{ keV})^2}& 0.1\text{ keV}<h\nu<1.5\text{ keV}\\
    \frac{10^7 \left(0.01\text{ keV}/h\nu\right)^2\sqrt{1.5\text{ keV}/h\nu}}{(1+20\cdot(T/T_\text{ keV})^{1.5})} + \frac{10^5\left(1.5\text{ keV}/h\nu\right)^{2.5}}{1+1000\cdot(T/T_\text{ keV})^2}&
    h\nu >1.5\text{ keV}.
    \end{cases}
\end{equation}

The computational region is a square 3.8 cm on each side, divided into 380 equally spaced cells with reflecting boundaries. A black body source, $Q(r)=B(0.5\text{ keV})\exp{(-18.7\cdot r^3)}$, is turned on at time $t=0$, and remains on for the entire duration of the simulation. A constant time step of $\Delta t = 10^{-13}$ s is used, $2\cdot 10^6$ new particles are created each time step, and we limit the total number of particles to be $2.5\cdot 10^7$.

\begin{figure} 
(a)\includegraphics*[width=7.5cm]{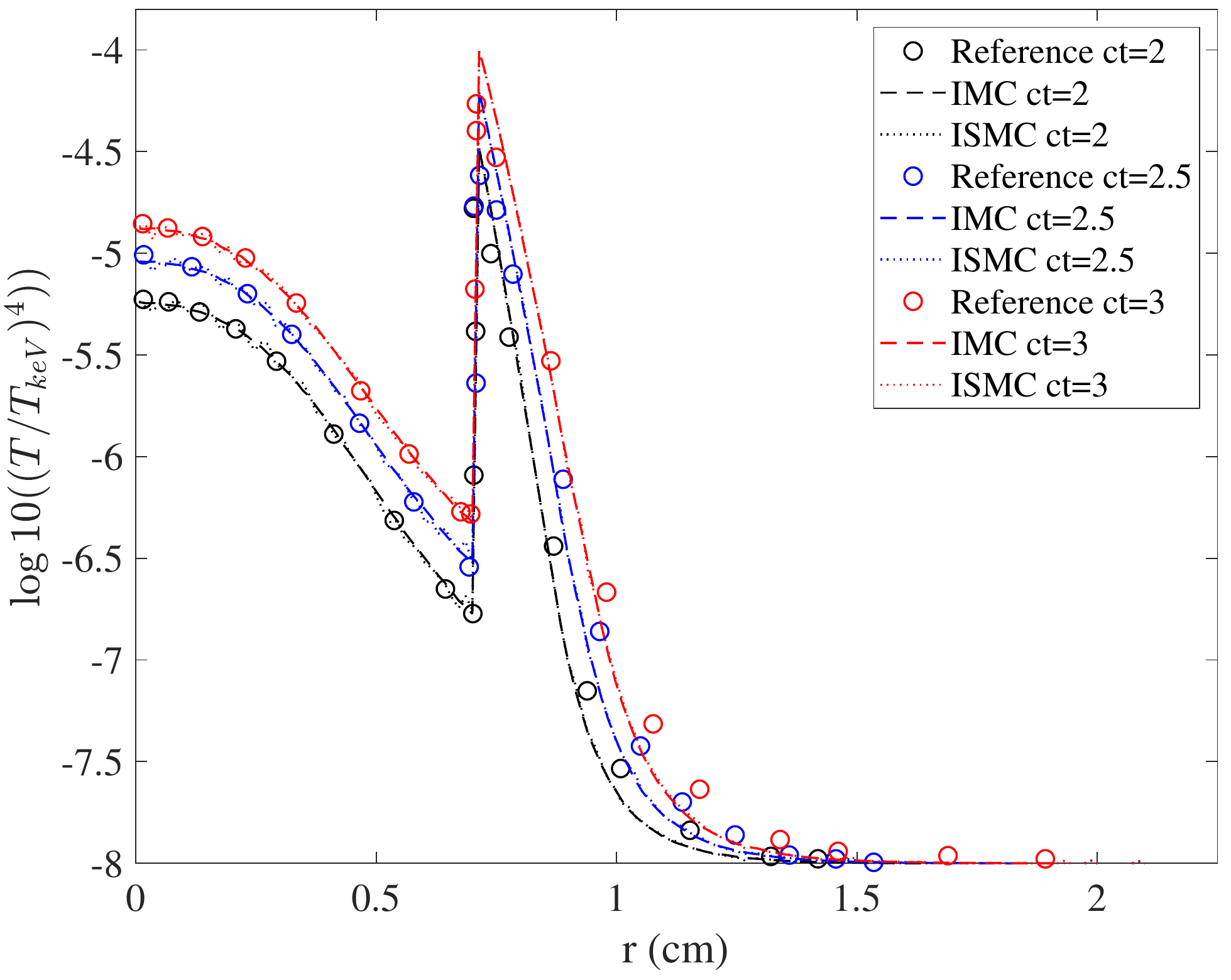}
(b)\includegraphics*[width=7.5cm]{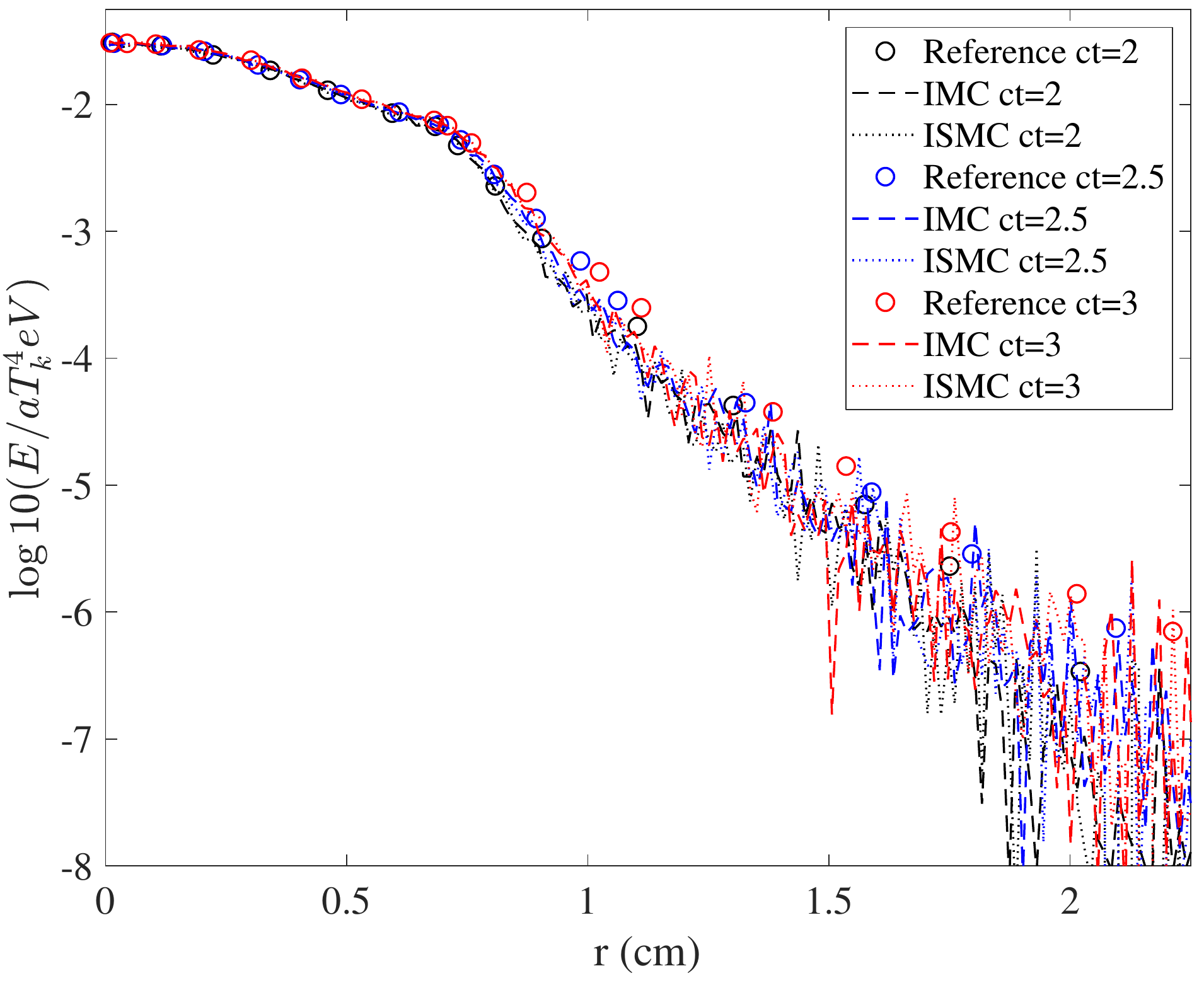}
\caption{(a) The material temperature at different times for the \citet{olson2020} 2D test problem. (b) The radiation energy density at different times for the \citet{olson2020} 2D test problem.}
\label{fig:olson20202D}
\end{figure}

First, in Fig.~\ref{fig:olson20202D_rad}(c-d) we show a colormap of the radiation energy density in the entire domain as well as a reference from~\cite{olson2020} (Fig.~\ref{fig:olson20202D_rad}(b)). Both MC methods agree very well with each other through the computational domain, and a good qualitative agreement with the $P_N$ reference data, including all geometrical patterns. Next, Fig.~\ref{fig:olson20202D} shows slices along the diagonal for (a) the material temperature and (b) the radiation energy density, along with reference $P_N$ data from~\cite{olson2020}. Again, there is a good agreement between all methods in the range $0\leqslant r\leqslant \sqrt{0.5}$. Farther away from the origin, the sharp discontinuity in the material properties causes a small difference between the MC methods and the reference solution. 

\section{Summary}
\label{discussion}
The ISMC method presented recently by \citet{ISMC} is a significant milestone in the field of MC radiation transfer. While the SMC method that was presented in \citet{SMC} was very successful in eliminating teleportation errors, its explicit nature imposed very severe time step restrictions. The implicitization presented in~\cite{ISMC} transformed the potential usability of the SMC method.

We have shown the major problem of teleportation using classic IMC implementation, both in finite spatial resolution and time step, in various optically thick materials in two-dimensional problems. As a result, the heat wave penetrates too deeply inside the opaque material, causing an increased ``leakage of energy'' from the optically-thin materials/vacuum. The ISMC algorithm is immune to this issue in all dimensions, and can converge faster than IMC. We have shown that the ISMC method is stable even with very large time steps, and suffers from no teleportation errors.

We benchmark the ISMC algorithm in a large number of well-known 1D and 2D problems, both gray and multi-frequency problems. In all of the problems, the ISMC tends to yield converged solutions much faster than IMC in both spatial resolution and larger time steps. In optically thin problems the results are comparable to IMC, while for opaque problems and/or problems with large spatial gradients in the opacity, in which standard IMC struggles with and requires high spatial resolution, the benefit of ISMC is clear and it is more efficient and requires significantly less spatial resolution. ISMC also has the added advantage, that with the absence of external source terms, the total number of particles is fixed, and no population control methods are needed. The run time of ISMC is comparable to that of IMC (at an equal number of particles/photons), both in optically thin and optically thick problems.

We note that in optically thin regions, standard IMC has less statistical noise since it gives a smoother material temperature profile at a given particle number. This main drawback of ISMC arises from the discretization of the emission and absorption processes, that introduce statistical noise, that can be reduced by increasing the number of particles.

\citet{implicit-time} have shown that various Monte-Carlo algorithms for radiation transport have a varying level of implicitness, that affects the stability of the solution regarding to large time steps. For example, even the basic IMC algorithm, is not fully implicit, where the opacity and the heat capacity ($\beta$) is determined from the values of the temperature at the beginning of the time step. Thus, a well-known instability occurs using (extreme) large time-steps, called "maximum principle violation", where the temperature is higher than the boundary temperature (and at later times, causes oscillatory solutions)~\cite{implicit-time,four_decades}. \citet{implicit-time} have showed that this instability disappears when using a fully implicit scheme (which is of course more complicated, due the non-linear dependence of the opacity and heat-capacity in the temperature). Here, the ISMC implementation is also not fully implicit, taking the opacity and $\zeta$ at the beginning of the time-step.
Thus, in all of the above benchmarks, we have found that the magnitude of the time step that led to maximum principle violations in IMC, was comparable to the one that gave rise to the same instability in ISMC, \emph{i.e.} empirically, IMC and ISMC have very similar tolerance for maximum principle violations.

Overall we show that the ISMC is a very viable method for MC radiation transport, and has significant advantages over IMC in optically thick regions and regions with large spatial opacity gradients.

\section*{Acknowledgement}
We would like to thank Gaël Poëtte for taking the time to read and comment on this paper, as well as fruitful discussions, and for amazing presentation on the ISMC algorithm in the \nth{26} International Conference on Transport Theory (ICTT-26) in Sep. 2019, Paris.
\bibliographystyle{cas-model2-names}
\bibliography{main.bib}
\end{document}